\documentclass[12pt,preprint]{aastex}
\usepackage{graphics,graphicx}
\usepackage{subfigure}
\usepackage{epsfig}
\usepackage{appendix}
\usepackage{lscape}
\usepackage{longtable}
\usepackage{url}
\usepackage{bigstrut,bigdelim,multirow}

\shorttitle{Short GRBs catalog of {\em Fermi} data}
\shortauthors{Lu et al}
\slugcomment{}
\begin{document}

\title{{\em Fermi}/GBM Short Gamma-ray Burst Catalog and Case Study for GRB 170817A/GW 170817}
\author{Rui-Jing Lu$^{\ast}$, Shen-Shi Du, Ji-Gui Cheng, Hou-Jun L\"{u}$^{\ast}$, Hai-Ming Zhang, Lin Lan, and En-Wei Liang$^{\ast}$} \affil{Guangxi Key Laboratory for Relativistic Astrophysics, Department of Physics, Guangxi University, Nanning 530004, China; luruijing@gxu.edu.cn; lhj@gxu.edu.cn; lew@gxu.edu.cn}

\begin{abstract}
Motivating by the discovery of association between GW 170817 and sGRB 170817A, we present a comprehensive analysis for sGRBs observed with Fermi/GBM in 9 operation years and study the properties of sGRB 170817A -like events. We derive a catalog of 275 typical sGRBs and 48 sGRB 170817A-like weak events from the GBM data of 2217 GRBs. We visibly identify two patterns of their light curve, single episode (Pattern I, 61\% of the SGRBs) and multiple episodes (Pattern II, 39\% of the SGRBs). Their duration distribution shows a tentative bimodal feature. Their spectra can be fitted with a cutoff power-law model, except for 4 sGRBs, and the spectral indices normally distribute at $\Gamma=0.69\pm 0.40$. Their $E_p$ values show a tentative bimodal distribution with peaks at 145 keV and 500 keV. No correlation among $T_{90}$, $E_p$, and $\Gamma$ is found. GRB 170817A is a soft, weak sGRB with $ E_{p}=124\pm 106$ keV, $L_{\rm iso}=(5.67\pm4.65)\times10^{46}\rm ~erg~s^{-1}$, and $E_{\rm iso}=(3.23\pm2.65)\times10^{46}\rm ~erg$. It follows the $E_{\rm iso}-E_{\rm p}$ relation of typical short GRBs. Its lightcurve is of Pattern II. Two lightcurve patterns, together with the potential two components in the $E_{\rm p}$ and $T_{90}$ distributions, we suspect that the current sample may include two distinct types of sGRBs from different progenitors. sGRB 170817A-like events may be from NS-NS mergers and those sGRBs with a Pattern I lightcurve may be from another distinct type of compact binary.
\end{abstract}


\section{Introduction}
It is believed that the progenitors of short gamma-ray bursts
(sGRBs) are coalescence of binaries of compact objects (Nakar 2007), i.e.,
neutron star$-$neutron star mergers (NS$-$NS, Pacz\'ynski 1986;
Eichler et al. 1989), or neutron star$-$black hole mergers
(NS$-$BH, Pacz\'ynski 1991). This was supported by observations
of their afterglows and host galaxies in the {\em Swift}
mission era (Gehrels et al. 2005; Barthelmy et al. 2005; Berger
et al. 2005; Fong et al. 2010). The coalescence of two compact
stars, on the other hand, may also power high frequency
gravitational wave (GW), which was directly detected by the
Laser Interferometer Gravitational wave Observatory (LIGO) on
September 14, 2015 (GW 150914£» Abbott et al. 2016a,b). During
the operation of advanced LIGO/Virgo in the last two years,
four GW events were convincingly detected, i.e., GWs 150914,
151226, 170104 and 170814), and one plausible candidate, LVT
151012, was also reported. All of those GW events are believed
to be from mergers of black hole (BH) binary with tens of solar
masses (Abbott et al. 2016a,b, 2017a, b).

GW, together electromagnetic (EM) radiation, now presents a new
probe for studying the universe. Searching for possible EM
counterparts of GW events is one of the hottest topic since the
discovery of GW 150914 (Abbott et al. 2016a). The
progenitor of GW 150914 is a BH-BH binary. It is highly
debating whether such a binary system can produce an EM
counterpart (Zhang et al. 2016; Connaughton et al. 2016; Perna
et al. 2016; Zhang 2016). Differing from merger of a BH-BH
system, prominent EM counterparts of NS-NS and NS-BH mergers
are highly expected (Metzger \& Berger 2012, for review), such
as short GRBs of their relativistic jets, multiple wavelength
afterglows when their jets propagate into surrounding medium
(Berger 2014, for review), sGRB-less X-ray counterpart (Sun et
al. 2017), and kilo/merger-nova in the optical/infrared band
(Li \& Pacz\'ynski 1998; Metzger et al. 2010; Yu et al. 2013;
Berger et al. 2013; Tanvir et al. 2013; Yang et al. 2015; Gao
et al. 2017). It was also proposed that mergers of double neutron stars may be also
related to the fast radio bursts (FRBs), which is a radio
transient with duration of milliseconds at cosmological
distances (Lorimer et al. 2007; Thornton et al. 2013; Totani
2013; Zhang 2014; Keane et al. 2016; Zhang 2016; Dai et al.
2017). Searching for connections between GW candidates and
SGRBs have been done by the LIGO collaboration team since the
operation of LIGO (Abadie et al. 2010, 2011, 2012; Abbott et
al. 2016c).

Very interestingly, aLIGO/Virgo recently detected GW 170817,
which was proposed to be produced by a NS-NS meger at a
distance away from us about 40 Mpc (Abbott et al. 2017c). Its
EM counterpart was also detected in the X-ray, optical, and
radio bands (Troja et al. 2017; Nicholl et al. 2017; Fong et
al. 2017; Hallinan et al. 2017), and a weak sGRB 170817A
(Goldstein et al. 2017) may be associated with GW 170817.
Limits on the MeV band of this event was also made by Hard X-ray Modulation Telescope(HXMT; Li
et al. 2017).

This paper dedicates to present a systematical analysis on the sGRB observed with {\em Fermi}/GBM since its operation in 2008 and investigate whether GRB 170817-like
events are distinct from typical sGRBs. Our data reduction are
presented in \S 2. We report an Fermi/GBM sGRB catalog of 9
operation years with our criteria and present its statistical
properties in \S 3. In \S 4, we compare the properties of GRB
170817A with the global sGRB catalog and make deep search for
such kind of weak sGRB events in the GBM data archive.
conclusions are drawn in section 5 with some discussions.
Throughout the paper, a concordance cosmology with parameters
$H_0=71~\rm km~s^{-1}~Mpc^{-1}$, $\Omega_M=0.30$, and
$\Omega_{\Lambda}=0.70$ is adopted.

\section{Data reduction and Sample Selection}
The {\em Fermi} satellite was launched in June 2008 and has
been operated for more than 9 years. There are two instruments
onboard the {\em Fermi} satellite. One is the Gamma-ray Burst
Monitor (GBM; Meegan et al. 2009), which has 12 sodium iodide
(NaI) and two bismuth germanate (BGO) scintillation detectors which
covers an energy band from 8 keV to 40 MeV. Another one is the Large
Area Telescope (LAT; Atwood et al. 2009), which has an energy
coverage from 20 MeV to 300 GeV. LAT Low
Energy data (LLE; 30-100 MeV) that are produced from a
non-standard LAT analysis by the means of increasing the
effective area of the LAT at low energy 30-100 MeV are also available for some GRBs.

We download the GBM and LAT data as well as the LLE data for
all GRBs from the public science support center at the official
{\em Fermi} web
site\footnote{ftp://legacy.gsfc.nasa.gov/fermi/data/gbm/bursts.}
and {\em Fermi} Archive
FTP\footnote{ftp://legacy.gsfc.nasa.gov/fermi/data/lat/triggers/.}.
A {\em PYTHON} code was developed to extract the
energy-dependent lightcurves and spectra by using the package
$gtBurst$\footnote{http://sourceforge.net/projects/gtburst/.}.
We separate the NaI and BGO detectors into two energy bands,
respectively (e.g., [8, 50] keV and [50, 1000] keV for NaI;
[250, 1000] keV and [1000, 40000] keV for BGO). We employ the
Bayesian Block algorithm to identify the lighjtcurves. Our
procedure is described as following:

(a) Firstly, we extract lightcurve with a time-bin of 64
millisecond to identify a possible signal in different energy
bands in the time interval of [-50, 200] s in order to search for possible precursors and extended emission of a sGRB, where the GBM
trigger time is set as 0. The number that controls blocks created in the \emph{gtburstfit} tool (Scargle 2013)
is normally adopted as 3 for picking up a weaker signal in our
analysis.

(b) If a possible Bayesian block can be identified, we
calculate the $T_{90}$ of the bursts in the
NaI energy band by using the Bayesian block method. Then, we
adopt a 16 ms time-bin, instead of the 64 ms time-bin, to  extract
the lightcurve again in the time interval [-2, 5]s for deriving the details of the temporal
structures of these sGRB. In case of that possible precursor and/or extended emission episodes is found, the time interval is accordingly broadened to include these structures.

(c) We extract the spectra of these events in the burst duration with the NaI and BGO
data (as well as LLE data if available) by using the
\emph{gtBurst} tool. {\tt XSPEC} is used to perform joint
spectral fits of the GBM and LAT data. The statistic {\tt
PGSTAT} is adopted to judge the goodness of the spectral
fits. A cutoff power-law model (CPL) is adopted as the
primary model in our fits, which is written as,
\begin{eqnarray}
N(E) = A\cdot E^{-\alpha}\exp(-\frac{E}{E_{\rm p}}).
\end{eqnarray}
In case that the CPL model cannot give reasonable fit to the high energy data, the Band function (Band) is used to fit the spectra (Band et
al. 1993),
\begin{eqnarray}
N(E)=B(E)=A\cases{(\frac{E}{100 \rm~keV})^{\alpha}\rm exp(-\frac{E}{E_0}),
            & $E<E_b$ \cr
[\frac{(\alpha-\beta)E_0}{100 \rm~keV}]^{\alpha-\beta} \rm exp(\beta-\alpha)(\frac{E}{100 \rm~keV})^{\beta},
            & $E\geq E_b$ \cr            }
\end{eqnarray}
where $\alpha$ and $\beta$ are the low and high energy photon
spectral indices, respectively; $E_b$ is the break energy in
the spectrum, and the peak energy ($E_{p}$) of spectrum is
related to $E_b$ by \begin{eqnarray} E_p = (2+\alpha) E_{b}=
(2+\alpha) E_{b}/(\alpha-\beta).
\end{eqnarray}
An extra power-law component ($N(E) = A \cdot
E^{-\lambda}$) are also needed to fit the spectra for few sGRBs that were observed with LAT.

During the 9 year operation (from August 2011 to August 2017),
2217 GRBs were triggered by {\em Fermi}/GBM. GRB duration is
energy band dependent (e.g., Qin et al. 2013). We adopt a
criterion of $T_{90}<2$ seconds in the GBM band for making our sGRB
sample similar to that in the BATSE sGRB sample. There are 275 GRBs are included. They are
listed in Table 1. Among them 7 sGRBs are detected by both GBM and LAT. The percentage of sGRBs is 13\% of the
GBM GRBs. It is less than that in the Burst and Transient Source Experiment (BASTE) catalog, which is
$\sim$(20-25)\%. This may be due to our selection effect since
we calculate the $T_{90}$ in the GBM band (8-1000 KeV), whereas
it is normally calculated in the 50-300 KeV band for BATSE GRBs. The
duration in a lower energy band tends to be longer than that in
a harder energy band (e.q., Zhang et al. 2007; Qin et al. 2013; L\"{u} et al.
2014). Therefore some GBM GRBs with $T_{90}<5$ seconds may be also classified into the
sGRB group in the 50-300 keV band.

\section{The sGRBs Catalog and Statistical Properties}

\subsection{Spacial Distribution and the log N-log P Curve}
\begin{figure}
\includegraphics[origin=c,angle=0,scale=1,width=1.00\textwidth,height=1.0\textheight]{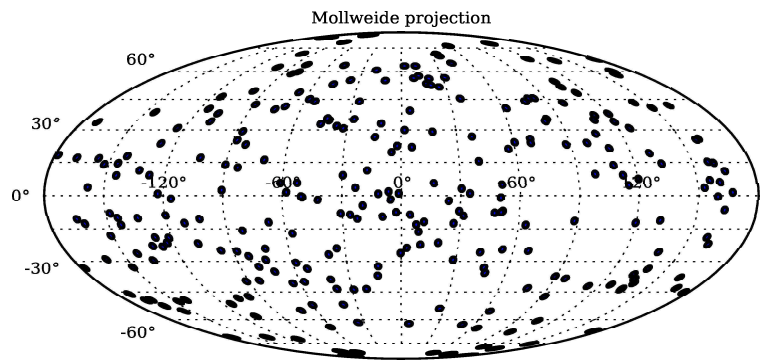}
\caption{Skymap of all short GRBs in our sample.}
\label{fig:skymapGRB}
\hfill
\end{figure}

\begin{figure}
\includegraphics[origin=c,angle=0,scale=1,width=0.8\textwidth,height=0.6\textheight]{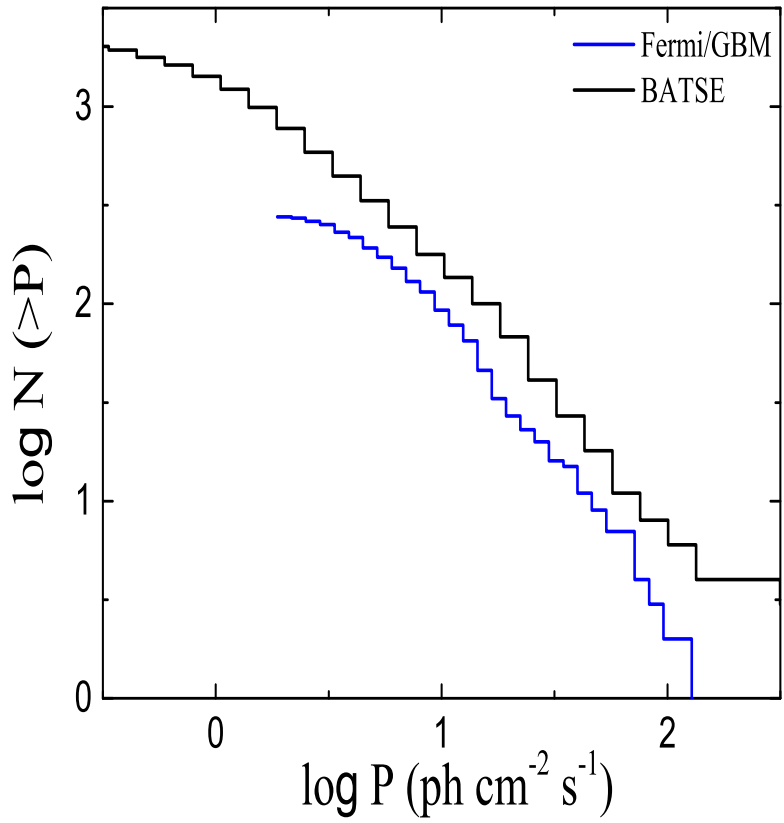}
\caption{Cumulative distribution of $\rm log~P$ for BATSE data (black line) and our sample (blue line).}
\label{fig:LogNP}
\end{figure}

Figure \ref{fig:skymapGRB} shows that the spacial distribution of the sGRB in our sample. It is found that the
distribution is isotropic. This is consistent with that observed with BATSE (e.g., Meegan et al. 1992). Figure
\ref{fig:LogNP} shows the $\rm \log N-\log P$ curve for our sample, where $\rm P$ is the photon flux in the units of
$\rm photons~cm^{-2}~s^{-1}$ in the GBM band. Note that the threshold of GBM is $\rm 0.7~photons~cm^{-2}~s^{-1}$. The
difference between GBM and {\em Fermi} at low energy may be due to the flux truncation effect and the low trigger
probability of sGRBs with a flux close to the GBM threshold (e.g. Qin et al. 2010).

\subsection{Lightcurves and Duration Distribution}
\begin{figure*}
\includegraphics[origin=c,angle=0,scale=0.6,width=0.400\textwidth,height=0.4\textheight]{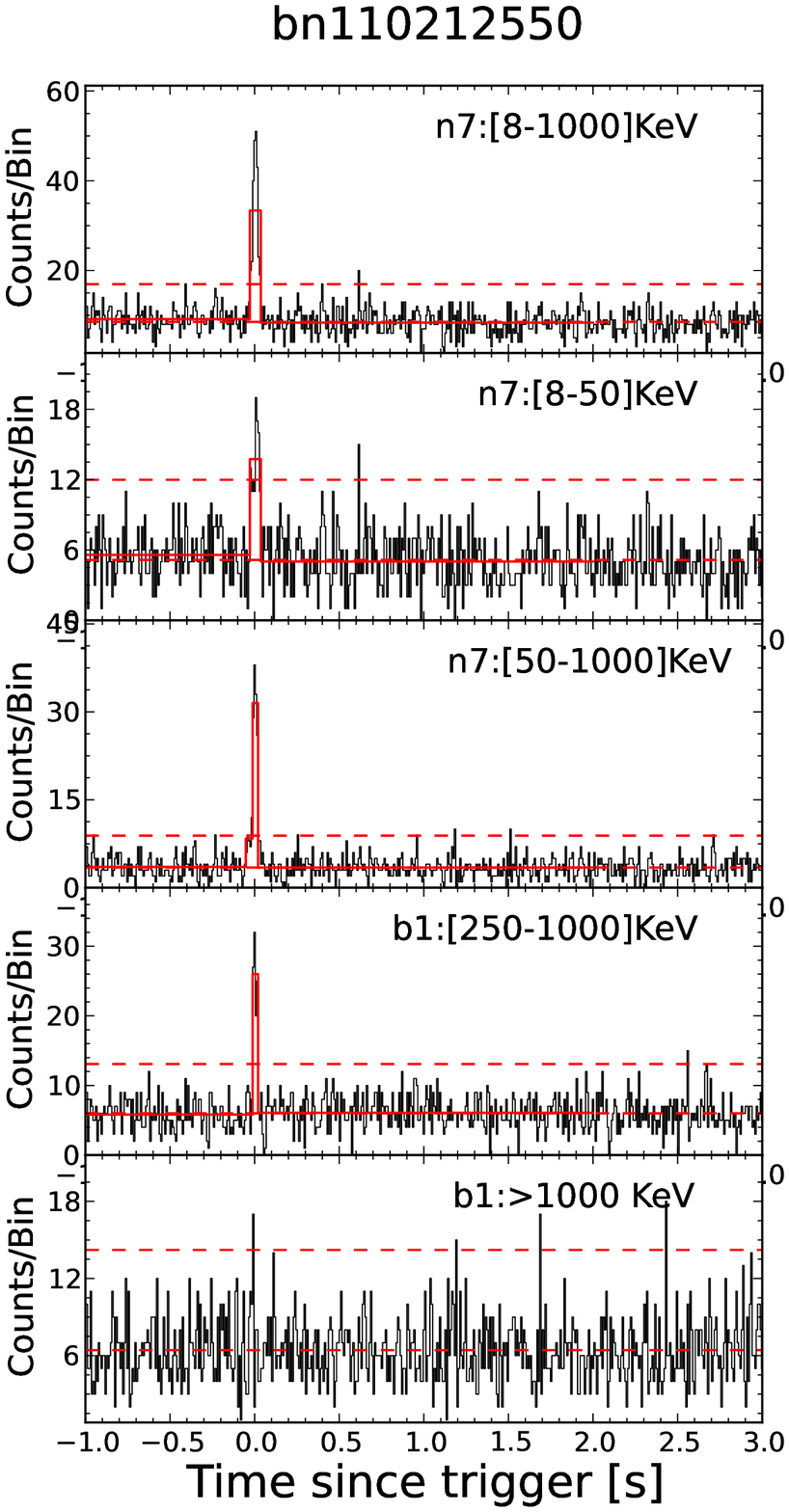}
\includegraphics[origin=c,angle=-90,scale=0.6,width=0.400\textwidth,height=0.4\textheight]{./f3a1SP.eps}
\includegraphics[origin=c,angle=0,scale=0.6,width=0.400\textwidth,height=0.4\textheight]{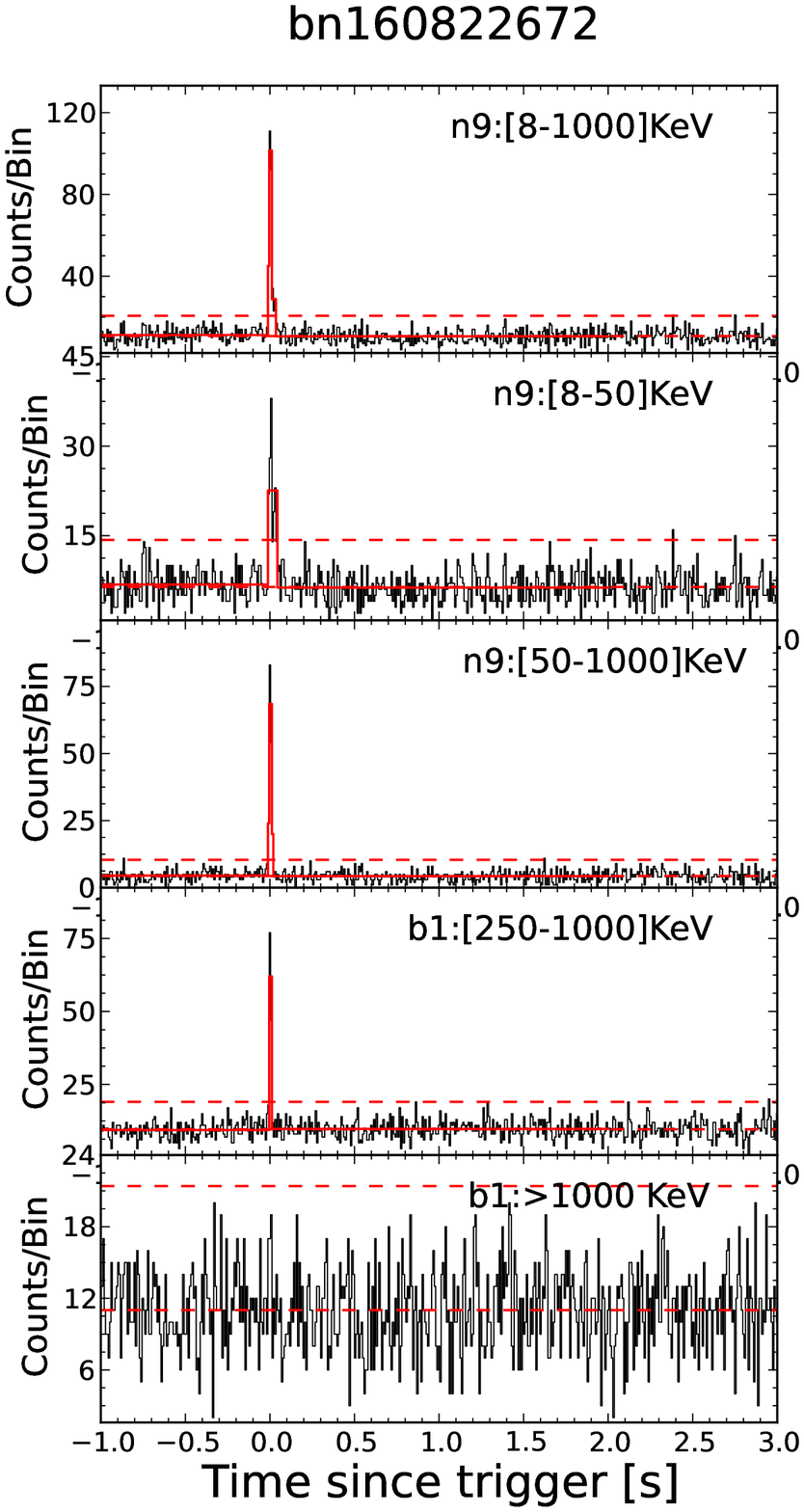}
\hfill
\includegraphics[origin=c,angle=-90,scale=0.6,width=0.400\textwidth,height=0.4\textheight]{./f3a2SP.eps}
\caption{(a): Examples of Pattern I SGRB lightcurves and spectra together with our Bayesian block analysis (red blocks in the left panels) and spectral fits (solid line in the right panels ). The horizontal dash lines in the left panels 3$\sigma$ signal over background emission.}
\label{fig:LCexample}

\end{figure*}

\begin{figure*}
\includegraphics[origin=c,angle=0,scale=0.6,width=0.4\textwidth,height=0.4\textheight]{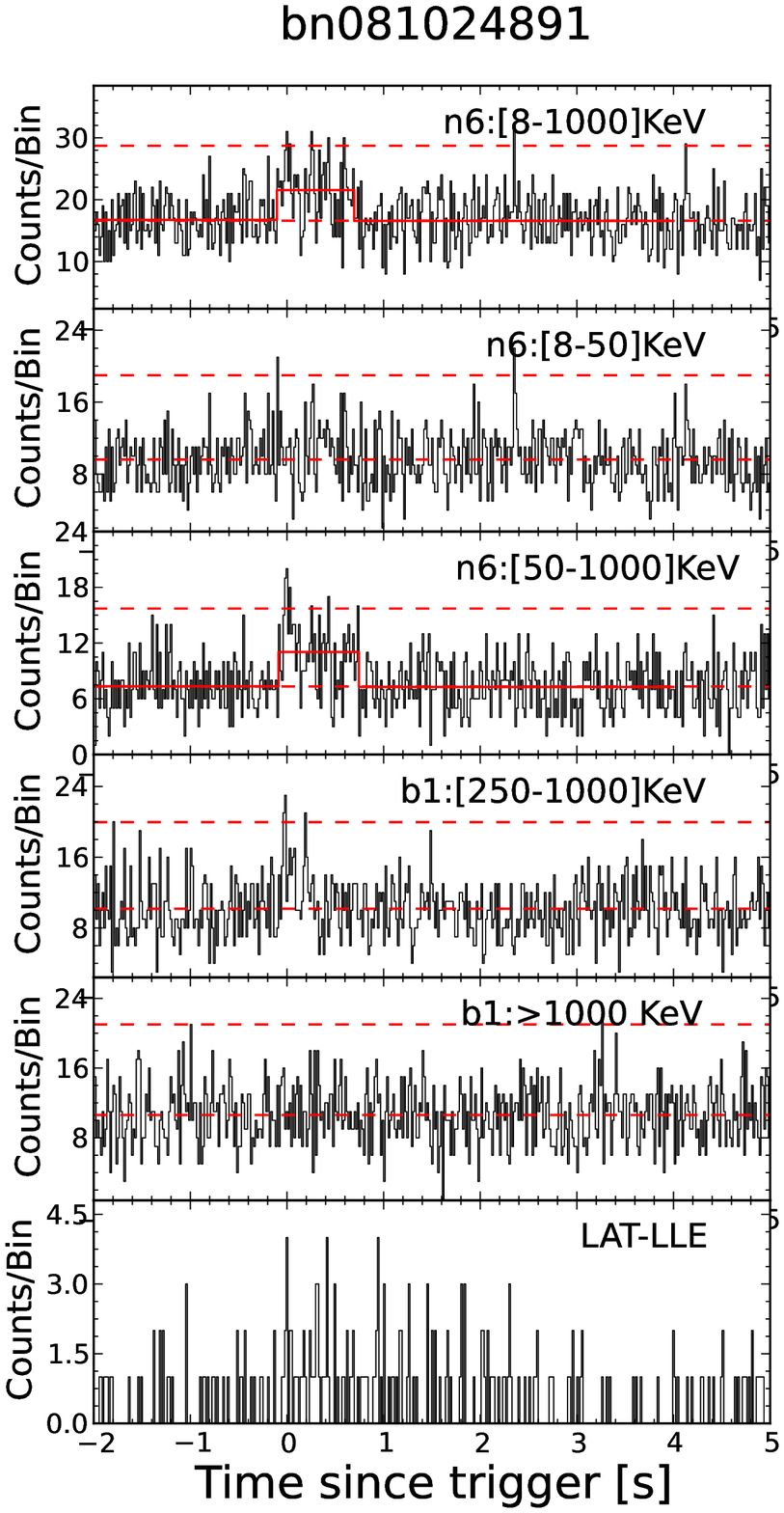}
\includegraphics[origin=c,angle=-90,scale=0.6,width=0.4\textwidth,height=0.4\textheight]{./f3dSP.eps}
\includegraphics[origin=c,angle=0,scale=0.6,width=0.4\textwidth,height=0.4\textheight]{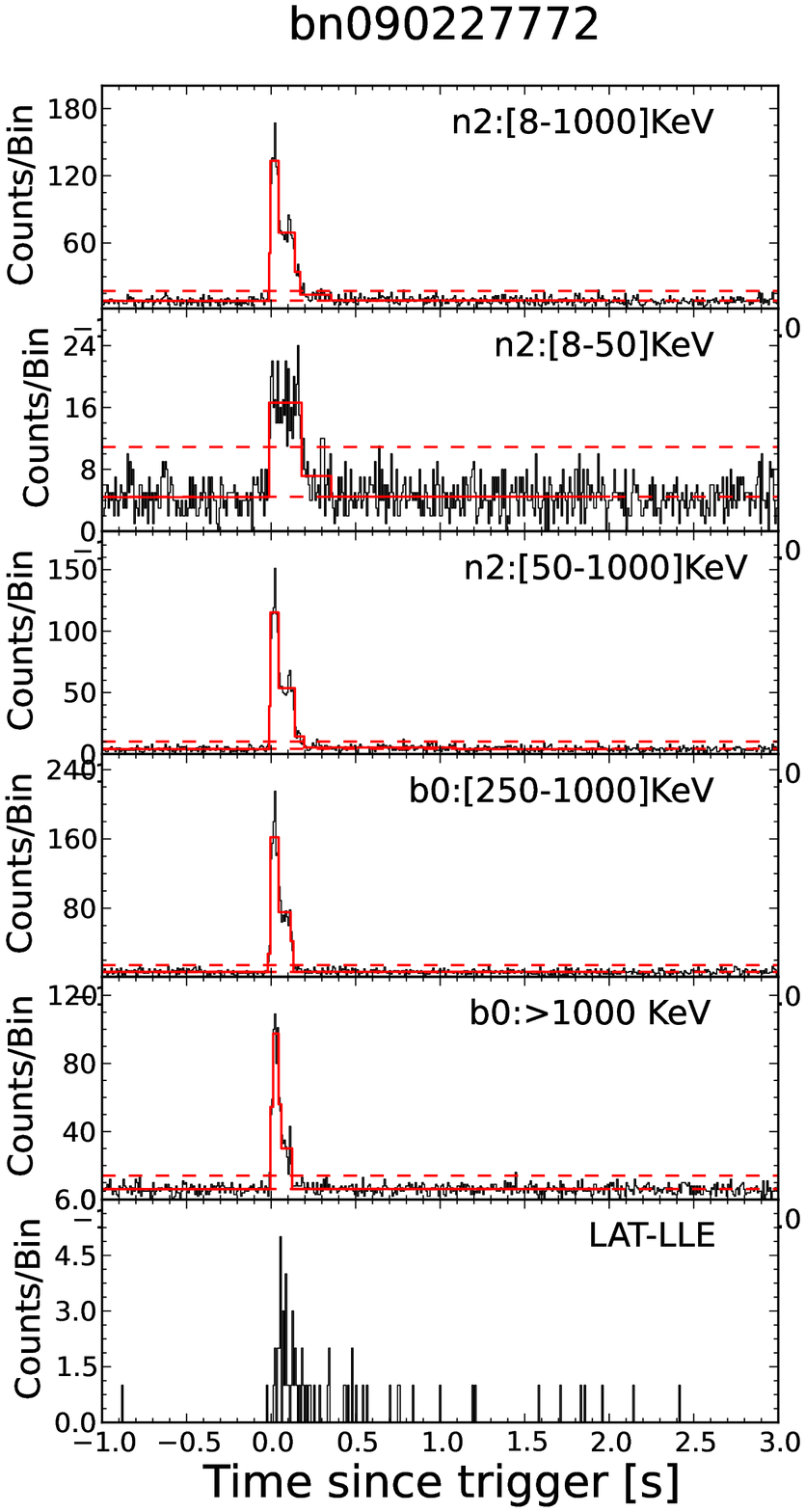}
\hfill
\includegraphics[origin=c,angle=-90,scale=0.6,width=0.4\textwidth,height=0.4\textheight]{./f3eSP.eps}
\caption{Lightcurves and spectra of 7 sGRBs observed by both GBM and LAT. Symbols are the same as Fig. \ref{fig:LCexample}.}
\label{fig:GBMLLE}

\end{figure*}

The lightcurves of all sGRBs in our sample 
usually are highly variable. We visibly identify two patterns of these lightcurves. One is dominated by a single episode (Pattern I). 61\% of the sGRBs (168 out of the 275 sGRB) are of this group. Figure \ref{fig:LCexample} (a) show two examples of this lightcurve pattern. These lightcurves usually rapidly increase
and drop, being dominated by one main block. The other one is composed of several multiple separated episodes or distinct pulses (Pattern II). About 39\% of the sGRBs (107 sGRBs) are of this group. The episodes or pulses of most sGRB in this group are connected blocks (83 out of 107 sGRBs). These sGRBs usually weaker than that of the Pattern I, but their durations tend to be longer than that of the Pattern I events. Figure \ref{fig:LCexample} (b) shows two examples of the Pattern II lightcurves.  Among the 107 sGRBs, 21 sGRBs have a lightcurve that is composed of some bright, separated blocks. We find only two cases (GRBs 081216 and 090510) that show a well-separated precursor structure in
their lightcurves. They are shown in Figure \ref{fig:GBMLLE}.

\begin{figure}
\includegraphics[angle=0,scale=0.6]{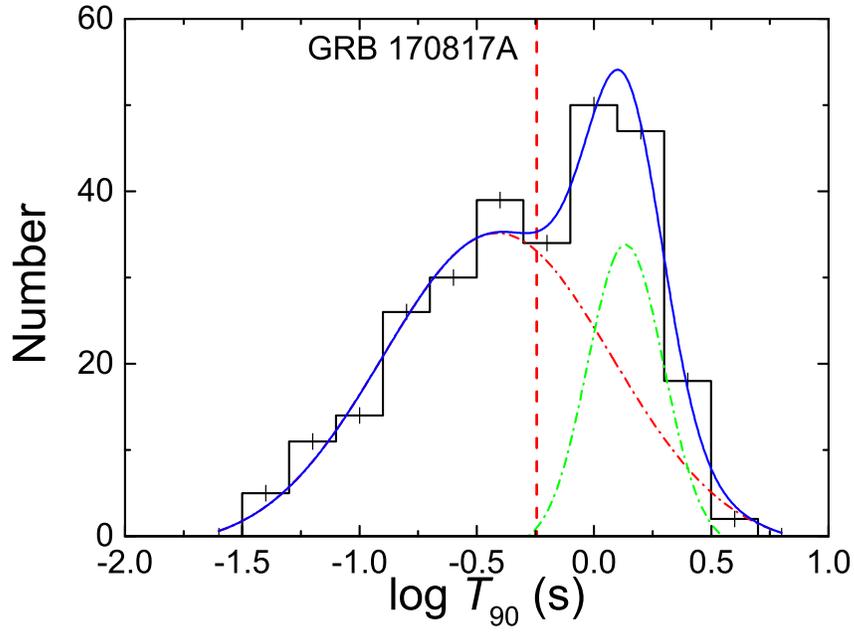}
\caption{Distributions of $T_{\rm {90}}$ for our sample. The solid and dashed lines are the fit with a two-component Gaussian function. GRB 170817A is marked with a vertical dash line.}
\label{fig:T90}
\hfill
\end{figure}

The $T_{90}$ distribution of the sGRBs are shown in Figure \ref{fig:T90}. It is very broad, spreading from tens of
millisecond to 2 seconds. The sharp cutoff at the long $T_{90}$ end would be due to our sample selection effect. A
tentative bimodal distribution is observed. Our fit with a model of two-component of Gaussian function yields two peaks at $\log T_{90}/{\rm s}=-0.41\pm 0.49$ and $\log T_{90}/{\rm s}=0.13\pm 0.16$. Note that the visible bimodal distribution and empirical multi-Gaussian function fit  depend on the bin size for making the distribution. We test the statistics of the bimodality with a method proposed by Ashman et al. (1994)\footnote{This method is based on a Gaussian mixture model with parameters estimated by maximum likelihood estimation using the expectation-maximization algorithm. It is widely used for population and classification studies (e.g., Knigge et al.2011; L\"{u} et al. 2010). The details of this method please refer to McLachlan \& Basford (1988) and McLachlan \& Peel (2000). We use the code of Ashman et al. (1994) who applied this technique to detect and measure the bimodality of astronomical data sets.}.  The significance of bimodality is evaluated with a chance probability $P_k$. A smaller $P_k$ rejects one single Gaussian distribution for the data with higher confidence. Conventionally, $P_k<0.05$ rejects one Gaussian component in the data. We get $P_k=0.006$, marginally suggesting two Gaussian components in the data.

\subsection{Spectral Properties}
The preferred spectral model in our analysis is the CPL model.
Except for GRBs 090227, 090510, 110529, 130310, and 160709, the
time-integrated spectra of all sGRBs in our sample are adequate
fitted with this model. Seven sGRBs, i.e., 081024, 090227,
090510,110529, 130310,141222, and 160709, were also detected
with LAT. Their lightcurves and spectra together with our fits
are shown in Figure \ref{fig:LCexample}. It is found that an
extra power-law component is required to fit the spectra of
sGRBs 090510 and 160709 (see also Ackermann et al. 2010; Zhang
et al. 2011). The fits with the Band function for GRBs 090222,
110529, 130310 are significantly improved over the CPL model.
Therefore, we adopt the Band function to fit the spctra of the three sGRBs.
The spectra of GRBs 081024 and 141222 in the GBM and LAT bands
still can be fitted with the CPL model. The spectral parameters
derived from our fits are reported in Table 1 and 2. The
distribution of reduced $\chi^2$ of our fits are shown in
Figure \ref{fig:goodness}.

\begin{figure}
\includegraphics[angle=0,scale=0.4]{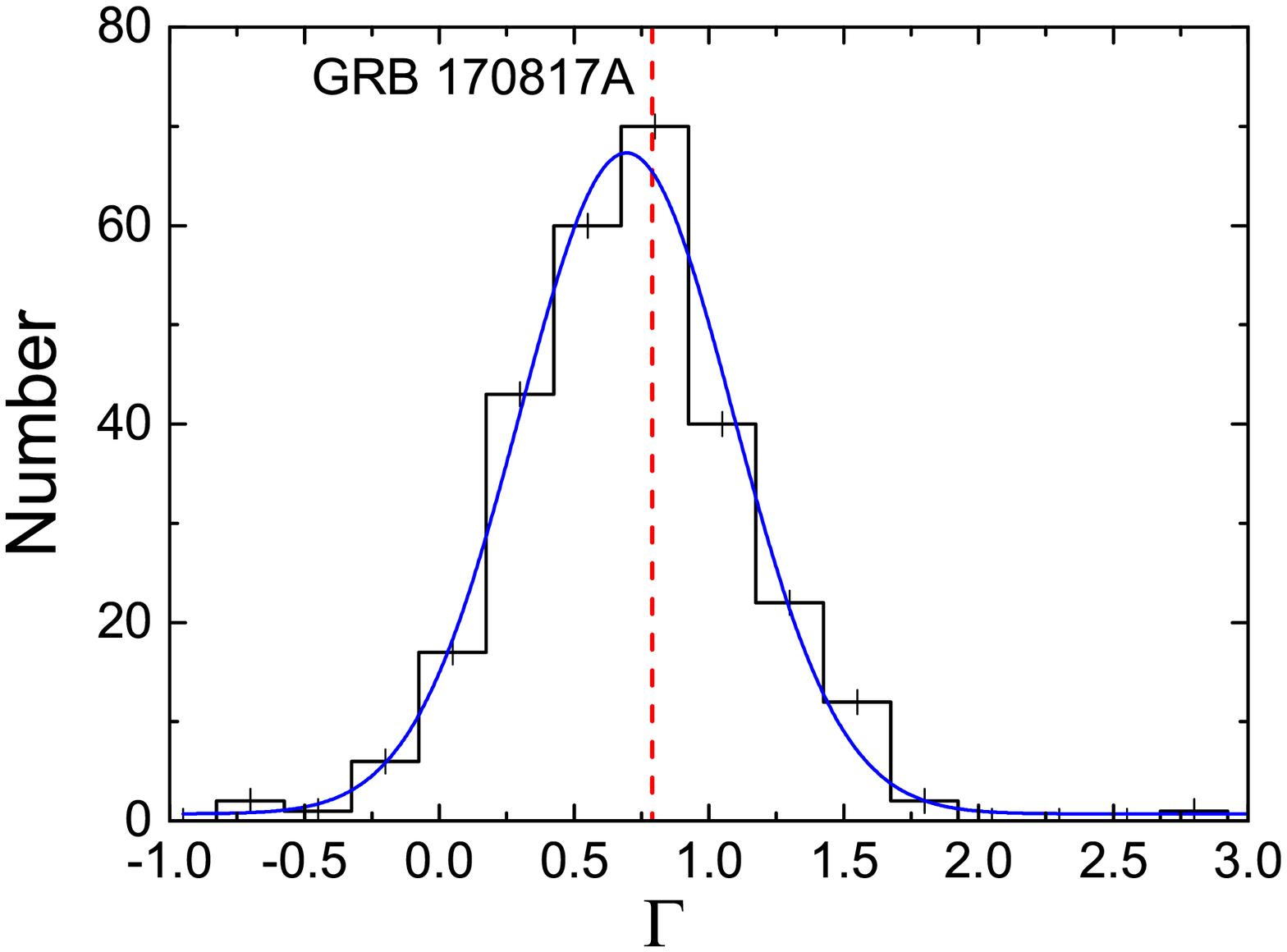}
\includegraphics[angle=0,scale=0.4]{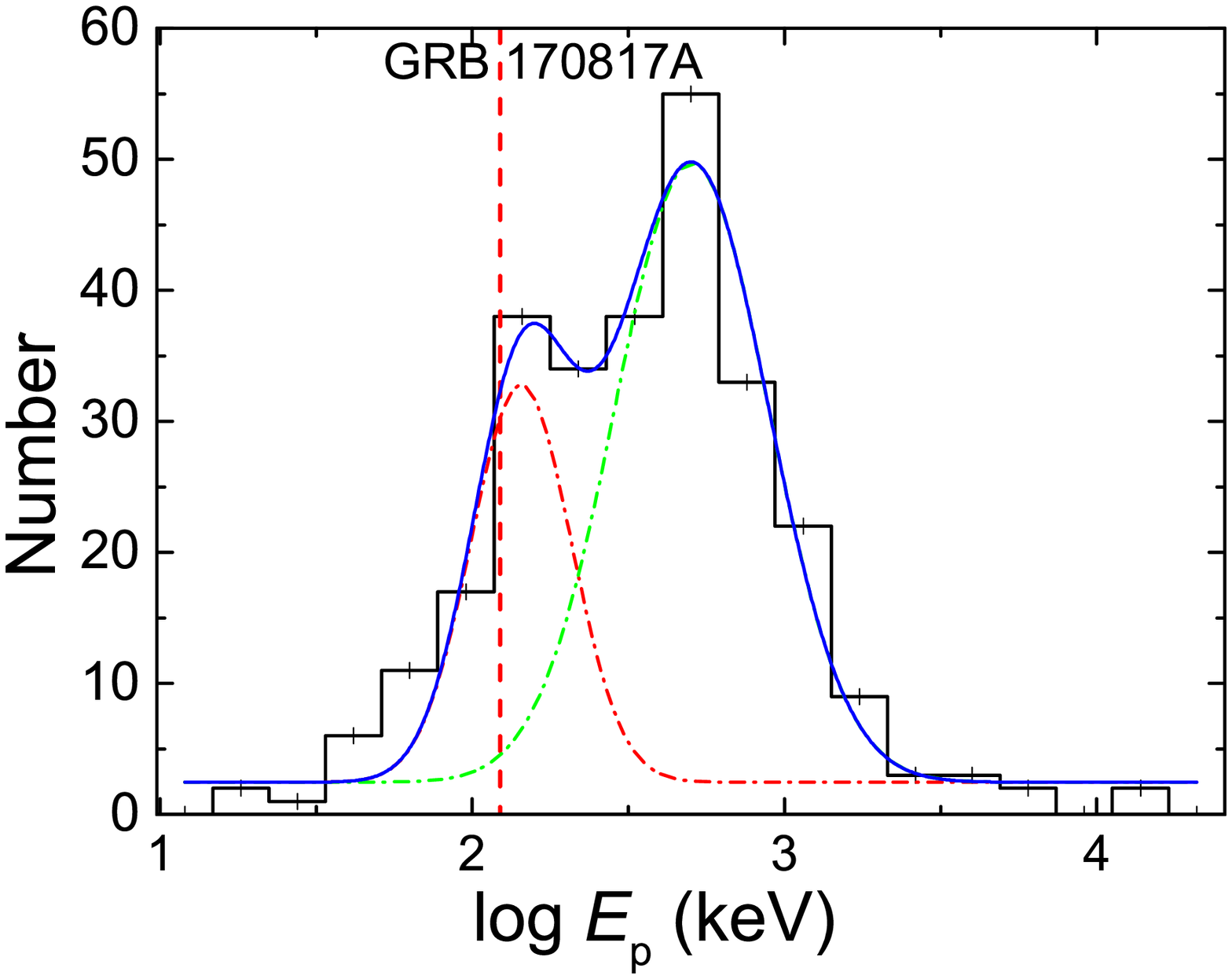}
\includegraphics[angle=0,scale=0.4]{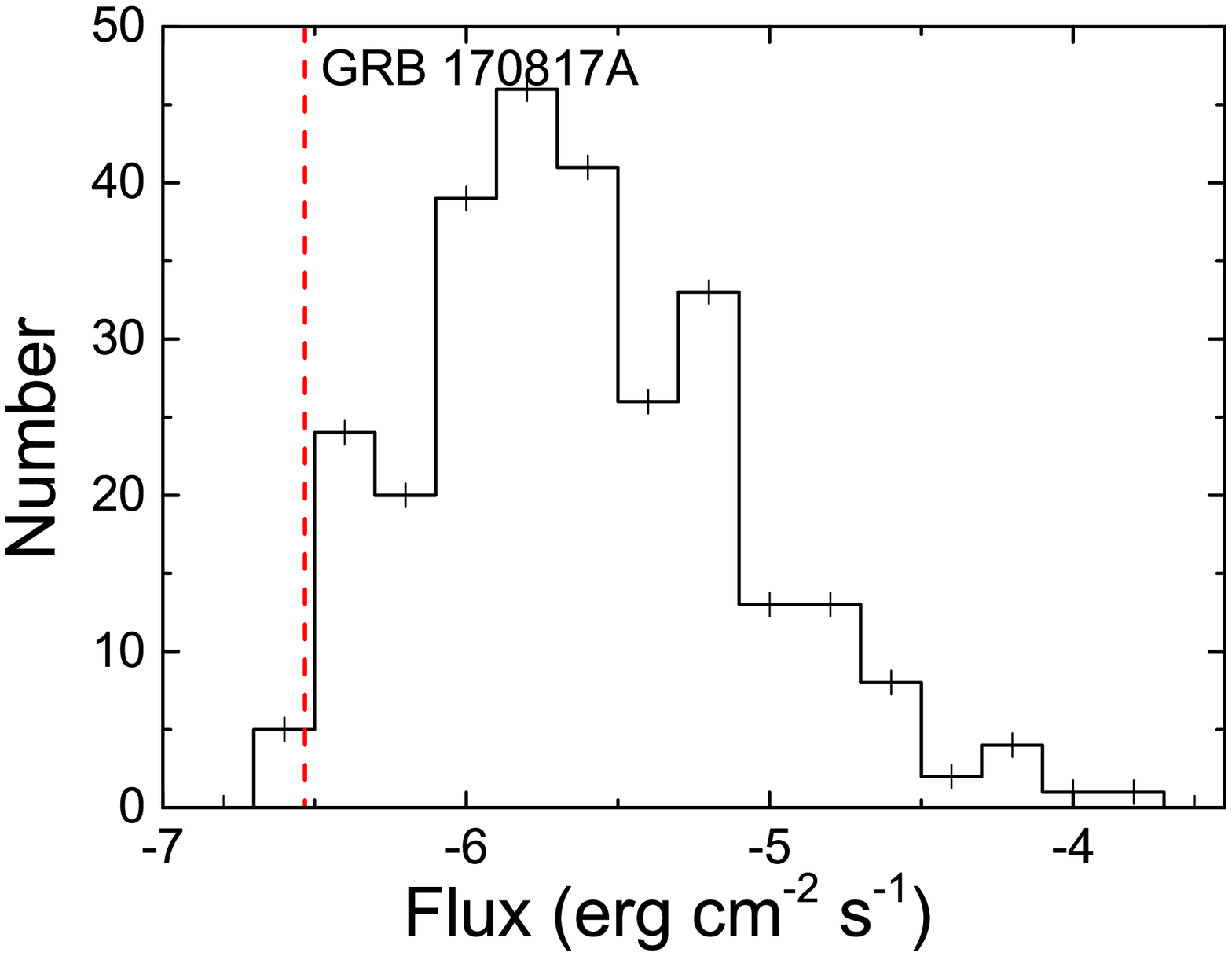}
\caption{Distributions of $E_{\rm p}$, Photon Index ($\Gamma$), and flux ($F_p$) for our sample. GRB 170817A is marked with a vertical dash line. The solid and dashed lines are the fit with a Gaussian function or a two-component Gaussian function.}
\label{fig:Distribution}
\hfill
\end{figure}

\begin{figure}
\includegraphics[angle=0,scale=0.4]{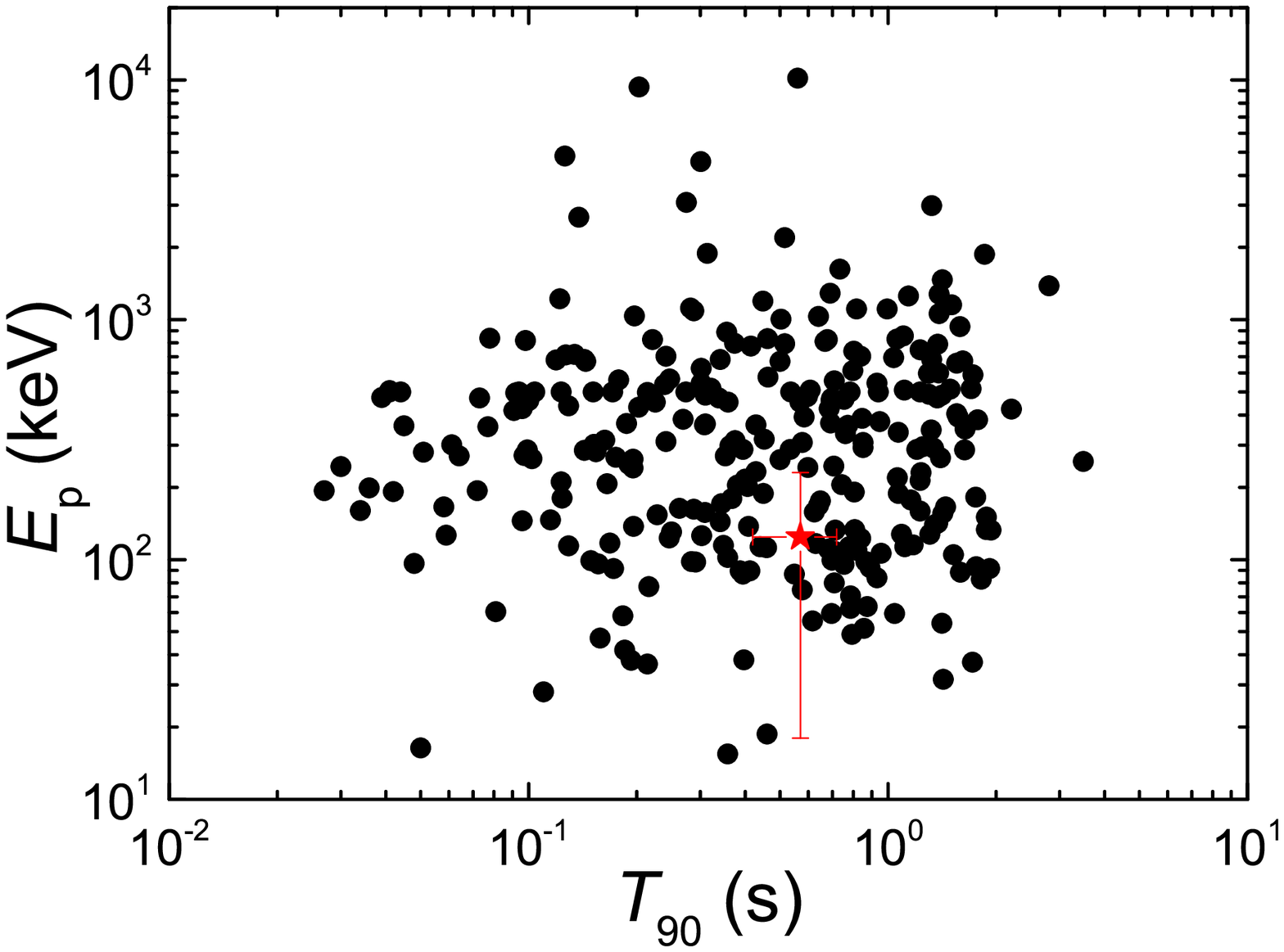}
\includegraphics[angle=0,scale=0.4]{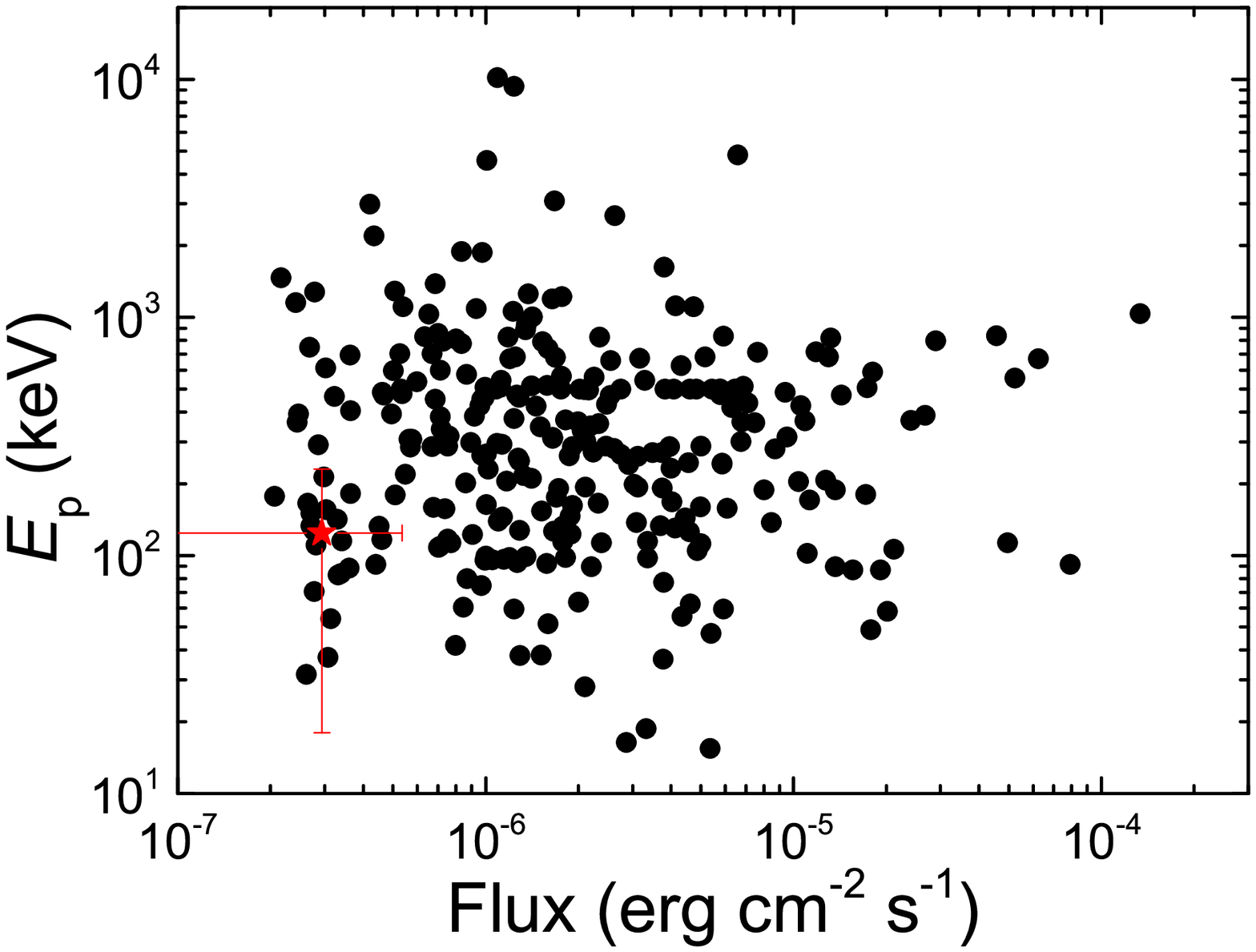}
\includegraphics[angle=0,scale=0.4]{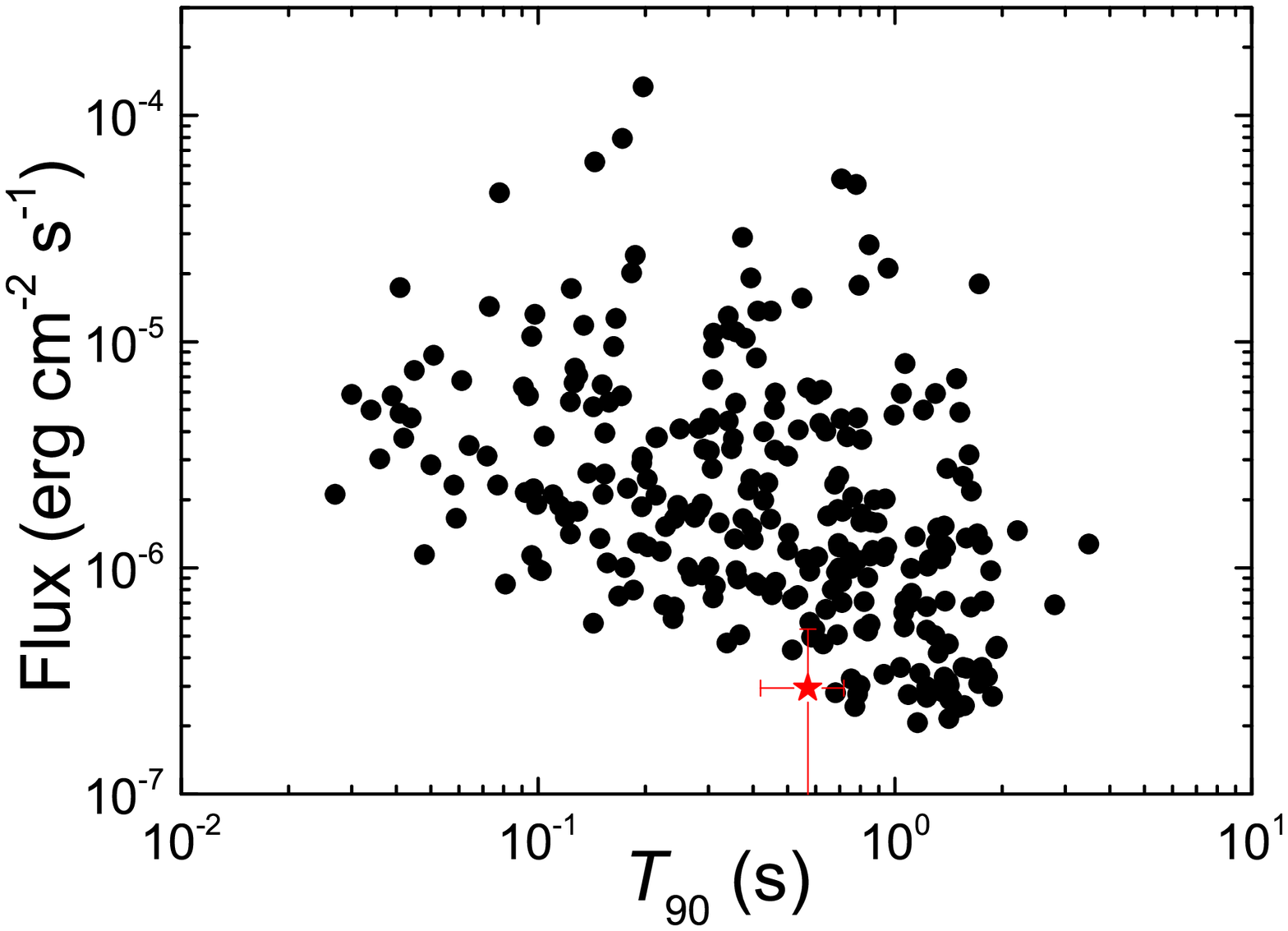}
\caption{sGRBs of our sample in the $E_{\rm p}-T_{90}$, $E_{\rm p}-{\rm Flux}$, and ${\rm Flux}-T_{\rm 90}$ planes.  GRB 170817A is marked with a red star.}
\label{fig:Relationship}
\hfill
\end{figure}

Figure \ref{fig:Distribution} shows the distributions of
$\Gamma$, $E_{\rm p}$, and peak flux ($F_p$) in the GBM energy
band (8 keV-40 MeV) for the sGRBs. The $\Gamma$ distribution is
normal, which is $\Gamma=0.69\pm 0.02$. The $E_p$
distribution ranges from tens keV to several thousands keV. It
shows a tentative signature of bimodality, with peaks at $\sim
140$ keV and $\sim 500$ keV. Similar signature is found in the HETE-2 GRB sample (Liang \& Dai 2004). Statistical test with the method proposed by Ashman et al. (1994) yields $P_k=0.007$. The distribution of $F_p$ ranges from $\sim 2\times 10^{-6}$ to
$10^{-4}$ erg s$^{-1}$ cm$^{-2}$. The cutoff feature in the
low-$F_p$ end is caused by the GBM threshold. Figure
\ref{fig:Relationship} shows the sGRBs in the $E_{p}-F_p$,
$E_{p}-T_{90}$ and $F_p-T_{90}$ planes. We do not find
any statistical correlation among them.

\section{GRB 170817A in Comparison with typical sGRBs}
GW 170817 is proposed from merger of a NS-NS binary in nearby galaxy NGC 4933 (Abbott et
al. 2017c). sGRB 170817A may be connected with GW 170817. It was
triggered by both {\em Fermi}/GBM and INTEGRAL at
$T_0$=12:41:06.47 UT on 17 August 2017 (Goldstein et al.
2017; Savchenko et al. 2017). It is closed to NGC 4933
with a distance of $24^{\circ}$. Taking NGC 4993 as its host
galaxy, one has the physical distance about $~40$ Mpc (Levan et
al. 2017).

\begin{figure}
\includegraphics[origin=c,angle=0,scale=1,width=0.500\textwidth,height=0.5\textheight]{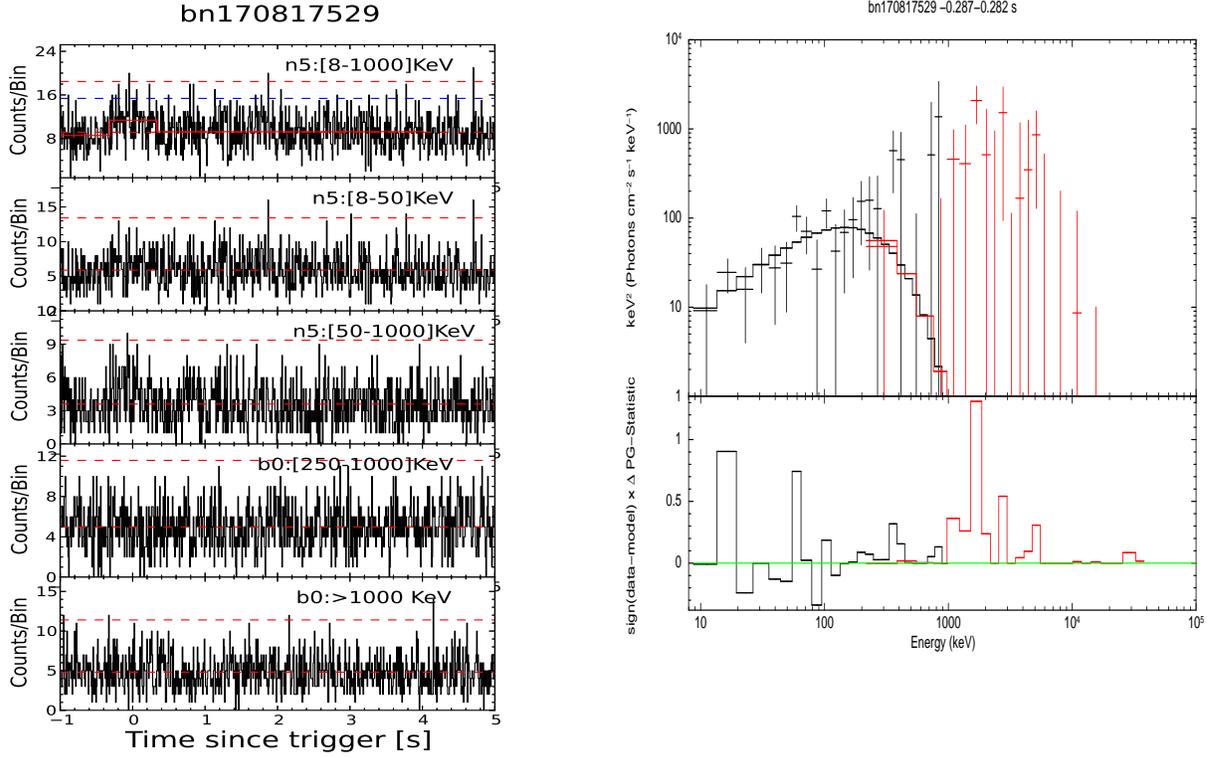}
\includegraphics[origin=c,angle=-90,scale=1,width=0.500\textwidth,height=0.5\textheight]{./f8b.eps}
\caption{Lightcurve ({\em left}) and spectrum ({\em right}) of GRB 170817A. The solid line is the Bayesian Blacks. The horizontal red and blue dash lines are 3$\sigma$ and 2$\sigma$ signal over the background emission, respectively. A cutoff power-law is invoked to fit the spectrum.}
\label{fig:GRB170817A}
\hfill
\end{figure}

We extract the lightcurve and spectrum of GRB 170817A by using
our {\em PYTHON} code. A weak signal can be identified
using the Bayesian Block method within $3\sigma$ confidence.
Its duration is $0.57\pm 0.15$ s, being consistent with that
reported by the {\em Fermi} team within error bar ($T_{90}\sim
0.647$ s; Goldstein et al. 2017). The lightcurves in different
energy bands are shown in Figure \ref{fig:GRB170817A}. Its
spectrum can be fit with the CPL model, which yields $E_{\rm
p}=124\pm 106$ keV and $\Gamma=0.79\pm 0.54$, with a reduced
$\chi^2=1.05$ in 242 degree of freedom. The fitting result is
also shown in Figure \ref{fig:GRB170817A}. We compare the
properties of this events with the global sGRBs in our sample
by marking this event in Figure \ref{fig:T90} and
\ref{fig:Distribution}. Its duration is right at the gap of the
tentative bimodal distribution. Its $\Gamma$ is consistent with
the typical $\Gamma$ value of the sGRBs. Its peak flux is at
the lowest end of the $F_p$ distribution, and its $E_p$ is
right at the low-energy peak of the tentative bimodal $E_p$
distribution of the global sGRB sample.

The peak flux and fluence in the energy band 8 keV to 40 MeV
are $(2.94\pm 2.41)\times10^{-7}$ erg cm$^{-2}$ s$^{-1}$ and
$(1.67\pm 1.37)\times10^{-7}$ erg cm$^{-2}$. If one adopt the
distance of $~40$ Mpc, the peak luminosity and isotropic energy
are corresponding to $(5.67\pm4.65)\times10^{46}$ erg s$^{-1}$
and $(3.23\pm2.65)\times10^{46}$ erg, respectively. Therefore,
it is definitely an extremely low-energy, low-luminosity, soft
sGRBs comparing to typical sGRBs (e.g., Zhang et al. 2017).
Figure \ref{fig:EpEiso} show this event in the $E_{\rm \gamma,
iso}-E_{\rm p}$ plane in order to examine whether it follows the $E_{\rm p}-E_{\gamma, \rm iso}$ relation of typical sGRBs. We find that it falls into the low-luminosity end, but is still within the
2$\sigma$ range of this relation.

\begin{figure}
\includegraphics[origin=c,angle=0,scale=1,width=0.500\textwidth,height=0.6\textheight]{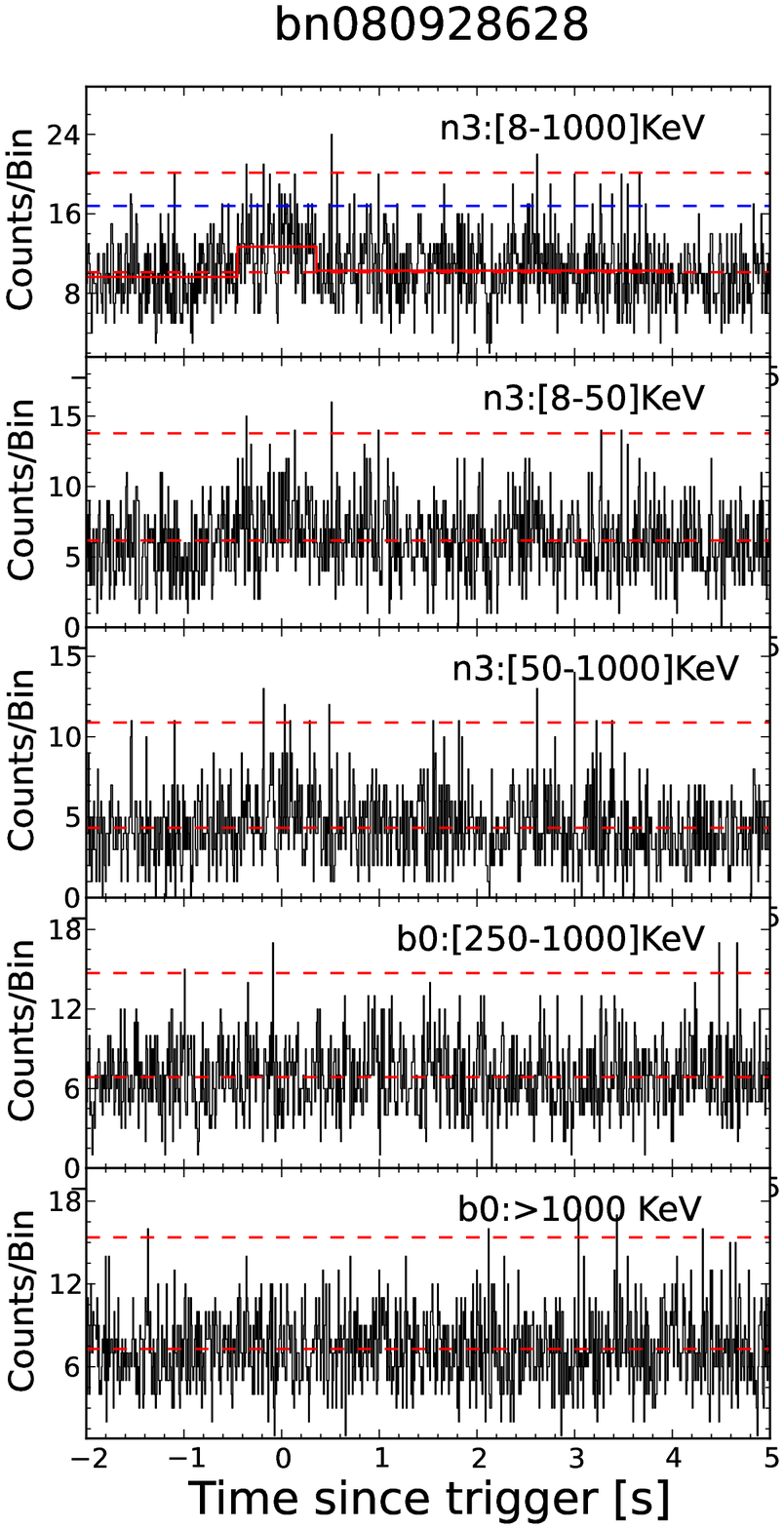}
\includegraphics[origin=c,angle=0,scale=1,width=0.500\textwidth,height=0.6\textheight]{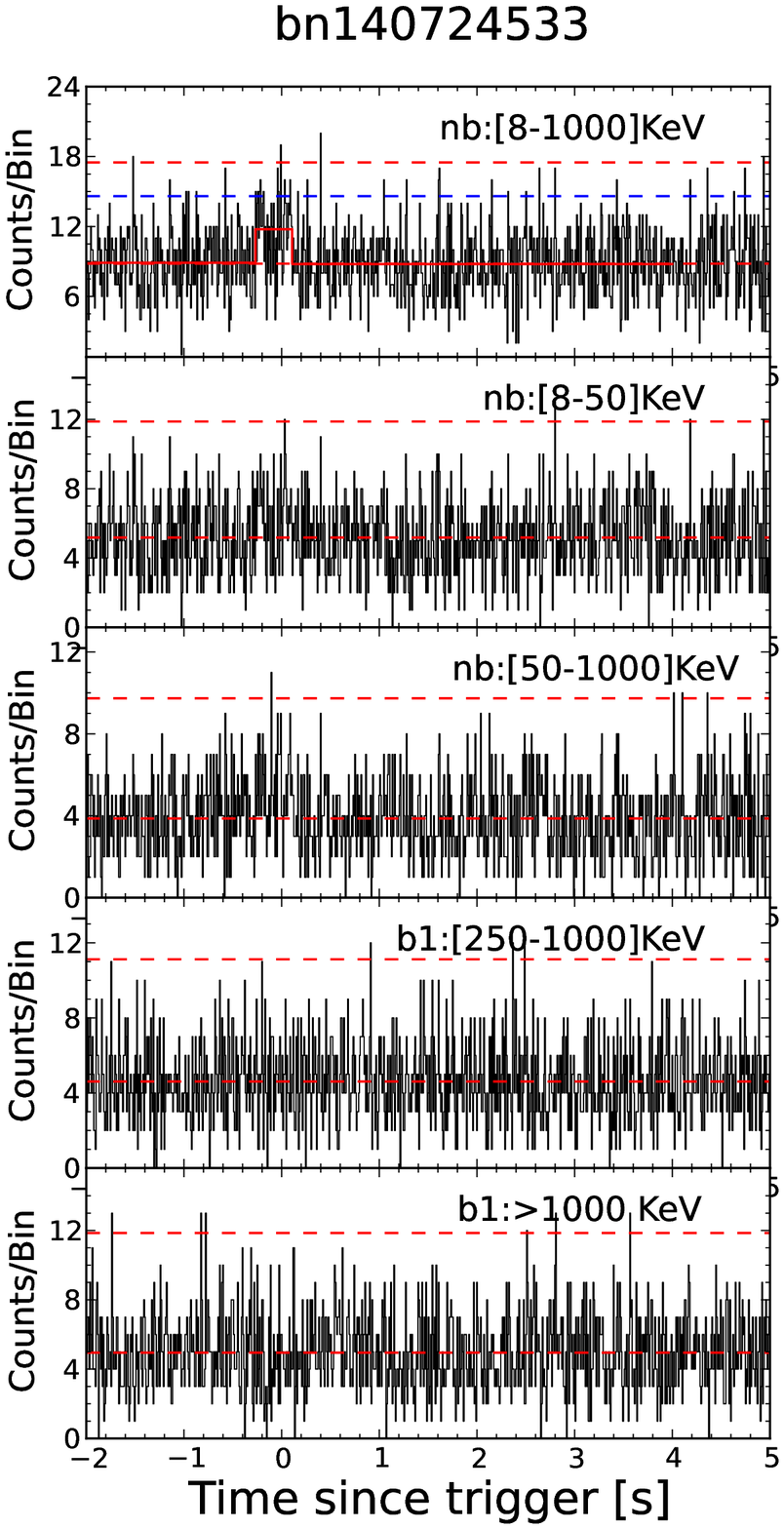}
\caption{Examples of light curves for 48 weak sGRBs obtained from our deep search with the same criterion for analyzing the lightcurve of GRB 170817A.}
\label{fig:GWLC}
\hfill
\end{figure}

Association of GW 170817 and GRB 170817A is of great interest.
The lightcurve of GRB 170817A is similar to that of Pattern II
in our sample. We have identified 83 similar sGRBs in the our
sample. However, GRB 170817A is even weaker than these sGRBs.
We adopt the same criterion for identifying the GRB 170817A to
make further search for similar events in the GBM data archive.
48 similar events are obtained. They are listed in Table 3. We show two examples of the lightcurves in Figure \ref{fig:GWLC}
Since the sinal is weak in these events, their spectra is highly uncertain. We therefore do not make spectral analysis for these events.

\section{Conclusions and Discussion}
We have present a comprehensive temporal and spectral analysis for the GRB data observed with {\em Fermi} in 9 year operation and dedicated
to study the properties of sGRBs and GRB 170817A-like events.
 Our results are summarized as
the following:
\begin{itemize}
 \item We obtain a catalog of 275 sGRBs. The skymap of these sGRBs are isotropic distribution, and the $\log N-\log P$ distribution is     consistent with that observed by BATSE.
\item Two light curve patterns are identified from the
    sGRB sample, i.e., single episode (Pattern I,
    61\% of the sGRBs in our sample), multiple episodes (Pattern II, 39\% of the sGRBs). Precursor is
    only seen in two GRBs in our sample. This is consistent with the observations with {\em Swift} (Hu et al. 2014). Their duration distribution show a tentative bimodal feature.
\item The spectra of the sGRBs are fitted with the CPL model, except for 5 sGRBs. Their $E_{\rm p}$ distributions show a tentative bimodal feature, peaking at 150 KeV and 500 keV. The $\Gamma$
    distribution is well fit with a Gaussian fucntion, yielding $\Gamma=0.69\pm0.02$.
    We do not find any correlation among $T_{90}$,
    $E_{\rm p}$, and $\Gamma$. An extra power-law component, which dominates the spectrum in the LAT band, is required to fit the spectra of GRBs 090510 and 160709.
\item GRB 170817A is a soft and weak sGRB. We get $T_{90}=0.57\pm 0.15$ s, $\rm E_{\rm p}=124\pm
    106$ keV. Its $\Gamma$ value is the same as the typical one
    of sGRBs. It is an extremely low-luminosity, low-energy sGRBs, but it still follows the $E_{\rm iso}-E_p$
    relation.
\item The lightcurve of GRB 170817A is similar to that of
    the Pattern II in our sample. With the criterion for
    identifying the lightcurve of sGRB 170817A, we also
    search for similar events and find 48 cases in the GBM GRB
    data archive in the past 9 years. Together with the 83 events of the Pattern II group in our typical sGRB sample, the GRB 170817A-like events
    are 131. This is comparable to that in the Patter I (168 sGRBs).
\end{itemize}

Motivating by the NS-NS merger origin of GRB 170817A and GW 170817, we are interested in searching for possible signatures of different types of sGRBs in current sample. Lightcurve pattern may give clues to the sGRB progenitors and central engines (e.g., Dichiara et al. 2013). The most favorable progenitors of sGRBs are coalescence of the NS-NS binary (NPacz\'ynski 1986; Eichler et al. 1989)
and NS-BH binary (Pacz\'ynski 1991). The two lightcurve patterns identified in this analysis, together with the tentative two-components in the $E_p$ and $T_{90}$ distributions, may imply two distinct types of progenitors. The lightcurve of GRB 170817A is grouped in to the Pattern II, which characterized with some weak pulses or episodes. Note that several distinct emission episodes may be due to late activity of the GRB central engines (e.g., Burrows et al. 2005; Fan \& Wei 2005; Zhang et al. 2006; Dai et al. 2006). This was convinced by the observed late X-ray flares post the prompt gamma-rays of some sGRBs (such as GRB 050724, Berger et al. 2005). With the fact that the progenitor of GRB 170817A is a NS-NS binary based on the observations of GW 170817 (Abbott et al. 2017c), one may suspect the progenitors of the sGRB with a Patter II lightcurve may be NS-NS mergers. sGRBs with a Pattern I lightcurve, on the other hand, may be from BH-NS mergers. The rapid increase and decrease feature of these lightcurves may suggest that they are produced when the NS was rapidly swallowed by the BH in a BH-NS binary (see Nakar 2007; Berger 2014 for reviews). As reported in Hu et al. (2014), precursors are rarely seen in the lightcurves of sGRBs. A precursor was detected only for two sGRBs out of the 275 typical GRBs. This likely suggests that most mergers of compact stars would be sudden events without precursors, particularly for mergers that produce sGRBs with a Patter I lightcurve.

\begin{figure}
\includegraphics[angle=0,scale=0.6]{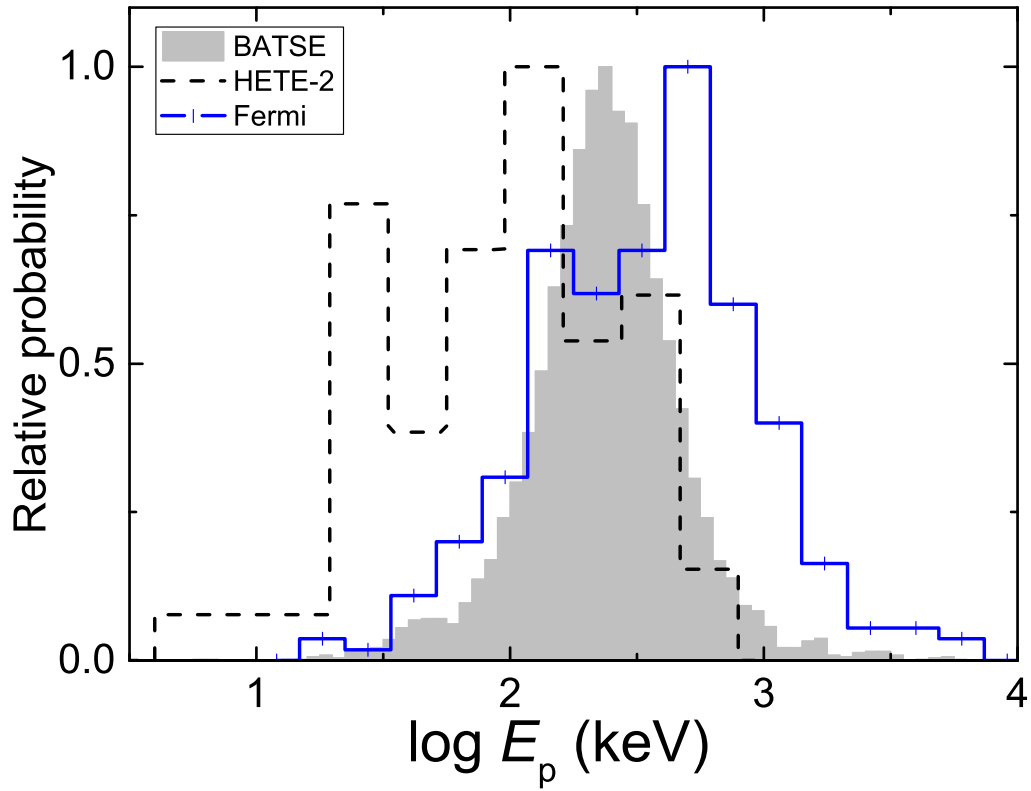}
\caption{Comparison of the $E_p$ distributions among our sGRB sample, a bright long GRB sample observed with BATSE (taken from Preece et al. 2000), and a HETE-2 GRB sample taken from Liang \& Dai 2004).}
\label{fig:EP}
\hfill
\end{figure}

The $E_p$ distribution of our sGRB sample is broad, ranging from tens of keV to thousands of keVs. We compare the $E_p$
distribution with a sample of bright, long GRBs observed with BATSE (Preece et al. 2000) and a ample of GRBs observed with HETE-2 (Liang \& Dai 2004) in Figure \ref{fig:EP}. We find that the $E_p$ values of our sGRB sample spread almost the same range as that of the HETE-2 sample\footnote{HETE-2 is sensitive in 2-400 keV band. The duration of HETE-2 GRBs in this energy band  is usually larger than 2. So, one cannot pick up sGRBs with $T_{90}<2$ seconds (e.g., Qin et al. 2013). We therefore use the global sample of HETE-2 GRBs for comparison.}, and is much broader than the bright BATSE sample. This fact suggests that the narrow $E_p$ distribution of BATSE GRB is due to the instrument selection effect and sample selection effect. The $E_p$ distribution of the HETE-2 sample moves to a lower energy end, spreading in the range of 20-300 keV. This is resulted from the energy band of HETE-2, which is sensitive in the 2-400 keV band. The energy band of GBM is from 8keV to 40 MeV, but is sensitive only in the 8-1000 keV band. Its $E_p$ distribution of much broader than that of the BATSE and HETE-2 GRB samples. If the bimodal $E_p$ distribution makes sense, the high $E_p$ component of the sGRBs have an $E_p$ much larger than that of bright long GRBs observed with BATSE. The low $E_p$ component of the sGRBs are comparable to the bright long GRBs. The observed long-soft and short-hard feature in BATSE sample would be due to selection effect.

\begin{figure}
\includegraphics[origin=c,angle=0,scale=1,width=1.0\textwidth,height=1.0\textheight]{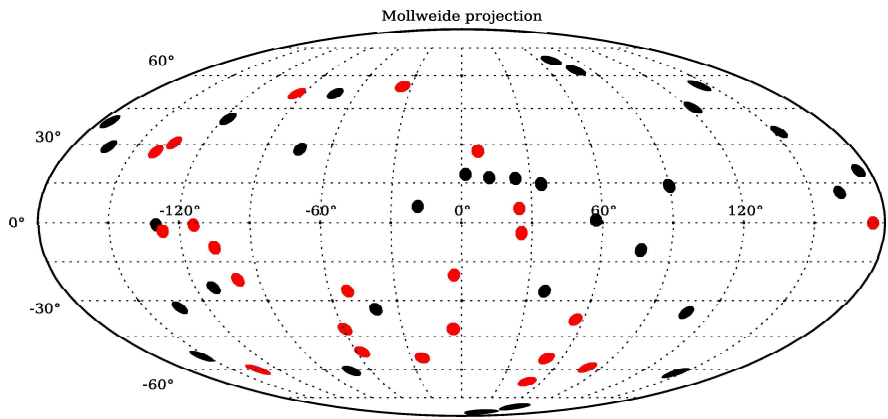}
\caption{Comparison of the sky distributions of very short GRBs ($T_{90}<100$ ms; black circles) and FRBs (red circles).}
\label{fig:skymapFRB}
\hfill
\end{figure}

It was proposed that FRBs may be originated from compact star mergers (Thornton et al. 2013; Totani
2013; Zhang 2014; Dai et al. 2017). This may imply a
possible connection between FRBs and sGRBs. Note that the durations of 23 FRBs are found to be less than 100
milliseconds so far (Lorimer et al. 2007; Keane et al. 2016). The durations of 31 sGRBs in our catalog are less than
100 milliseconds. We explore whether those extremely short GRBs are associated with FRBs in spatial distribution.
Figure \ref{fig:skymapFRB} shows these sGRBs and FRBs, but we do not find any association of the FRBs and these sGRBs.

\acknowledgments

We acknowledge the use of the public data from the {\em Fermi} data archive. We thank Bing Zhang, Zi-Gao Dai, Xiang-Yu Wang, Xue-Feng Wu, Bin-Bin Zhang ,and Jian-Yan Wei for helpful discussion. This work is supported by the National Basic Research Program (973 Programme) of China 2014CB845800, the National Natural Science Foundation of China (Grant No.11533003, 11603006, 11363002 and U1731239), the Guangxi Science Foundation (grant No. 2016GXNSFCB380005 and 2014GXNSFAA118011). The One-Hundred-Talents Program of Guangxi colleges, the high level innovation team and outstanding scholar program in Guangxi colleges, Scientific Research Foundation of Guangxi University (grant no XGZ150299), and special funding for Guangxi distinguished professors (Bagui Yingcai \& Bagui Xuezhe).




\begin{deluxetable}{cccccccccc}
\tabletypesize{\tiny}
\tablecaption{The spectral parameters of GRBs observed by GBM for our sample.}
\tablewidth{0pt}
\tabcolsep=2.5pt
\tablehead{ \colhead{GRB} & \colhead{$T_{\rm {90}}$ (s)} & \colhead{ PhotoIndex } & \colhead{$E_{\rm {cut}}$ (keV)} & \colhead{$F (\rm {erg~cm^{-2}~s^{-1}})$} &  \colhead{ $P (\rm {Photos~cm^{-2}~s^{-1}})$} &\colhead{PGS/dof} }
 \startdata
\hline
bn080815917 & 0.939$\pm$0.098 &  1.36 $\pm$ 0.02 & 4821.55 $\pm$ 1054.47 &  2.02E-06$\pm$3.06E-07 & 4.55E+00$\pm$1.40E-01 & 294/239 \\
bn080817720 & 3.497$\pm$0.122 &  0.90 $\pm$ 0.20 & 500.00 $\pm$ 621.61 &  1.28E-06$\pm$1.29E-06 & 4.91E+00$\pm$4.92E+00 & 239/242 \\
bn080905499 & 0.274$\pm$0.149 &  0.13 $\pm$ 0.09 & 256.09 $\pm$ 30.08 &  1.76E-06$\pm$3.16E-07 & 4.68E+00$\pm$7.55E-01 & 243/241 \\
bn080919790 & 0.100$\pm$0.216 &  1.10 $\pm$ 0.43 & 460.11 $\pm$ 762.29 &  9.86E-07$\pm$1.29E-07 & 5.51E+00$\pm$7.66E-01 & 223/243 \\
bn081024245 & 0.203$\pm$0.214 &  0.90 $\pm$ 0.41 & 500.00 $\pm$ 1216.71 &  1.24E-06$\pm$4.96E-07 & 4.80E+00$\pm$1.11E+00 & 242/239 \\
bn081024891 & 0.734$\pm$0.139 &  1.23 $\pm$ 0.09 & 9346.16 $\pm$ 1589.80 &  3.80E-06$\pm$1.24E-06 & 3.56E+00$\pm$8.86E-01 & 325/287 \\
bn081101491 & 0.156$\pm$0.149 &  -0.06 $\pm$ 0.55 & 96.09 $\pm$ 43.29 &  1.05E-06$\pm$1.61E-07 & 6.00E+00$\pm$7.41E-01 & 205/240 \\
bn081105614 & 0.058$\pm$0.237 &  -0.73 $\pm$ 1.02 & 165.95 $\pm$ 116.94 &  2.32E-06$\pm$1.71E-06 & 5.06E+00$\pm$3.61E+00 & 216/244 \\
bn081107321 & 1.617$\pm$0.039 &  0.13 $\pm$ 0.04 & 36.68 $\pm$ 1.02 &  6.31E-07$\pm$1.43E-08 & 9.51E+00$\pm$1.85E-01 & 302/240 \\
bn081119184 & 0.288$\pm$0.160 &  1.04 $\pm$ 0.04 & 1623.04 $\pm$ 370.35 &  1.91E-06$\pm$3.91E-07 & 3.80E+00$\pm$2.30E-01 & 267/241 \\
bn081122614 & 0.048$\pm$0.130 &  1.11 $\pm$ 0.47 & 96.28 $\pm$ 44.46 &  1.15E-06$\pm$9.96E-08 & 1.75E+01$\pm$5.15E-01 & 186/242 \\
bn081204517 & 0.221$\pm$0.111 &  0.51 $\pm$ 0.35 & 162.26 $\pm$ 88.75 &  1.18E-06$\pm$4.70E-08 & 6.99E+00$\pm$4.50E-01 & 202/240 \\
bn081216531 & 0.957$\pm$0.047 &  0.56 $\pm$ 0.06 & 827.82 $\pm$ 111.26 &  2.11E-05$\pm$3.11E-06 & 3.11E+01$\pm$3.27E+00 & 266/241 \\
bn081223419 & 0.400$\pm$0.062 &  0.41 $\pm$ 0.06 & 106.23 $\pm$ 6.89 &  1.33E-06$\pm$8.20E-08 & 1.02E+01$\pm$5.37E-01 & 253/242 \\
bn081226509 & 0.195$\pm$0.118 &  0.49 $\pm$ 0.07 & 262.62 $\pm$ 29.20 &  1.87E-06$\pm$1.82E-07 & 7.06E+00$\pm$5.44E-01 & 257/242 \\
bn081229187 & 0.388$\pm$0.136 &  0.29 $\pm$ 0.07 & 498.44 $\pm$ 60.67 &  2.20E-06$\pm$3.26E-07 & 3.67E+00$\pm$4.41E-01 & 277/240 \\
bn081230871 & 0.708$\pm$0.116 &  0.10 $\pm$ 0.09 & 216.49 $\pm$ 21.86 &  8.69E-07$\pm$1.57E-07 & 2.65E+00$\pm$4.31E-01 & 261/240 \\
bn090108020 & 0.649$\pm$0.047 &  0.44 $\pm$ 0.12 & 79.85 $\pm$ 7.05 &  1.70E-06$\pm$3.26E-07 & 1.69E+01$\pm$3.11E+00 & 235/243 \\
bn090113778 & 0.463$\pm$0.156 &  0.57 $\pm$ 0.27 & 175.51 $\pm$ 69.82 &  8.65E-07$\pm$6.88E-07 & 5.07E+00$\pm$3.68E+00 & 253/242 \\
bn090117640 & 1.061$\pm$0.049 &  1.10 $\pm$ 0.11 & 51.61 $\pm$ 4.91 &  5.49E-07$\pm$5.68E-08 & 1.18E+01$\pm$1.26E+00 & 270/240 \\
bn090219074 & 0.516$\pm$0.178 &  0.54 $\pm$ 0.51 & 137.51 $\pm$ 81.14 &  7.26E-07$\pm$1.09E-07 & 5.08E+00$\pm$5.86E-01 & 230/240 \\
bn090227772 & 0.197$\pm$0.027 &  -0.39 $\pm$ 0.03 & 1156.32 $\pm$ 78.83 &  1.34E-04$\pm$5.97E-06 & 9.04E+01$\pm$2.21E+00 & 286/287 \\
bn090305052 & 0.808$\pm$0.060 &  0.43 $\pm$ 0.03 & 576.57 $\pm$ 28.04 &  3.70E-06$\pm$1.17E-07 & 6.37E+00$\pm$1.46E-01 & 251/242 \\
bn090316311 & 1.879$\pm$0.093 &  0.93 $\pm$ 0.30 & 133.41 $\pm$ 75.95 &  2.70E-07$\pm$2.44E-07 & 2.87E+00$\pm$2.47E+00 & 242/239 \\
bn090316311 & 1.879$\pm$0.093 &  0.93 $\pm$ 0.30 & 133.41 $\pm$ 75.95 &  2.70E-07$\pm$2.44E-07 & 2.87E+00$\pm$2.47E+00 & 242/239 \\
bn090328713 & 0.127$\pm$0.086 &  0.83 $\pm$ 0.03 & 1035.26 $\pm$ 97.33 &  7.65E-06$\pm$4.75E-07 & 1.43E+01$\pm$4.71E-01 & 259/240 \\
bn090510016 & 0.846$\pm$0.036 &  -0.68 $\pm$ 0.08 & 2987.80 $\pm$ 449.33 &  2.68E-05$\pm$3.18E-06 & 1.16E+01$\pm$7.13E-01 & 420/313 \\
bn090514726 & 1.141$\pm$0.050 &  0.46 $\pm$ 0.04 & 150.52 $\pm$ 7.89 &  1.38E-06$\pm$4.59E-08 & 8.21E+00$\pm$2.29E-01 & 252/241 \\
bn090518080 & 1.295$\pm$0.093 &  1.33 $\pm$ 0.03 & 387.54 $\pm$ 54.87 &  5.01E-07$\pm$3.30E-08 & 4.55E+00$\pm$1.39E-01 & 219/242 \\
bn090610648 & 1.399$\pm$0.105 &  0.70 $\pm$ 0.02 & 1256.67 $\pm$ 69.60 &  2.74E-06$\pm$1.36E-07 & 3.41E+00$\pm$7.61E-02 & 228/242 \\
bn090617208 & 0.129$\pm$0.086 &  0.56 $\pm$ 0.04 & 713.49 $\pm$ 76.55 &  7.10E-06$\pm$5.93E-07 & 1.19E+01$\pm$5.60E-01 & 246/241 \\
bn090620901 & 1.108$\pm$0.140 &  0.63 $\pm$ 0.06 & 596.78 $\pm$ 92.08 &  9.93E-07$\pm$1.54E-07 & 2.19E+00$\pm$2.07E-01 & 256/239 \\
bn090907808 & 0.805$\pm$0.067 &  0.30 $\pm$ 0.05 & 265.51 $\pm$ 15.84 &  1.73E-06$\pm$8.42E-08 & 5.28E+00$\pm$2.21E-01 & 243/242 \\
bn090924625 & 0.154$\pm$0.124 &  0.51 $\pm$ 0.24 & 437.17 $\pm$ 224.17 &  3.95E-06$\pm$4.15E-07 & 9.66E+00$\pm$5.29E-01 & 228/240 \\
bn090927422 & 0.640$\pm$0.127 &  1.30 $\pm$ 0.05 & 508.60 $\pm$ 118.55 &  6.54E-07$\pm$8.00E-08 & 4.79E+00$\pm$2.31E-01 & 249/240 \\
bn091126333 & 0.195$\pm$0.123 &  0.50 $\pm$ 0.27 & 287.09 $\pm$ 128.85 &  2.91E-06$\pm$1.14E-07 & 1.03E+01$\pm$3.72E-01 & 226/239 \\
bn091219462 & 1.933$\pm$0.077 &  1.27 $\pm$ 0.18 & 293.65 $\pm$ 174.24 &  4.50E-07$\pm$2.55E-07 & 4.33E+00$\pm$1.79E+00 & 234/241 \\
bn100117879 & 0.284$\pm$0.118 &  0.17 $\pm$ 0.10 & 191.09 $\pm$ 22.42 &  1.82E-06$\pm$3.24E-07 & 6.61E+00$\pm$1.01E+00 & 252/243 \\
bn100118100 & 1.550$\pm$0.128 &  0.62 $\pm$ 0.02 & 1032.15 $\pm$ 63.18 &  2.54E-06$\pm$1.37E-07 & 3.31E+00$\pm$7.55E-02 & 263/243 \\
bn100212588 & 0.786$\pm$0.191 &  0.82 $\pm$ 0.09 & 97.95 $\pm$ 11.75 &  2.77E-07$\pm$3.10E-08 & 3.27E+00$\pm$3.34E-01 & 275/243 \\
bn100216422 & 0.119$\pm$0.203 &  0.02 $\pm$ 0.16 & 241.22 $\pm$ 33.64 &  1.68E-06$\pm$7.82E-07 & 4.27E+00$\pm$1.91E+00 & 255/239 \\
bn100223110 & 0.124$\pm$0.074 &  0.23 $\pm$ 0.05 & 679.68 $\pm$ 43.68 &  1.72E-05$\pm$1.33E-06 & 1.98E+01$\pm$1.51E+00 & 207/241 \\
bn100301068 & 0.064$\pm$0.157 &  0.35 $\pm$ 0.35 & 270.73 $\pm$ 145.13 &  3.48E-06$\pm$2.13E-07 & 1.10E+01$\pm$6.25E-01 & 213/240 \\
bn100328141 & 0.536$\pm$0.078 &  0.57 $\pm$ 0.03 & 659.64 $\pm$ 50.13 &  4.08E-06$\pm$2.48E-07 & 7.52E+00$\pm$1.87E-01 & 256/239 \\
bn100406758 & 1.338$\pm$0.110 &  0.13 $\pm$ 0.08 & 70.63 $\pm$ 4.86 &  2.85E-07$\pm$2.34E-08 & 2.48E+00$\pm$1.86E-01 & 256/243 \\
bn100411516 & 0.597$\pm$0.173 &  0.94 $\pm$ 0.48 & 499.92 $\pm$ 1282.11 &  5.35E-07$\pm$2.43E-07 & 2.21E+00$\pm$5.72E-01 & 244/240 \\
bn100516369 & 0.543$\pm$0.181 &  0.26 $\pm$ 0.64 & 67.35 $\pm$ 39.79 &  3.02E-07$\pm$5.13E-08 & 3.02E+00$\pm$4.36E-01 & 268/241 \\
bn100525744 & 0.504$\pm$0.133 &  0.38 $\pm$ 0.34 & 291.61 $\pm$ 156.88 &  1.42E-06$\pm$1.00E-07 & 4.32E+00$\pm$3.74E-01 & 319/240 \\
bn100612545 & 0.598$\pm$0.073 &  0.08 $\pm$ 0.04 & 477.38 $\pm$ 22.79 &  5.86E-06$\pm$2.52E-07 & 8.09E+00$\pm$2.75E-01 & 239/242 \\
bn100616773 & 0.375$\pm$0.142 &  1.08 $\pm$ 0.04 & 1003.44 $\pm$ 195.44 &  1.65E-06$\pm$2.26E-07 & 5.16E+00$\pm$2.52E-01 & 247/240 \\
bn100625773 & 0.250$\pm$0.070 &  0.38 $\pm$ 0.06 & 242.15 $\pm$ 19.17 &  4.12E-06$\pm$3.00E-07 & 1.49E+01$\pm$1.00E+00 & 272/242 \\
bn100706693 & 0.247$\pm$0.170 &  0.13 $\pm$ 0.48 & 313.07 $\pm$ 183.65 &  1.76E-06$\pm$2.45E-07 & 3.86E+00$\pm$5.42E-01 & 216/239 \\
bn100714686 & 0.338$\pm$0.073 &  1.21 $\pm$ 0.03 & 130.22 $\pm$ 8.86 &  4.66E-07$\pm$1.33E-08 & 6.67E+00$\pm$1.28E-01 & 294/240 \\
bn100717446 & 0.930$\pm$0.109 &  0.74 $\pm$ 0.05 & 566.28 $\pm$ 65.62 &  1.12E-06$\pm$1.11E-07 & 3.03E+00$\pm$1.65E-01 & 264/240 \\
bn100811108 & 0.344$\pm$0.064 &  -0.08 $\pm$ 0.04 & 471.85 $\pm$ 23.15 &  1.13E-05$\pm$5.80E-07 & 1.37E+01$\pm$6.10E-01 & 236/239 \\
bn100827455 & 0.091$\pm$0.089 &  0.36 $\pm$ 0.17 & 417.99 $\pm$ 110.89 &  6.30E-06$\pm$4.36E-06 & 1.34E+01$\pm$8.09E+00 & 220/240 \\
bn100916779 & 0.196$\pm$0.100 &  -0.27 $\pm$ 0.15 & 104.88 $\pm$ 9.97 &  3.09E-06$\pm$9.20E-07 & 1.40E+01$\pm$3.85E+00 & 240/240 \\
bn100929916 & 0.358$\pm$0.080 &  0.46 $\pm$ 0.17 & 544.89 $\pm$ 175.65 &  5.36E-06$\pm$3.42E-06 & 1.01E+01$\pm$5.55E+00 & 230/240 \\
bn101026034 & 0.061$\pm$0.119 &  0.16 $\pm$ 0.31 & 300.52 $\pm$ 120.53 &  6.75E-06$\pm$3.14E-07 & 1.60E+01$\pm$7.84E-01 & 198/243 \\
bn101031625 & 0.363$\pm$0.108 &  0.67 $\pm$ 0.06 & 170.79 $\pm$ 17.35 &  8.94E-07$\pm$6.56E-08 & 5.99E+00$\pm$2.79E-01 & 242/241 \\
bn101101744 & 0.797$\pm$0.107 &  -0.18 $\pm$ 0.08 & 15.48 $\pm$ 0.62 &  3.02E-07$\pm$1.34E-08 & 7.51E+00$\pm$3.21E-01 & 256/239 \\
bn101104810 & 0.677$\pm$0.100 &  0.05 $\pm$ 0.05 & 298.33 $\pm$ 18.70 &  2.35E-06$\pm$1.58E-07 & 4.98E+00$\pm$2.81E-01 & 250/243 \\
bn101129652 & 0.358$\pm$0.075 &  -0.06 $\pm$ 0.05 & 612.97 $\pm$ 30.29 &  1.11E-05$\pm$6.69E-07 & 1.06E+01$\pm$6.13E-01 & 279/241 \\
bn101129726 & 0.641$\pm$0.064 &  0.78 $\pm$ 0.03 & 825.67 $\pm$ 73.09 &  4.02E-06$\pm$2.55E-07 & 8.41E+00$\pm$2.14E-01 & 240/243 \\
bn101208203 & 0.584$\pm$0.167 &  -0.36 $\pm$ 0.13 & 102.17 $\pm$ 9.80 &  4.94E-07$\pm$1.28E-07 & 2.16E+00$\pm$5.33E-01 & 256/242 \\
bn101213849 & 1.751$\pm$0.072 &  1.04 $\pm$ 0.06 & 89.86 $\pm$ 8.51 &  3.63E-07$\pm$1.88E-08 & 5.49E+00$\pm$2.21E-01 & 295/240 \\
bn101216721 & 1.701$\pm$0.028 &  0.99 $\pm$ 0.03 & 168.24 $\pm$ 7.76 &  1.41E-06$\pm$2.51E-08 & 1.37E+01$\pm$2.05E-01 & 276/241 \\
bn101224227 & 0.314$\pm$0.180 &  0.97 $\pm$ 0.37 & 392.79 $\pm$ 715.15 &  8.34E-07$\pm$7.32E-08 & 4.21E+00$\pm$3.58E-01 & 218/242 \\
bn110112934 & 0.241$\pm$0.155 &  0.87 $\pm$ 0.06 & 515.02 $\pm$ 94.88 &  1.65E-06$\pm$2.09E-07 & 5.88E+00$\pm$3.51E-01 & 250/241 \\
bn110131780 & 0.301$\pm$0.168 &  1.33 $\pm$ 0.05 & 1892.91 $\pm$ 1119.94 &  1.01E-06$\pm$4.76E-07 & 3.62E+00$\pm$4.38E-01 & 265/241 \\
bn110212550 & 0.051$\pm$0.085 &  0.40 $\pm$ 0.18 & 280.25 $\pm$ 76.47 &  8.72E-06$\pm$4.80E-06 & 2.83E+01$\pm$1.39E+01 & 205/240 \\
bn110307972 & 0.686$\pm$0.121 &  0.45 $\pm$ 0.06 & 310.24 $\pm$ 30.63 &  9.56E-07$\pm$9.32E-08 & 3.00E+00$\pm$1.93E-01 & 256/240 \\
bn110409179 & 0.149$\pm$0.140 &  -0.26 $\pm$ 0.54 & 98.86 $\pm$ 41.07 &  1.35E-06$\pm$2.30E-07 & 6.54E+00$\pm$1.00E+00 & 229/240 \\
bn110509475 & 0.352$\pm$0.114 &  1.29 $\pm$ 0.02 & 4567.26 $\pm$ 692.87 &  3.73E-06$\pm$4.21E-07 & 7.03E+00$\pm$1.74E-01 & 252/241 \\
bn110526715 & 0.441$\pm$0.071 &  0.77 $\pm$ 0.04 & 426.70 $\pm$ 33.04 &  2.38E-06$\pm$1.03E-07 & 8.58E+00$\pm$2.48E-01 & 272/242 \\
bn110529034 & 0.395$\pm$0.043 &  -0.77 $\pm$ 0.08 & 857.74 $\pm$ 212.80 &  1.91E-05$\pm$2.59E-06 & 2.66E+01$\pm$2.04E+00 & 232/284 \\
bn110605780 & 1.412$\pm$0.151 &  0.88 $\pm$ 0.33 & 270.13 $\pm$ 194.15 &  4.62E-07$\pm$2.00E-08 & 2.79E+00$\pm$1.45E-01 & 230/240 \\
bn110703557 & 1.374$\pm$0.058 &  1.14 $\pm$ 0.07 & 112.98 $\pm$ 14.24 &  3.29E-07$\pm$2.44E-08 & 4.76E+00$\pm$2.71E-01 & 243/240 \\
bn110716018 & 1.058$\pm$0.052 &  0.72 $\pm$ 0.06 & 86.79 $\pm$ 6.01 &  6.34E-07$\pm$3.03E-08 & 7.47E+00$\pm$2.97E-01 & 261/241 \\
bn110717180 & 0.045$\pm$0.109 &  0.56 $\pm$ 0.21 & 361.77 $\pm$ 131.17 &  7.47E-06$\pm$6.47E-06 & 2.32E+01$\pm$1.89E+01 & 203/240 \\
bn110728056 & 0.783$\pm$0.142 &  0.56 $\pm$ 0.06 & 483.80 $\pm$ 59.10 &  1.08E-06$\pm$1.39E-07 & 2.60E+00$\pm$2.52E-01 & 265/241 \\
bn110819665 & 1.317$\pm$0.048 &  1.04 $\pm$ 0.03 & 219.81 $\pm$ 13.36 &  1.50E-06$\pm$4.20E-08 & 1.29E+01$\pm$2.98E-01 & 279/239 \\
bn111011094 & 0.154$\pm$0.085 &  0.51 $\pm$ 0.24 & 180.43 $\pm$ 62.98 &  2.60E-06$\pm$1.90E-06 & 1.40E+01$\pm$9.72E+00 & 210/240 \\
bn111022854 & 0.129$\pm$0.141 &  0.77 $\pm$ 0.09 & 281.90 $\pm$ 55.16 &  1.78E-06$\pm$3.00E-07 & 8.92E+00$\pm$1.12E+00 & 228/242 \\
bn111024896 & 0.288$\pm$0.135 &  0.10 $\pm$ 0.12 & 141.64 $\pm$ 17.88 &  9.32E-07$\pm$2.07E-07 & 4.24E+00$\pm$8.56E-01 & 234/240 \\
bn111103948 & 0.303$\pm$0.123 &  0.72 $\pm$ 0.05 & 829.05 $\pm$ 77.88 &  4.58E-06$\pm$4.19E-07 & 8.57E+00$\pm$6.58E-01 & 220/242 \\
bn111221739 & 0.413$\pm$0.045 &  0.74 $\pm$ 0.02 & 1087.50 $\pm$ 41.34 &  1.37E-05$\pm$3.70E-07 & 2.08E+01$\pm$4.92E-01 & 265/242 \\
bn120101354 & 0.168$\pm$0.186 &  0.24 $\pm$ 0.60 & 114.31 $\pm$ 77.34 &  7.52E-07$\pm$1.27E-07 & 4.65E+00$\pm$6.13E-01 & 186/240 \\
bn120118898 & 0.753$\pm$0.051 &  0.88 $\pm$ 0.09 & 125.95 $\pm$ 17.11 &  9.93E-07$\pm$1.11E-07 & 1.04E+01$\pm$9.45E-01 & 260/240 \\
bn120210650 & 1.382$\pm$0.060 &  1.63 $\pm$ 0.11 & 1196.14 $\pm$ 1532.06 &  7.13E-07$\pm$7.17E-07 & 6.36E+00$\pm$6.36E+00 & 252/242 \\
bn120222021 & 0.890$\pm$0.031 &  0.60 $\pm$ 0.05 & 89.88 $\pm$ 5.02 &  1.58E-06$\pm$6.36E-08 & 1.64E+01$\pm$6.46E-01 & 281/242 \\
bn120302722 & 0.516$\pm$0.166 &  0.21 $\pm$ 0.57 & 95.47 $\pm$ 62.27 &  4.34E-07$\pm$7.87E-08 & 2.89E+00$\pm$4.77E-01 & 243/242 \\
bn120323507 & 0.450$\pm$0.010 &  1.49 $\pm$ 0.03 & 600.95 $\pm$ 83.16 &  1.37E-05$\pm$6.29E-07 & 1.38E+02$\pm$2.95E+00 & 267/238 \\
bn120327418 & 0.225$\pm$0.187 &  0.11 $\pm$ 0.77 & 92.66 $\pm$ 67.49 &  6.86E-07$\pm$1.93E-07 & 5.08E+00$\pm$8.44E-01 & 223/240 \\
bn120331055 & 0.110$\pm$0.078 &  0.39 $\pm$ 0.52 & 28.08 $\pm$ 8.81 &  2.10E-06$\pm$1.40E-07 & 4.41E+01$\pm$2.44E+00 & 173/241 \\
bn120410585 & 0.178$\pm$0.130 &  -0.93 $\pm$ 0.21 & 116.91 $\pm$ 15.50 &  2.25E-06$\pm$8.03E-07 & 6.15E+00$\pm$2.08E+00 & 268/240 \\
bn120415891 & 0.214$\pm$0.162 &  1.03 $\pm$ 0.17 & 2203.59 $\pm$ 1845.33 &  3.77E-06$\pm$3.82E-06 & 8.40E+00$\pm$8.43E+00 & 217/240 \\
bn120429003 & 0.838$\pm$0.132 &  0.71 $\pm$ 0.35 & 188.99 $\pm$ 104.45 &  5.26E-07$\pm$4.54E-08 & 3.40E+00$\pm$2.20E-01 & 237/238 \\
bn120519721 & 0.787$\pm$0.053 &  0.35 $\pm$ 0.03 & 452.42 $\pm$ 23.68 &  4.62E-06$\pm$1.67E-07 & 9.04E+00$\pm$2.42E-01 & 265/241 \\
bn120524134 & 0.241$\pm$0.083 &  0.41 $\pm$ 0.43 & 36.60 $\pm$ 11.72 &  6.70E-07$\pm$4.03E-08 & 1.18E+01$\pm$6.13E-01 & 214/242 \\
bn120603439 & 0.302$\pm$0.081 &  0.73 $\pm$ 0.15 & 702.98 $\pm$ 279.14 &  4.31E-06$\pm$2.15E-06 & 9.52E+00$\pm$3.79E+00 & 249/241 \\
bn120604220 & 1.424$\pm$0.122 &  0.52 $\pm$ 0.42 & 62.49 $\pm$ 28.38 &  2.61E-07$\pm$1.79E-08 & 3.35E+00$\pm$1.65E-01 & 263/239 \\
bn120605453 & 2.205$\pm$0.036 &  1.29 $\pm$ 0.03 & 704.99 $\pm$ 80.25 &  1.46E-06$\pm$7.11E-08 & 8.67E+00$\pm$2.33E-01 & 274/239 \\
bn120608489 & 0.275$\pm$0.144 &  0.88 $\pm$ 0.37 & 628.18 $\pm$ 620.08 &  1.67E-06$\pm$1.62E-07 & 5.17E+00$\pm$4.55E-01 & 223/242 \\
bn120609580 & 1.089$\pm$0.113 &  -0.39 $\pm$ 0.10 & 31.72 $\pm$ 1.74 &  2.76E-07$\pm$2.99E-08 & 3.54E+00$\pm$3.59E-01 & 285/242 \\
bn120612687 & 0.171$\pm$0.091 &  0.54 $\pm$ 0.06 & 560.75 $\pm$ 74.92 &  5.77E-06$\pm$5.64E-07 & 1.17E+01$\pm$7.54E-01 & 234/242 \\
bn120619884 & 0.500$\pm$0.193 &  0.30 $\pm$ 0.10 & 423.96 $\pm$ 71.10 &  1.20E-06$\pm$4.02E-07 & 2.65E+00$\pm$8.55E-01 & 257/241 \\
bn120624309 & 0.709$\pm$0.029 &  0.85 $\pm$ 0.01 & 3081.29 $\pm$ 81.57 &  5.24E-05$\pm$1.04E-06 & 3.97E+01$\pm$6.24E-01 & 247/243 \\
bn120716712 & 0.742$\pm$0.049 &  0.65 $\pm$ 0.07 & 127.37 $\pm$ 10.52 &  1.17E-06$\pm$7.96E-08 & 9.67E+00$\pm$5.98E-01 & 248/241 \\
bn120811014 & 0.341$\pm$0.058 &  0.22 $\pm$ 0.03 & 669.75 $\pm$ 30.77 &  1.30E-05$\pm$5.68E-07 & 1.50E+01$\pm$6.72E-01 & 232/241 \\
bn120814201 & 0.696$\pm$0.107 &  0.67 $\pm$ 0.04 & 555.88 $\pm$ 52.15 &  2.54E-06$\pm$1.81E-07 & 6.33E+00$\pm$3.22E-01 & 238/243 \\
bn120817168 & 0.078$\pm$0.049 &  0.60 $\pm$ 0.05 & 836.69 $\pm$ 60.04 &  4.57E-05$\pm$4.04E-06 & 7.00E+01$\pm$6.01E+00 & 223/240 \\
bn120827216 & 1.756$\pm$0.113 &  0.16 $\pm$ 0.06 & 205.17 $\pm$ 12.83 &  1.27E-06$\pm$7.54E-08 & 4.28E+00$\pm$2.08E-01 & 273/243 \\
bn120830297 & 1.043$\pm$0.056 &  0.34 $\pm$ 0.02 & 683.78 $\pm$ 28.70 &  5.92E-06$\pm$1.97E-07 & 7.66E+00$\pm$1.72E-01 & 284/242 \\
bn120831901 & 0.202$\pm$0.139 &  0.50 $\pm$ 0.08 & 469.51 $\pm$ 86.44 &  2.47E-06$\pm$5.00E-07 & 5.66E+00$\pm$7.84E-01 & 257/242 \\
bn120907017 & 1.818$\pm$0.074 &  0.66 $\pm$ 0.28 & 93.20 $\pm$ 34.90 &  3.31E-07$\pm$2.83E-07 & 3.53E+00$\pm$2.96E+00 & 223/242 \\
bn120914144 & 1.228$\pm$0.085 &  0.35 $\pm$ 0.09 & 59.67 $\pm$ 4.45 &  2.98E-07$\pm$2.58E-08 & 3.50E+00$\pm$2.95E-01 & 271/241 \\
bn120915000 & 0.429$\pm$0.124 &  0.25 $\pm$ 0.07 & 431.51 $\pm$ 44.46 &  1.99E-06$\pm$2.39E-07 & 3.66E+00$\pm$3.23E-01 & 272/242 \\
bn121004211 & 1.388$\pm$0.100 &  0.64 $\pm$ 0.07 & 82.78 $\pm$ 7.41 &  2.78E-07$\pm$2.18E-08 & 3.18E+00$\pm$2.19E-01 & 281/242 \\
bn121005340 & 1.416$\pm$0.117 &  0.39 $\pm$ 0.12 & 226.13 $\pm$ 38.22 &  5.11E-07$\pm$1.38E-07 & 1.98E+00$\pm$4.53E-01 & 202/242 \\
bn121012724 & 0.348$\pm$0.080 &  0.50 $\pm$ 0.04 & 364.45 $\pm$ 27.92 &  3.36E-06$\pm$1.76E-07 & 9.57E+00$\pm$4.12E-01 & 211/240 \\
bn121014638 & 0.263$\pm$0.166 &  1.08 $\pm$ 0.48 & 1278.94 $\pm$ 3511.43 &  1.01E-06$\pm$2.98E-07 & 2.37E+00$\pm$5.55E-01 & 219/240 \\
bn121023322 & 0.459$\pm$0.071 &  0.84 $\pm$ 0.02 & 1107.48 $\pm$ 82.65 &  5.01E-06$\pm$2.58E-07 & 9.02E+00$\pm$2.08E-01 & 262/240 \\
bn121116459 & 0.820$\pm$0.127 &  -0.50 $\pm$ 0.14 & 114.56 $\pm$ 11.03 &  7.08E-07$\pm$2.23E-07 & 2.54E+00$\pm$7.52E-01 & 239/242 \\
bn121119579 & 1.590$\pm$0.065 &  1.35 $\pm$ 0.04 & 163.40 $\pm$ 18.15 &  3.60E-07$\pm$1.65E-08 & 5.22E+00$\pm$1.54E-01 & 303/240 \\
bn121122564 & 1.419$\pm$0.091 &  1.27 $\pm$ 0.30 & 112.30 $\pm$ 64.53 &  3.03E-07$\pm$2.19E-07 & 4.92E+00$\pm$3.15E+00 & 305/240 \\
bn121124606 & 0.027$\pm$0.191 &  0.10 $\pm$ 0.65 & 193.96 $\pm$ 139.05 &  2.11E-06$\pm$5.00E-07 & 7.13E+00$\pm$1.39E+00 & 187/240 \\
bn121127914 & 0.165$\pm$0.075 &  0.19 $\pm$ 0.18 & 500.00 $\pm$ 240.40 &  1.27E-05$\pm$1.26E-05 & 1.88E+01$\pm$1.83E+01 & 251/240 \\
bn121211574 & 1.565$\pm$0.139 &  0.79 $\pm$ 0.36 & 109.06 $\pm$ 58.24 &  2.46E-07$\pm$9.67E-09 & 2.63E+00$\pm$8.36E-02 & 236/242 \\
bn121211695 & 1.717$\pm$0.084 &  0.57 $\pm$ 0.12 & 88.33 $\pm$ 12.77 &  3.08E-07$\pm$6.03E-08 & 3.15E+00$\pm$5.61E-01 & 283/241 \\
bn130112286 & 2.805$\pm$0.057 &  0.63 $\pm$ 0.04 & 155.63 $\pm$ 9.27 &  6.87E-07$\pm$2.87E-08 & 4.76E+00$\pm$1.41E-01 & 263/242 \\
bn130112353 & 0.803$\pm$0.081 &  0.61 $\pm$ 0.04 & 392.50 $\pm$ 32.03 &  1.59E-06$\pm$8.78E-08 & 4.93E+00$\pm$1.72E-01 & 257/241 \\
bn130127743 & 0.030$\pm$0.159 &  -0.09 $\pm$ 0.65 & 243.90 $\pm$ 155.34 &  5.85E-06$\pm$1.77E-06 & 1.34E+01$\pm$3.03E+00 & 194/242 \\
bn130204484 & 0.041$\pm$0.127 &  0.74 $\pm$ 0.31 & 500.00 $\pm$ 742.40 &  4.83E-06$\pm$1.39E-06 & 1.44E+01$\pm$2.25E+00 & 224/240 \\
bn130217688 & 1.229$\pm$0.098 &  -0.13 $\pm$ 0.13 & 37.28 $\pm$ 3.25 &  2.68E-07$\pm$4.31E-08 & 3.42E+00$\pm$5.33E-01 & 278/241 \\
bn130307126 & 0.310$\pm$0.070 &  0.76 $\pm$ 0.02 & 1383.95 $\pm$ 83.95 &  1.09E-05$\pm$5.15E-07 & 1.40E+01$\pm$3.65E-01 & 268/240 \\
bn130310840 & 0.778$\pm$0.017 &  -1.04 $\pm$ 0.02 & 2671.11 $\pm$ 261.03 &  4.96E-05$\pm$3.33E-06 & 5.19E+01$\pm$1.19E+00 & 315/284 \\
bn130416770 & 0.098$\pm$0.082 &  0.70 $\pm$ 0.07 & 817.67 $\pm$ 90.54 &  1.32E-05$\pm$1.97E-06 & 2.41E+01$\pm$3.50E+00 & 262/242 \\
bn130504314 & 0.374$\pm$0.032 &  0.34 $\pm$ 0.06 & 740.79 $\pm$ 63.29 &  2.90E-05$\pm$2.82E-06 & 3.50E+01$\pm$2.60E+00 & 268/238 \\
bn130515056 & 0.158$\pm$0.083 &  -0.05 $\pm$ 0.09 & 206.80 $\pm$ 16.11 &  5.40E-06$\pm$7.88E-07 & 1.50E+01$\pm$2.09E+00 & 235/241 \\
bn130518551 & 1.612$\pm$0.050 &  0.70 $\pm$ 0.02 & 750.94 $\pm$ 51.48 &  3.17E-06$\pm$1.57E-07 & 6.30E+00$\pm$1.35E-01 & 277/241 \\
bn130617564 & 0.693$\pm$0.114 &  0.72 $\pm$ 0.25 & 367.86 $\pm$ 185.08 &  1.28E-06$\pm$1.28E-06 & 4.85E+00$\pm$4.84E+00 & 209/243 \\
bn130622615 & 1.174$\pm$0.094 &  0.76 $\pm$ 0.32 & 112.90 $\pm$ 55.28 &  3.41E-07$\pm$1.33E-08 & 3.46E+00$\pm$1.27E-01 & 253/240 \\
bn130628860 & 0.429$\pm$0.103 &  0.64 $\pm$ 0.04 & 798.31 $\pm$ 77.64 &  3.99E-06$\pm$2.89E-07 & 6.82E+00$\pm$2.77E-01 & 254/240 \\
bn130701761 & 1.490$\pm$0.042 &  0.40 $\pm$ 0.02 & 674.34 $\pm$ 22.13 &  6.86E-06$\pm$1.48E-07 & 9.75E+00$\pm$1.58E-01 & 249/241 \\
bn130706900 & 0.044$\pm$0.152 &  0.75 $\pm$ 0.28 & 499.94 $\pm$ 647.47 &  4.60E-06$\pm$1.18E-06 & 1.41E+01$\pm$1.80E+00 & 162/241 \\
bn130716442 & 0.696$\pm$0.116 &  0.62 $\pm$ 0.25 & 460.90 $\pm$ 278.73 &  1.24E-06$\pm$1.97E-08 & 3.35E+00$\pm$8.83E-02 & 198/242 \\
bn130722990 & 0.710$\pm$0.110 &  1.48 $\pm$ 0.63 & 115.00 $\pm$ 106.15 &  7.03E-07$\pm$1.01E-07 & 1.38E+01$\pm$1.41E+00 & 261/241 \\
bn130802730 & 0.042$\pm$0.178 &  0.16 $\pm$ 0.46 & 192.37 $\pm$ 109.70 &  3.75E-06$\pm$3.41E-07 & 1.35E+01$\pm$9.33E-01 & 185/241 \\
bn130804023 & 0.876$\pm$0.061 &  0.29 $\pm$ 0.05 & 232.12 $\pm$ 13.12 &  2.00E-06$\pm$8.65E-08 & 6.82E+00$\pm$2.56E-01 & 236/241 \\
bn130808253 & 0.175$\pm$0.107 &  0.29 $\pm$ 0.45 & 46.98 $\pm$ 17.44 &  1.00E-06$\pm$5.99E-08 & 1.37E+01$\pm$6.85E-01 & 218/242 \\
bn130912358 & 0.282$\pm$0.068 &  0.72 $\pm$ 0.04 & 514.06 $\pm$ 45.60 &  4.14E-06$\pm$2.11E-07 & 1.18E+01$\pm$4.40E-01 & 261/241 \\
bn130919173 & 0.930$\pm$0.086 &  0.31 $\pm$ 0.40 & 59.60 $\pm$ 23.27 &  3.38E-07$\pm$1.62E-08 & 3.85E+00$\pm$1.52E-01 & 275/240 \\
bn131004904 & 0.690$\pm$0.068 &  1.22 $\pm$ 0.24 & 108.07 $\pm$ 49.81 &  5.07E-07$\pm$3.03E-07 & 8.04E+00$\pm$4.35E+00 & 259/240 \\
bn131126163 & 0.134$\pm$0.053 &  0.03 $\pm$ 0.05 & 267.24 $\pm$ 16.28 &  1.18E-05$\pm$7.92E-07 & 2.73E+01$\pm$1.76E+00 & 238/243 \\
bn131128629 & 1.444$\pm$0.087 &  0.61 $\pm$ 0.06 & 63.64 $\pm$ 4.19 &  2.64E-07$\pm$1.14E-08 & 3.57E+00$\pm$1.33E-01 & 255/240 \\
bn131217108 & 0.624$\pm$0.092 &  0.51 $\pm$ 0.02 & 1116.99 $\pm$ 46.32 &  6.09E-06$\pm$2.31E-07 & 6.22E+00$\pm$1.56E-01 & 250/239 \\
bn131217506 & 0.630$\pm$0.103 &  0.90 $\pm$ 0.06 & 84.00 $\pm$ 7.32 &  4.60E-07$\pm$2.48E-08 & 6.47E+00$\pm$2.80E-01 & 257/239 \\
bn140105065 & 0.761$\pm$0.100 &  1.01 $\pm$ 0.14 & 1288.77 $\pm$ 847.17 &  2.05E-06$\pm$1.62E-06 & 4.64E+00$\pm$2.51E+00 & 242/239 \\
bn140129499 & 0.077$\pm$0.166 &  0.74 $\pm$ 0.32 & 357.68 $\pm$ 236.25 &  2.32E-06$\pm$2.14E-07 & 9.24E+00$\pm$5.48E-01 & 174/240 \\
bn140209313 & 1.638$\pm$0.015 &  0.87 $\pm$ 0.03 & 165.99 $\pm$ 5.21 &  2.19E-06$\pm$3.77E-08 & 1.86E+01$\pm$4.37E-01 & 299/240 \\
bn140211091 & 1.554$\pm$0.088 &  1.03 $\pm$ 0.04 & 157.79 $\pm$ 14.17 &  3.63E-07$\pm$1.75E-08 & 3.81E+00$\pm$1.22E-01 & 265/242 \\
bn140329272 & 0.039$\pm$0.131 &  0.65 $\pm$ 0.27 & 474.50 $\pm$ 258.65 &  5.77E-06$\pm$8.88E-07 & 1.58E+01$\pm$1.05E+00 & 223/244 \\
bn140428906 & 0.310$\pm$0.082 &  0.66 $\pm$ 0.05 & 718.29 $\pm$ 93.38 &  9.40E-06$\pm$8.59E-07 & 1.81E+01$\pm$1.00E+00 & 246/243 \\
bn140501139 & 0.190$\pm$0.154 &  0.95 $\pm$ 0.32 & 484.48 $\pm$ 461.27 &  1.29E-06$\pm$2.98E-07 & 5.51E+00$\pm$4.58E-01 & 229/241 \\
bn140506880 & 1.376$\pm$0.050 &  0.53 $\pm$ 0.17 & 116.86 $\pm$ 20.90 &  1.53E-06$\pm$5.84E-07 & 1.21E+01$\pm$4.06E+00 & 226/240 \\
bn140511095 & 0.228$\pm$0.126 &  0.78 $\pm$ 0.08 & 334.74 $\pm$ 52.77 &  1.52E-06$\pm$1.88E-07 & 6.76E+00$\pm$5.55E-01 & 302/242 \\
bn140605377 & 0.143$\pm$0.099 &  -0.06 $\pm$ 0.09 & 249.98 $\pm$ 22.91 &  5.16E-06$\pm$7.84E-07 & 1.18E+01$\pm$1.74E+00 & 214/242 \\
bn140610487 & 0.321$\pm$0.164 &  0.51 $\pm$ 0.09 & 404.78 $\pm$ 66.79 &  1.58E-06$\pm$3.00E-07 & 4.16E+00$\pm$5.87E-01 & 268/241 \\
bn140619475 & 0.707$\pm$0.107 &  0.18 $\pm$ 0.04 & 790.97 $\pm$ 39.93 &  4.56E-06$\pm$2.38E-07 & 4.30E+00$\pm$1.72E-01 & 272/241 \\
bn140619490 & 0.115$\pm$0.176 &  0.37 $\pm$ 0.48 & 146.26 $\pm$ 70.95 &  1.88E-06$\pm$2.36E-07 & 1.06E+01$\pm$1.00E+00 & 204/240 \\
bn140624423 & 0.034$\pm$0.098 &  0.51 $\pm$ 0.32 & 159.89 $\pm$ 71.22 &  4.98E-06$\pm$2.57E-07 & 2.96E+01$\pm$9.77E-01 & 172/240 \\
bn140626843 & 1.635$\pm$0.063 &  0.91 $\pm$ 0.18 & 153.36 $\pm$ 48.17 &  6.70E-07$\pm$3.07E-07 & 6.36E+00$\pm$2.35E+00 & 249/242 \\
bn140710537 & 0.359$\pm$0.145 &  -0.38 $\pm$ 0.15 & 127.45 $\pm$ 16.05 &  9.72E-07$\pm$3.31E-07 & 3.39E+00$\pm$1.10E+00 & 246/239 \\
bn140720158 & 0.059$\pm$0.173 &  0.05 $\pm$ 0.67 & 126.48 $\pm$ 84.04 &  1.66E-06$\pm$4.25E-07 & 8.05E+00$\pm$1.63E+00 & 202/243 \\
bn140807500 & 0.500$\pm$0.053 &  0.76 $\pm$ 0.03 & 516.97 $\pm$ 35.91 &  3.12E-06$\pm$1.23E-07 & 9.46E+00$\pm$2.25E-01 & 258/241 \\
bn140831374 & 1.371$\pm$0.085 &  0.99 $\pm$ 0.15 & 794.06 $\pm$ 510.61 &  1.26E-06$\pm$8.97E-07 & 3.93E+00$\pm$1.92E+00 & 296/239 \\
bn140901821 & 0.172$\pm$0.036 &  -0.05 $\pm$ 0.12 & 682.79 $\pm$ 57.83 &  7.91E-05$\pm$3.18E-05 & 6.85E+01$\pm$2.57E+01 & 291/241 \\
bn140930134 & 0.838$\pm$0.147 &  0.07 $\pm$ 0.08 & 245.18 $\pm$ 21.82 &  9.05E-07$\pm$1.27E-07 & 2.37E+00$\pm$2.94E-01 & 238/241 \\
bn141004973 & 1.069$\pm$0.050 &  1.45 $\pm$ 0.02 & 287.07 $\pm$ 27.53 &  7.18E-07$\pm$2.68E-08 & 8.90E+00$\pm$1.36E-01 & 256/240 \\
bn141011282 & 0.073$\pm$0.065 &  0.60 $\pm$ 0.06 & 471.61 $\pm$ 42.89 &  1.43E-05$\pm$1.29E-06 & 3.72E+01$\pm$2.82E+00 & 219/243 \\
bn141102536 & 1.856$\pm$0.072 &  0.69 $\pm$ 0.04 & 453.83 $\pm$ 37.83 &  9.73E-07$\pm$6.13E-08 & 2.95E+00$\pm$1.14E-01 & 293/242 \\
bn141105406 & 0.692$\pm$0.069 &  0.47 $\pm$ 0.04 & 261.44 $\pm$ 15.70 &  1.82E-06$\pm$7.66E-08 & 6.74E+00$\pm$2.58E-01 & 250/241 \\
bn141121414 & 1.411$\pm$0.142 &  0.37 $\pm$ 0.09 & 122.36 $\pm$ 12.02 &  3.14E-07$\pm$3.57E-08 & 2.06E+00$\pm$2.14E-01 & 250/243 \\
bn141122087 & 0.123$\pm$0.185 &  0.46 $\pm$ 0.48 & 210.41 $\pm$ 152.75 &  1.41E-06$\pm$2.34E-07 & 6.36E+00$\pm$7.39E-01 & 210/239 \\
bn141124277 & 0.560$\pm$0.163 &  0.64 $\pm$ 0.07 & 339.16 $\pm$ 38.95 &  1.09E-06$\pm$1.27E-07 & 4.01E+00$\pm$3.75E-01 & 275/240 \\
bn141126233 & 0.947$\pm$0.118 &  1.17 $\pm$ 0.18 & 1875.45 $\pm$ 2496.19 &  1.24E-06$\pm$1.23E-07 & 2.94E+00$\pm$6.12E-02 & 269/240 \\
bn141128962 & 0.185$\pm$0.108 &  0.73 $\pm$ 0.37 & 91.61 $\pm$ 46.29 &  7.98E-07$\pm$2.97E-08 & 9.14E+00$\pm$3.81E-01 & 211/243 \\
bn141202470 & 0.995$\pm$0.060 &  0.20 $\pm$ 0.03 & 370.82 $\pm$ 15.54 &  4.74E-06$\pm$1.55E-07 & 9.44E+00$\pm$2.65E-01 & 239/240 \\
bn141208632 & 0.096$\pm$0.230 &  0.44 $\pm$ 0.51 & 144.94 $\pm$ 101.12 &  1.14E-06$\pm$1.65E-07 & 6.83E+00$\pm$7.47E-01 & 230/241 \\
bn141213300 & 0.415$\pm$0.064 &  0.37 $\pm$ 0.10 & 54.30 $\pm$ 4.38 &  8.35E-07$\pm$8.27E-08 & 1.07E+01$\pm$9.15E-01 & 240/244 \\
bn141222298 & 1.724$\pm$0.017 &  1.51 $\pm$ 0.01 & 10168.70 $\pm$ 1104.26 &  1.80E-05$\pm$1.18E-06 & 4.50E+01$\pm$1.02E+00 & 343/287 \\
bn141230871 & 0.092$\pm$0.142 &  1.07 $\pm$ 0.30 & 494.56 $\pm$ 476.88 &  2.16E-06$\pm$4.47E-07 & 1.11E+01$\pm$9.27E-01 & 237/241 \\
bn150120123 & 1.503$\pm$0.129 &  1.16 $\pm$ 0.58 & 132.59 $\pm$ 127.31 &  2.42E-07$\pm$4.04E-08 & 3.25E+00$\pm$4.49E-01 & 243/240 \\
bn150128624 & 0.097$\pm$0.159 &  0.85 $\pm$ 0.41 & 271.80 $\pm$ 173.24 &  2.24E-06$\pm$2.35E-07 & 1.29E+01$\pm$8.17E-01 & 212/239 \\
bn150325696 & 0.072$\pm$0.131 &  0.09 $\pm$ 0.41 & 193.89 $\pm$ 88.14 &  3.12E-06$\pm$2.37E-07 & 1.04E+01$\pm$6.80E-01 & 224/242 \\
bn150728151 & 1.585$\pm$0.120 &  0.81 $\pm$ 0.29 & 376.38 $\pm$ 275.99 &  1.36E-06$\pm$9.37E-08 & 5.68E+00$\pm$2.83E-01 & 232/242 \\
bn150805746 & 0.772$\pm$0.131 &  0.72 $\pm$ 0.08 & 41.91 $\pm$ 3.18 &  2.44E-07$\pm$1.38E-08 & 4.69E+00$\pm$2.47E-01 & 244/239 \\
bn150810485 & 0.381$\pm$0.056 &  0.69 $\pm$ 0.08 & 1106.43 $\pm$ 214.16 &  1.04E-05$\pm$2.43E-06 & 1.44E+01$\pm$2.39E+00 & 254/238 \\
bn150811849 & 0.548$\pm$0.048 &  0.27 $\pm$ 0.02 & 778.26 $\pm$ 27.28 &  1.56E-05$\pm$4.58E-07 & 1.64E+01$\pm$4.41E-01 & 224/242 \\
bn150819440 & 1.067$\pm$0.018 &  1.16 $\pm$ 0.03 & 588.36 $\pm$ 69.94 &  8.01E-06$\pm$5.57E-07 & 4.26E+01$\pm$1.44E+00 & 294/240 \\
bn150906944 & 0.163$\pm$0.110 &  0.30 $\pm$ 0.23 & 363.16 $\pm$ 109.74 &  9.50E-06$\pm$8.45E-06 & 2.16E+01$\pm$1.90E+01 & 186/242 \\
bn150908408 & 0.607$\pm$0.160 &  0.99 $\pm$ 0.23 & 934.72 $\pm$ 1018.83 &  1.12E-06$\pm$1.12E-06 & 3.13E+00$\pm$3.13E+00 & 265/242 \\
bn150912600 & 0.356$\pm$0.128 &  0.18 $\pm$ 0.08 & 204.17 $\pm$ 19.64 &  1.34E-06$\pm$1.61E-07 & 4.65E+00$\pm$4.95E-01 & 259/243 \\
bn150922234 & 0.151$\pm$0.066 &  0.35 $\pm$ 0.06 & 314.73 $\pm$ 25.37 &  6.45E-06$\pm$5.82E-07 & 1.77E+01$\pm$1.31E+00 & 215/240 \\
bn150923297 & 0.081$\pm$0.146 &  0.28 $\pm$ 0.59 & 60.55 $\pm$ 34.37 &  8.46E-07$\pm$1.28E-07 & 9.39E+00$\pm$1.17E+00 & 231/241 \\
bn150923864 & 1.775$\pm$0.044 &  0.27 $\pm$ 0.07 & 86.87 $\pm$ 5.31 &  7.13E-07$\pm$4.38E-08 & 5.78E+00$\pm$3.11E-01 & 250/240 \\
bn151022577 & 0.309$\pm$0.147 &  0.72 $\pm$ 0.42 & 188.27 $\pm$ 157.36 &  7.37E-07$\pm$7.88E-08 & 4.81E+00$\pm$4.53E-01 & 227/239 \\
bn151212030 & 1.323$\pm$0.164 &  0.61 $\pm$ 0.09 & 137.72 $\pm$ 16.18 &  4.20E-07$\pm$5.40E-08 & 3.15E+00$\pm$3.63E-01 & 262/240 \\
bn151218857 & 0.535$\pm$0.120 &  1.32 $\pm$ 0.25 & 508.78 $\pm$ 673.29 &  7.53E-07$\pm$7.53E-07 & 5.75E+00$\pm$5.75E+00 & 285/240 \\
bn151222340 & 0.568$\pm$0.070 &  0.60 $\pm$ 0.02 & 886.68 $\pm$ 44.00 &  6.26E-06$\pm$2.30E-07 & 9.18E+00$\pm$2.31E-01 & 269/242 \\
bn151228129 & 0.302$\pm$0.122 &  0.21 $\pm$ 0.07 & 383.35 $\pm$ 36.10 &  3.28E-06$\pm$3.42E-07 & 6.45E+00$\pm$5.42E-01 & 269/242 \\
bn151229285 & 0.697$\pm$0.048 &  1.24 $\pm$ 0.06 & 156.82 $\pm$ 16.87 &  1.00E-06$\pm$4.97E-08 & 1.32E+01$\pm$5.06E-01 & 274/242 \\
bn151229486 & 0.152$\pm$0.169 &  0.75 $\pm$ 0.33 & 499.27 $\pm$ 566.15 &  2.12E-06$\pm$6.65E-07 & 6.49E+00$\pm$1.14E+00 & 229/242 \\
bn151231568 & 0.291$\pm$0.067 &  0.48 $\pm$ 0.05 & 287.48 $\pm$ 22.21 &  3.35E-06$\pm$2.05E-07 & 1.16E+01$\pm$6.44E-01 & 227/240 \\
bn160310291 & 1.347$\pm$0.060 &  1.12 $\pm$ 0.02 & 454.10 $\pm$ 40.70 &  1.10E-06$\pm$5.90E-08 & 6.50E+00$\pm$1.47E-01 & 304/239 \\
bn160406503 & 0.123$\pm$0.090 &  0.29 $\pm$ 0.08 & 301.72 $\pm$ 32.74 &  5.43E-06$\pm$7.28E-07 & 1.45E+01$\pm$1.72E+00 & 234/239 \\
bn160408268 & 0.341$\pm$0.076 &  0.70 $\pm$ 0.04 & 544.93 $\pm$ 44.75 &  4.45E-06$\pm$2.19E-07 & 1.18E+01$\pm$4.99E-01 & 277/243 \\
bn160411062 & 0.367$\pm$0.140 &  0.24 $\pm$ 0.10 & 99.34 $\pm$ 10.05 &  5.07E-07$\pm$8.00E-08 & 3.58E+00$\pm$4.78E-01 & 266/242 \\
bn160516237 & 1.248$\pm$0.124 &  0.63 $\pm$ 0.07 & 97.73 $\pm$ 5.82 &  1.09E-06$\pm$8.47E-08 & 1.09E+01$\pm$9.09E-01 & 248/240 \\
bn160519012 & 1.037$\pm$0.129 &  0.90 $\pm$ 0.35 & 139.29 $\pm$ 82.89 &  3.62E-07$\pm$2.14E-08 & 3.62E+00$\pm$1.57E-01 & 259/241 \\
bn160612842 & 0.096$\pm$0.090 &  0.29 $\pm$ 0.18 & 500.00 $\pm$ 251.93 &  1.05E-05$\pm$9.78E-06 & 1.74E+01$\pm$1.45E+01 & 249/243 \\
bn160714097 & 0.239$\pm$0.166 &  0.50 $\pm$ 0.52 & 143.72 $\pm$ 98.64 &  5.97E-07$\pm$6.61E-08 & 3.86E+00$\pm$3.15E-01 & 269/241 \\
bn160715298 & 0.050$\pm$0.058 &  0.13 $\pm$ 0.32 & 16.41 $\pm$ 2.10 &  2.86E-06$\pm$2.16E-06 & 7.66E+01$\pm$5.86E+01 & 170/240 \\
bn160720275 & 1.917$\pm$0.124 &  0.52 $\pm$ 0.11 & 179.56 $\pm$ 24.22 &  4.40E-07$\pm$9.33E-08 & 2.41E+00$\pm$4.56E-01 & 335/239 \\
bn160721806 & 0.216$\pm$0.102 &  0.24 $\pm$ 0.32 & 295.60 $\pm$ 110.10 &  3.78E-06$\pm$2.26E-07 & 9.79E+00$\pm$2.40E-01 & 209/241 \\
bn160726065 & 0.712$\pm$0.068 &  1.11 $\pm$ 0.03 & 694.85 $\pm$ 82.14 &  1.78E-06$\pm$1.09E-07 & 7.67E+00$\pm$1.96E-01 & 244/242 \\
bn160727971 & 0.754$\pm$0.104 &  1.11 $\pm$ 0.06 & 48.68 $\pm$ 3.33 &  3.23E-07$\pm$1.14E-08 & 7.21E+00$\pm$2.17E-01 & 294/242 \\
bn160804180 & 0.616$\pm$0.065 &  0.55 $\pm$ 0.03 & 537.49 $\pm$ 41.07 &  4.35E-06$\pm$2.40E-07 & 9.39E+00$\pm$2.84E-01 & 232/243 \\
bn160804968 & 0.094$\pm$0.121 &  0.52 $\pm$ 0.24 & 499.92 $\pm$ 357.96 &  5.78E-06$\pm$5.90E-06 & 1.26E+01$\pm$1.27E+01 & 231/243 \\
bn160806584 & 1.234$\pm$0.047 &  0.51 $\pm$ 0.05 & 91.59 $\pm$ 4.81 &  1.02E-06$\pm$4.38E-08 & 9.62E+00$\pm$3.54E-01 & 238/241 \\
bn160818198 & 1.227$\pm$0.064 &  0.88 $\pm$ 0.06 & 77.09 $\pm$ 5.68 &  5.30E-07$\pm$2.69E-08 & 7.72E+00$\pm$2.95E-01 & 266/242 \\
bn160818230 & 0.407$\pm$0.142 &  0.35 $\pm$ 0.11 & 132.66 $\pm$ 14.18 &  8.61E-07$\pm$1.44E-07 & 5.18E+00$\pm$7.58E-01 & 248/242 \\
bn160820496 & 0.307$\pm$0.092 &  0.67 $\pm$ 0.05 & 466.12 $\pm$ 47.15 &  2.75E-06$\pm$1.90E-07 & 7.86E+00$\pm$3.98E-01 & 250/241 \\
bn160821937 & 0.102$\pm$0.146 &  1.51 $\pm$ 0.30 & 425.56 $\pm$ 529.87 &  9.72E-07$\pm$9.80E-07 & 1.08E+01$\pm$1.08E+01 & 230/240 \\
bn160822672 & 0.041$\pm$0.069 &  0.72 $\pm$ 0.14 & 507.20 $\pm$ 169.53 &  1.73E-05$\pm$8.02E-06 & 4.98E+01$\pm$1.68E+01 & 228/236 \\
bn160826938 & 1.155$\pm$0.125 &  0.29 $\pm$ 0.11 & 55.42 $\pm$ 5.37 &  2.06E-07$\pm$3.00E-08 & 2.46E+00$\pm$3.33E-01 & 287/243 \\
bn160829334 & 0.246$\pm$0.140 &  -0.09 $\pm$ 0.44 & 230.29 $\pm$ 96.91 &  1.89E-06$\pm$2.65E-07 & 4.61E+00$\pm$5.65E-01 & 207/243 \\
bn160917479 & 1.113$\pm$0.073 &  0.85 $\pm$ 0.20 & 500.00 $\pm$ 538.04 &  7.71E-07$\pm$7.83E-07 & 2.75E+00$\pm$2.76E+00 & 273/242 \\
bn161015400 & 0.099$\pm$0.169 &  0.40 $\pm$ 0.10 & 286.09 $\pm$ 47.09 &  1.91E-06$\pm$3.76E-07 & 6.04E+00$\pm$1.04E+00 & 255/243 \\
bn161026373 & 0.122$\pm$0.144 &  0.61 $\pm$ 0.36 & 262.80 $\pm$ 161.27 &  1.77E-06$\pm$1.51E-07 & 7.73E+00$\pm$4.73E-01 & 230/242 \\
bn161128216 & 0.577$\pm$0.109 &  1.04 $\pm$ 0.06 & 201.72 $\pm$ 22.90 &  5.75E-07$\pm$4.01E-08 & 5.20E+00$\pm$2.77E-01 & 266/242 \\
bn161212652 & 0.577$\pm$0.160 &  0.83 $\pm$ 0.31 & 500.00 $\pm$ 793.00 &  9.67E-07$\pm$2.82E-07 & 3.31E+00$\pm$4.80E-01 & 233/241 \\
bn161218222 & 0.187$\pm$0.085 &  0.37 $\pm$ 0.03 & 1224.74 $\pm$ 55.18 &  2.41E-05$\pm$1.18E-06 & 1.87E+01$\pm$9.99E-01 & 255/240 \\
bn161227498 & 0.852$\pm$0.084 &  0.79 $\pm$ 0.18 & 177.43 $\pm$ 54.90 &  5.63E-07$\pm$2.79E-07 & 4.19E+00$\pm$1.71E+00 & 223/241 \\
bn161228405 & 1.104$\pm$0.070 &  0.53 $\pm$ 0.23 & 182.38 $\pm$ 62.36 &  6.99E-07$\pm$5.97E-07 & 3.84E+00$\pm$3.18E+00 & 203/240 \\
bn170111760 & 0.138$\pm$0.152 &  0.32 $\pm$ 0.42 & 347.38 $\pm$ 222.20 &  2.62E-06$\pm$2.14E-07 & 6.34E+00$\pm$4.64E-01 & 236/238 \\
bn170111815 & 0.453$\pm$0.103 &  0.77 $\pm$ 0.07 & 123.25 $\pm$ 12.38 &  7.59E-07$\pm$6.27E-08 & 7.29E+00$\pm$4.72E-01 & 273/242 \\
bn170124528 & 0.269$\pm$0.140 &  0.03 $\pm$ 0.49 & 112.87 $\pm$ 56.08 &  9.19E-07$\pm$1.16E-07 & 4.85E+00$\pm$5.40E-01 & 228/243 \\
bn170127067 & 0.144$\pm$0.039 &  -0.30 $\pm$ 0.05 & 369.80 $\pm$ 12.74 &  6.24E-05$\pm$4.58E-06 & 8.04E+01$\pm$5.30E+00 & 230/241 \\
bn170127634 & 0.397$\pm$0.106 &  0.46 $\pm$ 0.26 & 307.41 $\pm$ 141.10 &  1.51E-06$\pm$3.87E-08 & 4.83E+00$\pm$1.22E-01 & 227/242 \\
bn170205521 & 1.226$\pm$0.034 &  1.53 $\pm$ 0.17 & 74.82 $\pm$ 18.83 &  6.75E-07$\pm$1.83E-07 & 1.59E+01$\pm$3.52E+00 & 221/243 \\
bn170206453 & 1.300$\pm$0.019 &  0.62 $\pm$ 0.04 & 308.21 $\pm$ 20.22 &  5.89E-06$\pm$2.79E-07 & 2.28E+01$\pm$8.32E-01 & 277/240 \\
bn170208758 & 1.324$\pm$0.056 &  0.81 $\pm$ 0.06 & 317.21 $\pm$ 36.19 &  1.25E-06$\pm$1.06E-07 & 6.01E+00$\pm$4.05E-01 & 270/241 \\
bn170212034 & 1.387$\pm$0.139 &  0.10 $\pm$ 0.26 & 470.90 $\pm$ 116.47 &  1.23E-06$\pm$1.34E-07 & 1.74E+00$\pm$1.40E-01 & 294/238 \\
bn170219002 & 0.183$\pm$0.093 &  0.46 $\pm$ 0.05 & 671.89 $\pm$ 48.90 &  2.02E-05$\pm$1.65E-06 & 3.10E+01$\pm$2.28E+00 & 210/240 \\
bn170222209 & 1.202$\pm$0.049 &  0.10 $\pm$ 0.03 & 383.86 $\pm$ 14.93 &  5.00E-06$\pm$1.47E-07 & 8.72E+00$\pm$2.40E-01 & 238/242 \\
bn170302166 & 1.416$\pm$0.087 &  0.83 $\pm$ 0.08 & 38.15 $\pm$ 3.07 &  2.16E-07$\pm$1.24E-08 & 4.70E+00$\pm$2.37E-01 & 260/241 \\
bn170304003 & 0.193$\pm$0.083 &  0.60 $\pm$ 0.34 & 58.15 $\pm$ 20.04 &  1.29E-06$\pm$1.18E-06 & 1.85E+01$\pm$1.65E+01 & 211/242 \\
bn170305256 & 0.461$\pm$0.050 &  0.29 $\pm$ 0.05 & 159.22 $\pm$ 8.82 &  3.32E-06$\pm$1.69E-07 & 1.61E+01$\pm$7.87E-01 & 257/242 \\
bn170325331 & 0.104$\pm$0.147 &  0.66 $\pm$ 0.28 & 499.95 $\pm$ 571.18 &  3.82E-06$\pm$1.08E-06 & 1.02E+01$\pm$1.60E+00 & 227/241 \\
bn170506169 & 0.394$\pm$0.112 &  0.48 $\pm$ 0.26 & 488.72 $\pm$ 285.45 &  2.46E-06$\pm$5.45E-09 & 5.29E+00$\pm$2.37E-01 & 228/242 \\
bn170516327 & 0.143$\pm$0.140 &  1.36 $\pm$ 0.16 & 38.01 $\pm$ 7.05 &  5.69E-07$\pm$1.11E-07 & 1.63E+01$\pm$2.93E+00 & 204/239 \\
bn170604603 & 0.308$\pm$0.089 &  0.52 $\pm$ 0.04 & 683.61 $\pm$ 65.92 &  6.80E-06$\pm$5.20E-07 & 1.13E+01$\pm$4.51E-01 & 206/241 \\
bn170708046 & 0.126$\pm$0.051 &  0.91 $\pm$ 0.08 & 284.87 $\pm$ 29.71 &  6.59E-06$\pm$7.39E-07 & 3.98E+01$\pm$4.21E+00 & 239/240 \\
bn170709334 & 0.868$\pm$0.109 &  0.26 $\pm$ 0.06 & 288.49 $\pm$ 20.91 &  1.19E-06$\pm$9.48E-08 & 3.22E+00$\pm$2.05E-01 & 253/239 \\
bn170711713 & 0.462$\pm$0.114 &  0.66 $\pm$ 0.03 & 1464.51 $\pm$ 131.65 &  5.92E-06$\pm$5.68E-07 & 6.05E+00$\pm$2.31E-01 & 234/240 \\
bn170713996 & 0.681$\pm$0.084 &  0.09 $\pm$ 0.11 & 18.76 $\pm$ 1.21 &  2.81E-07$\pm$2.40E-08 & 6.78E+00$\pm$5.33E-01 & 229/240 \\
bn170726249 & 0.668$\pm$0.105 &  0.84 $\pm$ 0.05 & 287.63 $\pm$ 31.73 &  8.02E-07$\pm$6.28E-08 & 4.35E+00$\pm$1.85E-01 & 285/242 \\
bn170726794 & 0.214$\pm$0.155 &  0.59 $\pm$ 0.11 & 364.52 $\pm$ 67.83 &  2.09E-06$\pm$5.02E-07 & 6.74E+00$\pm$1.27E+00 & 263/239 \\
bn170728961 & 0.856$\pm$0.033 &  0.44 $\pm$ 0.06 & 98.03 $\pm$ 5.62 &  1.60E-06$\pm$9.50E-08 & 1.66E+01$\pm$1.02E+00 & 264/241 \\
bn170802638 & 0.409$\pm$0.043 &  0.79 $\pm$ 0.07 & 834.84 $\pm$ 158.45 &  8.46E-06$\pm$1.61E-06 & 1.77E+01$\pm$2.16E+00 & 241/239 \\
bn170803729 & 0.852$\pm$0.045 &  0.54 $\pm$ 0.07 & 110.42 $\pm$ 8.58 &  1.13E-06$\pm$8.50E-08 & 9.42E+00$\pm$6.11E-01 & 265/242 \\
bn170816599 & 1.520$\pm$0.052 &  0.51 $\pm$ 0.01 & 812.73 $\pm$ 25.67 &  4.87E-06$\pm$1.23E-07 & 6.76E+00$\pm$9.63E-02 & 255/239 \\
\hline
\enddata
\end{deluxetable}

\begin{deluxetable}{cccccccccc}
\tabletypesize{\tiny}
\tablecaption{The spectral parameters of GRBs observed by both GBM and LLE for our sample.}
\tablewidth{0pt}
\tabcolsep=2.5pt
\tablehead{ \colhead{GRB} & \colhead{Model} & \colhead{$T_{\rm {90}}$ (s)} & \colhead{$\alpha$} & \colhead{$\beta$} & \colhead{$E_{\rm p}$ (keV)} & \colhead{$\lambda$} &  \colhead{ $F (\rm {erg~cm^{-2}~s^{-1}})$} & \colhead{ $P (\rm {Photo~cm^{-2}~s^{-1}})$} &\colhead{PGS/dof} }
 \startdata
\hline
bn090227772 & band & 0.197$\pm$0.027 &  -0.39 $\pm$ 0.03 & -2.74 $\pm$ 0.07 & 1156.32 $\pm$ 78.83 &  -- & 1.34E-04$\pm$5.97E-06  & 9.04E+01$\pm$2.21E+00 & 286/287 \\
bn090510016 & band+pl & 0.846$\pm$0.036 &  -0.68 $\pm$ 0.08 & -3.42 $\pm$ 0.22 & 2987.80 $\pm$ 449.33 &  1.64 $\pm$ 1.64  & 2.68E-05$\pm$3.18E-06 & 1.16E+01$\pm$7.13E-01 & 420/313 \\
bn110529034 & band & 0.395$\pm$0.043 &  -0.77 $\pm$ 0.08 & -2.54 $\pm$ 0.10 & 857.74 $\pm$ 212.80 &  -- & 1.91E-05$\pm$2.59E-06  & 2.66E+01$\pm$2.04E+00 & 232/284 \\
bn130310840 & band & 0.778$\pm$0.017 &  -1.04 $\pm$ 0.02 & -2.60 $\pm$ 0.09 & 2671.11 $\pm$ 261.03 &  -- & 4.96E-05$\pm$3.33E-06  & 5.19E+01$\pm$1.19E+00 & 315/284 \\
bn160709826 & band+pl & 0.793$\pm$0.046 &  -0.40 $\pm$ 0.12 & -2.95 $\pm$ 0.09 & 1061.37 $\pm$ 191.32 &  2.71 $\pm$ 2.71  & 1.78E-05$\pm$1.98E+36 & 1.45E+01$\pm$4.53E+38 & 306/282 \\
\hline
\enddata
\end{deluxetable}

\begin{deluxetable}{cccc}
\tabletypesize{\tiny}
\tablecaption{GRB 170817A-like sGRBs obtained from our deep search with the same criterion as that for GRB 170817A}
\tablewidth{0pt}
\tabcolsep=2.5pt
\tablehead{ \colhead{GRB} &  \colhead{$T_{\rm {90}}$ (s)} & \colhead{ $T_{\rm start}$ }}
 \startdata
\hline
bn080724401 & 0.914$\pm$0.139 &  -0.201 \\
bn080913735 & 0.893$\pm$0.147 &  -0.688 \\
bn080928628 & 0.718$\pm$0.127 &  -0.413 \\
bn081003644 & 0.606$\pm$0.190 &  0.211 \\
bn081213173 & 0.056$\pm$0.196 &  -0.039 \\
bn081215880 & 1.083$\pm$0.130 &  -0.575 \\
bn090518244 & 2.123$\pm$0.061 &  -0.464 \\
bn090917661 & 1.962$\pm$0.099 &  -0.443 \\
bn100116897 & 1.959$\pm$0.115 &  -0.406 \\
bn100401297 & 1.147$\pm$0.143 &  -0.188 \\
bn100714672 & 0.225$\pm$0.140 &  -0.316 \\
bn110304071 & 1.521$\pm$0.090 &  -0.103 \\
bn110517453 & 0.198$\pm$0.190 &  -0.094 \\
bn110709862 & 1.889$\pm$0.143 &  -0.480 \\
bn130206817 & 0.451$\pm$0.233 &  2.193 \\
bn130620498 & 1.469$\pm$0.135 &  -0.382 \\
bn130627372 & 0.655$\pm$0.198 &  -0.512 \\
bn130705398 & 0.208$\pm$0.193 &  -0.036 \\
bn130802730 & 0.042$\pm$0.178 &  -0.050 \\
bn130924910 & 1.490$\pm$0.143 &  -0.602 \\
bn131123543 & 1.281$\pm$0.147 &  -0.313 \\
bn140105748 & 1.451$\pm$0.118 &  -0.985 \\
bn140110814 & 1.484$\pm$0.102 &  -1.052 \\
bn140216331 & 1.154$\pm$0.207 &  -0.292 \\
bn140304557 & 1.962$\pm$0.155 &  -0.577 \\
bn140406120 & 0.441$\pm$0.113 &  -0.281 \\
bn140521732 & 1.298$\pm$0.100 &  -0.151 \\
bn140716306 & 1.073$\pm$0.126 &  -0.560 \\
bn140724533 & 0.342$\pm$0.198 &  -0.259 \\
bn140912664 & 1.728$\pm$0.132 &  -1.148 \\
bn141102112 & 0.024$\pm$0.208 &  -0.035 \\
bn141105358 & 0.895$\pm$0.174 &  -0.106 \\
bn150911315 & 0.910$\pm$0.140 &  -0.376 \\
bn151014592 & 1.473$\pm$0.160 &  -0.604 \\
bn160101215 & 1.986$\pm$0.106 &  -0.018 \\
bn160225720 & 0.682$\pm$0.335 &  1.155 \\
bn160314473 & 0.113$\pm$0.195 &  -0.040 \\
bn160323293 & 0.362$\pm$0.170 &  3.695 \\
bn160816414 & 1.828$\pm$0.164 &  -0.509 \\
bn161106786 & 1.224$\pm$0.152 &  0.297 \\
bn161230298 & 0.082$\pm$0.169 &  -0.052 \\
bn170112970 & 1.564$\pm$0.147 &  -0.225 \\
bn170203486 & 0.236$\pm$0.178 &  -0.063 \\
bn170219110 & 0.490$\pm$0.158 &  -0.175 \\
bn170310417 & 1.390$\pm$0.170 &  -0.710 \\
bn170520202 & 1.819$\pm$0.137 &  -0.832 \\
bn170711019 & 1.637$\pm$0.129 &  -1.139 \\
\hline
\enddata
\end{deluxetable}

\clearpage

\begin{figure*}
\includegraphics[origin=c,angle=0,scale=0.6,width=0.4\textwidth,height=0.4\textheight]{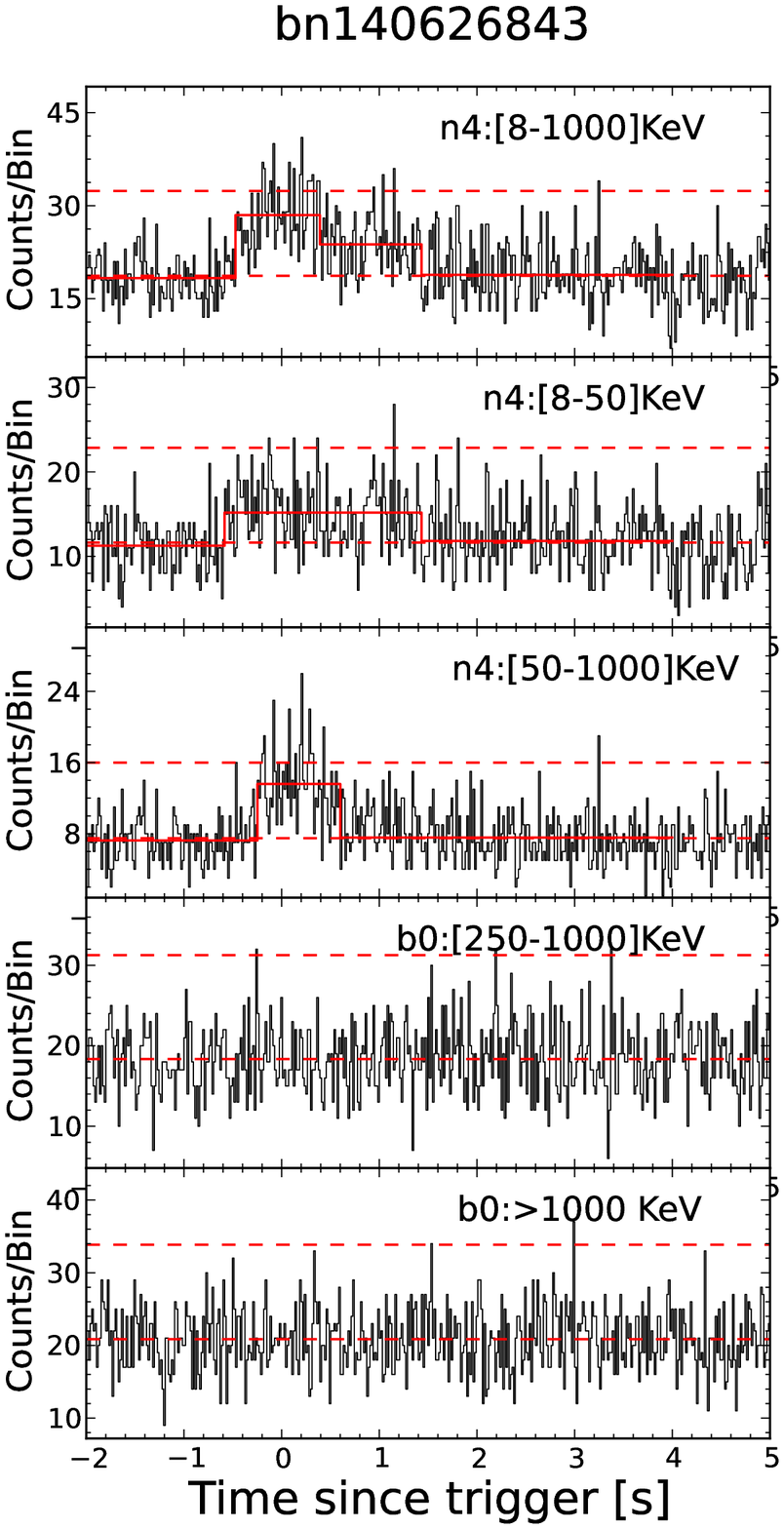}
\includegraphics[origin=c,angle=-90,scale=0.6,width=0.4\textwidth,height=0.4\textheight]{./f3b1SP.eps}
\includegraphics[origin=c,angle=0,scale=0.6,width=0.4\textwidth,height=0.4\textheight]{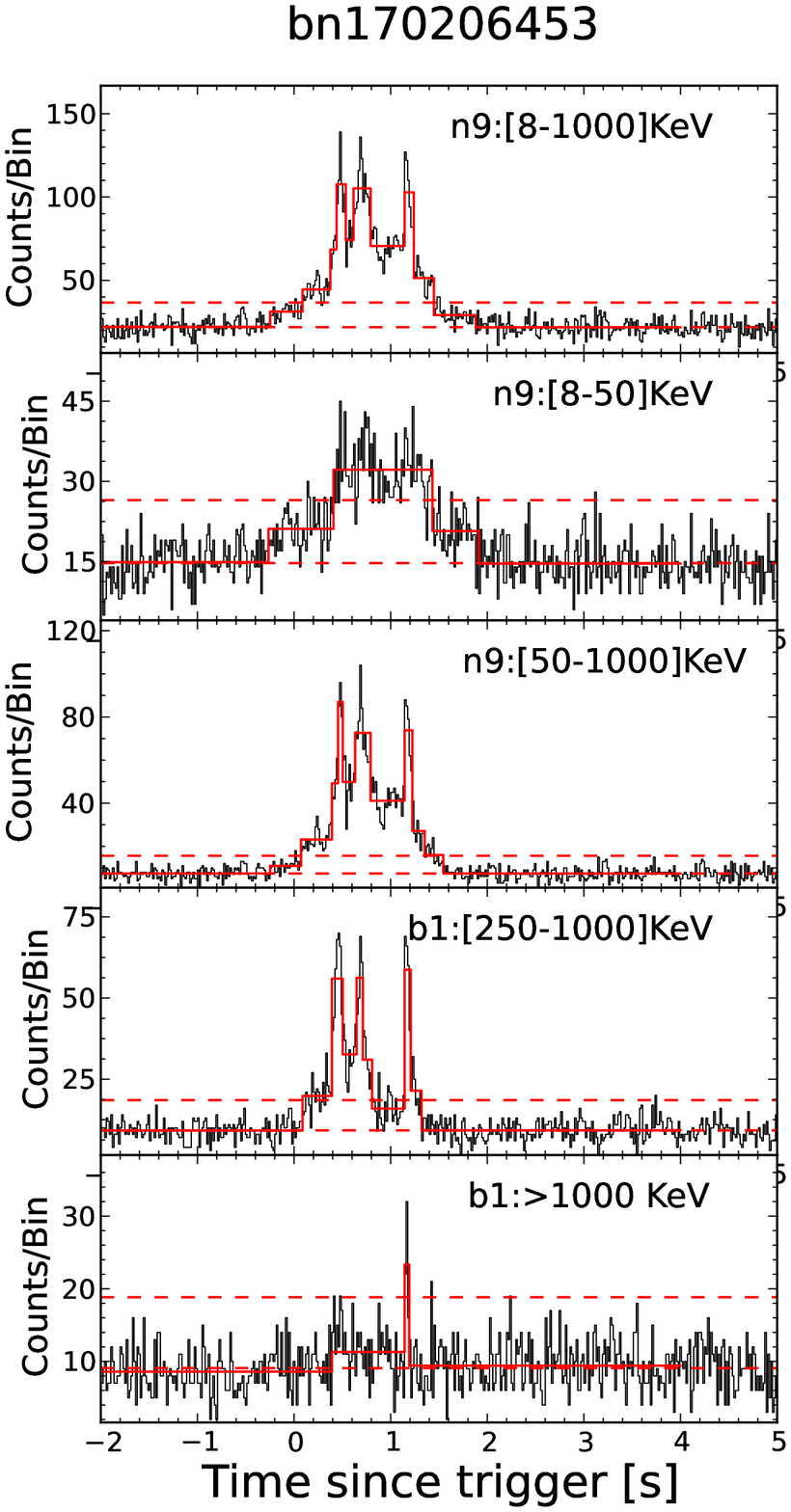}
\hfill
\includegraphics[origin=c,angle=-90,scale=0.6,width=0.4\textwidth,height=0.4\textheight]{./f3b2SP.eps}

\center{Fig. \ref{fig:LCexample}(b). Examples of Pattern II SGRB lightcurves and spectra together with our Bayesian block analysis. Symbols are the same as Fig. \ref{fig:LCexample}(a).}
\end{figure*}

\begin{figure*}
\includegraphics[origin=c,angle=0,scale=0.6,width=0.4\textwidth,height=0.4\textheight]{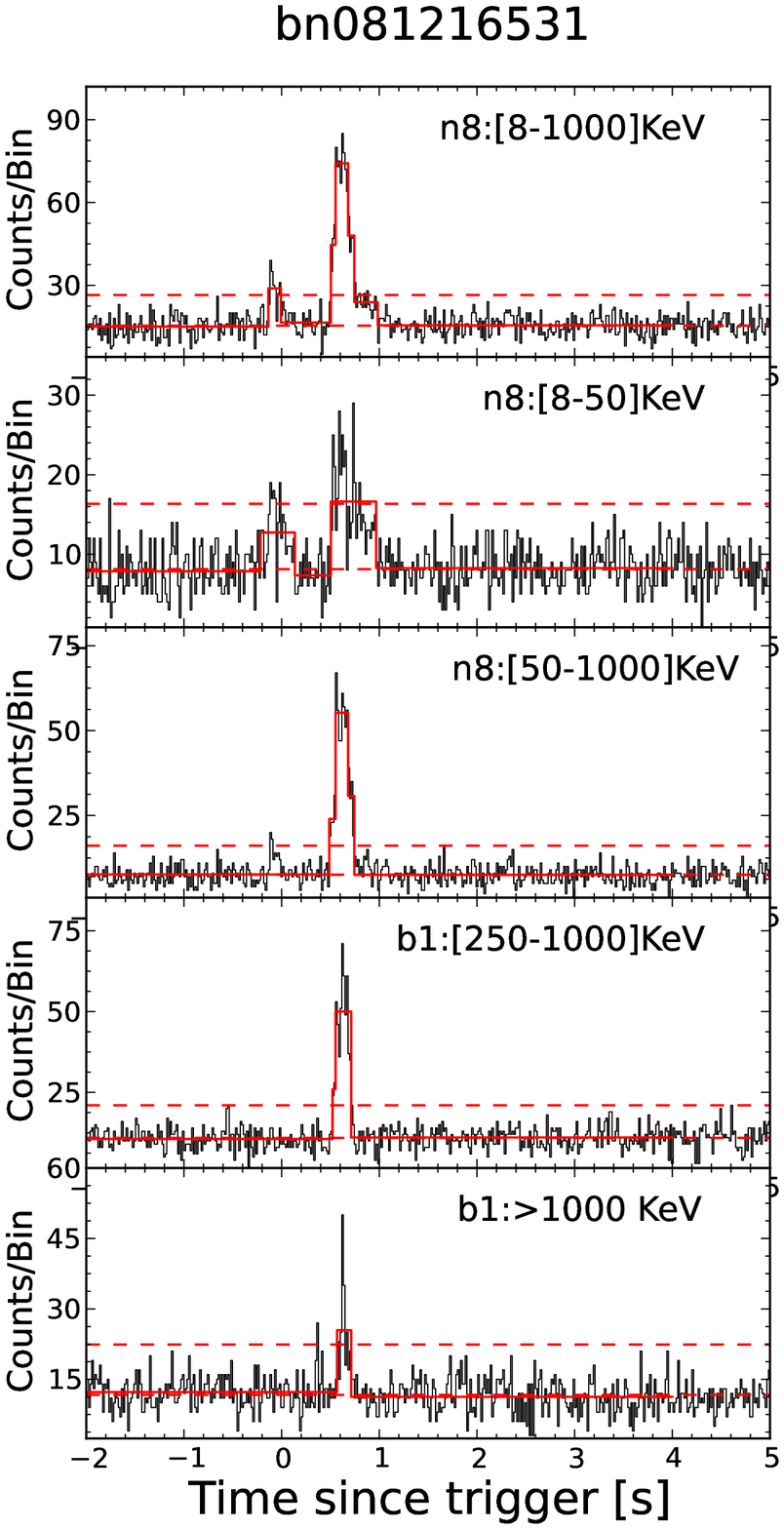}
\includegraphics[origin=c,angle=-90,scale=0.6,width=0.4\textwidth,height=0.4\textheight]{./f3c1SP.eps}
\includegraphics[origin=c,angle=0,scale=0.6,width=0.4\textwidth,height=0.4\textheight]{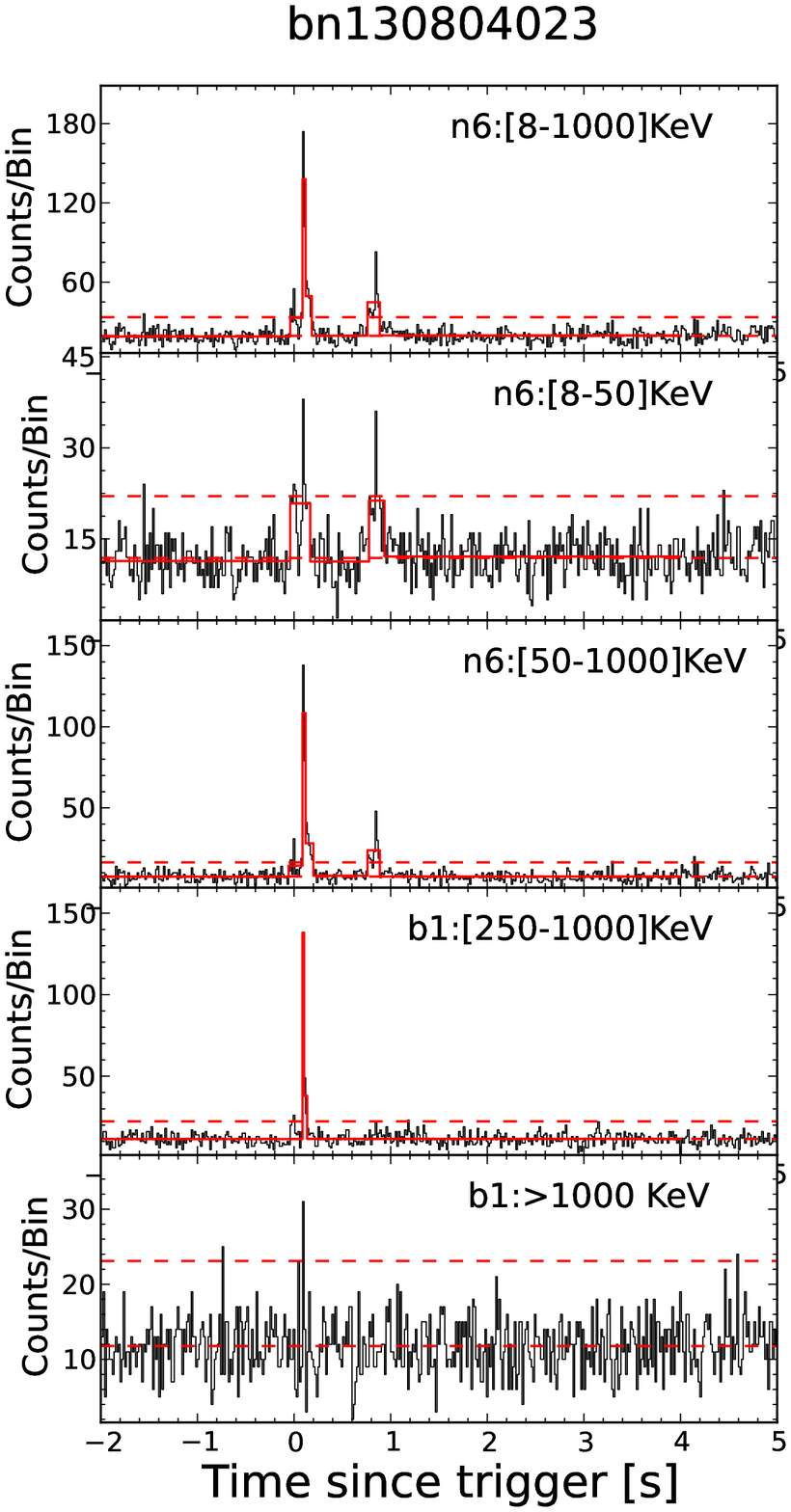}
\hfill
\includegraphics[origin=c,angle=-90,scale=0.6,width=0.4\textwidth,height=0.4\textheight]{./f3c2SP.eps}

\center{Fig. \ref{fig:LCexample}(b)---Continued.}
\end{figure*}

\begin{figure*}
\includegraphics[origin=c,angle=0,scale=0.6,width=0.4\textwidth,height=0.4\textheight]{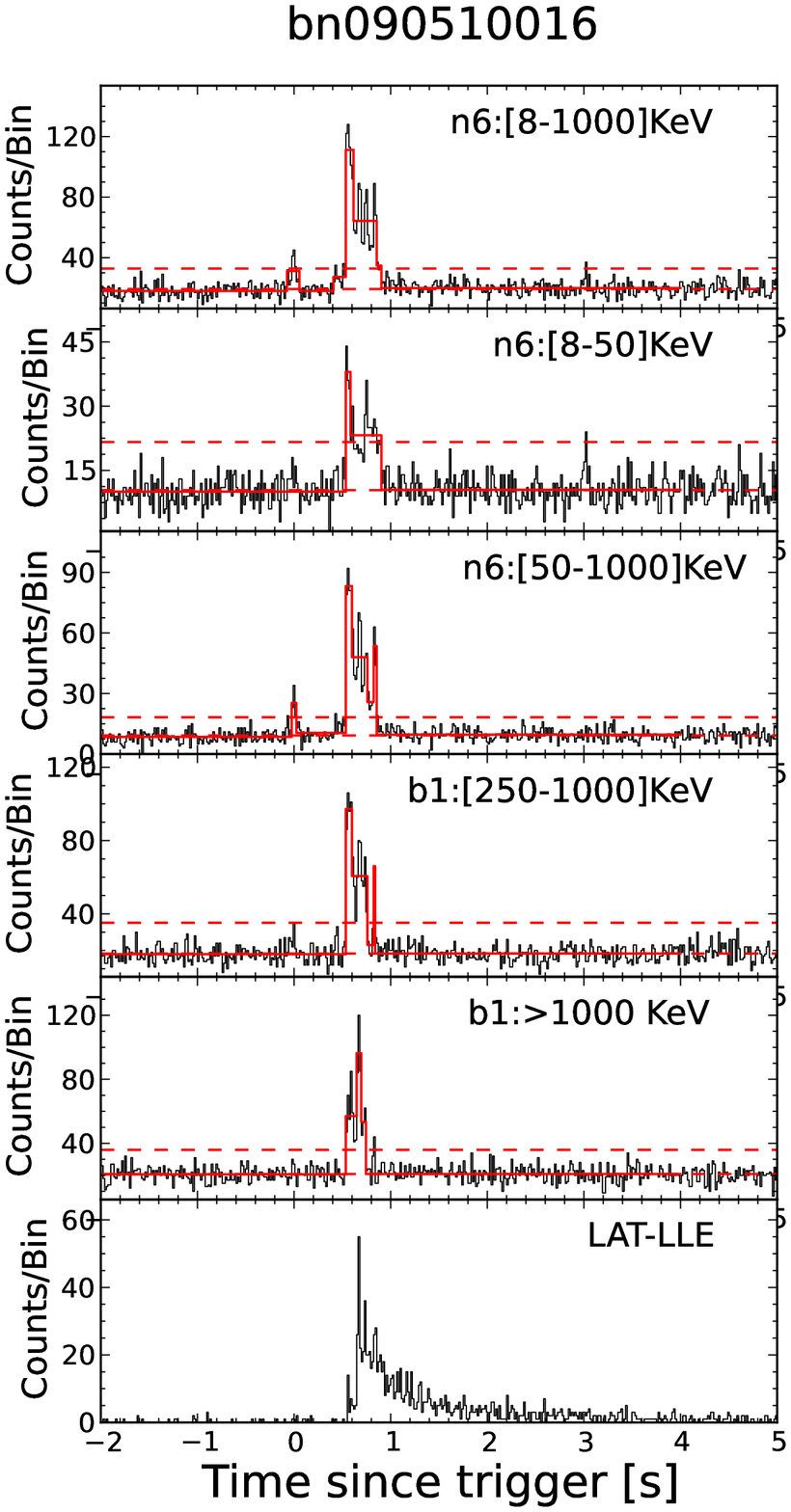}
\includegraphics[origin=c,angle=-90,scale=0.6,width=0.4\textwidth,height=0.4\textheight]{./f3fSP.eps}
\includegraphics[origin=c,angle=0,scale=0.6,width=0.4\textwidth,height=0.4\textheight]{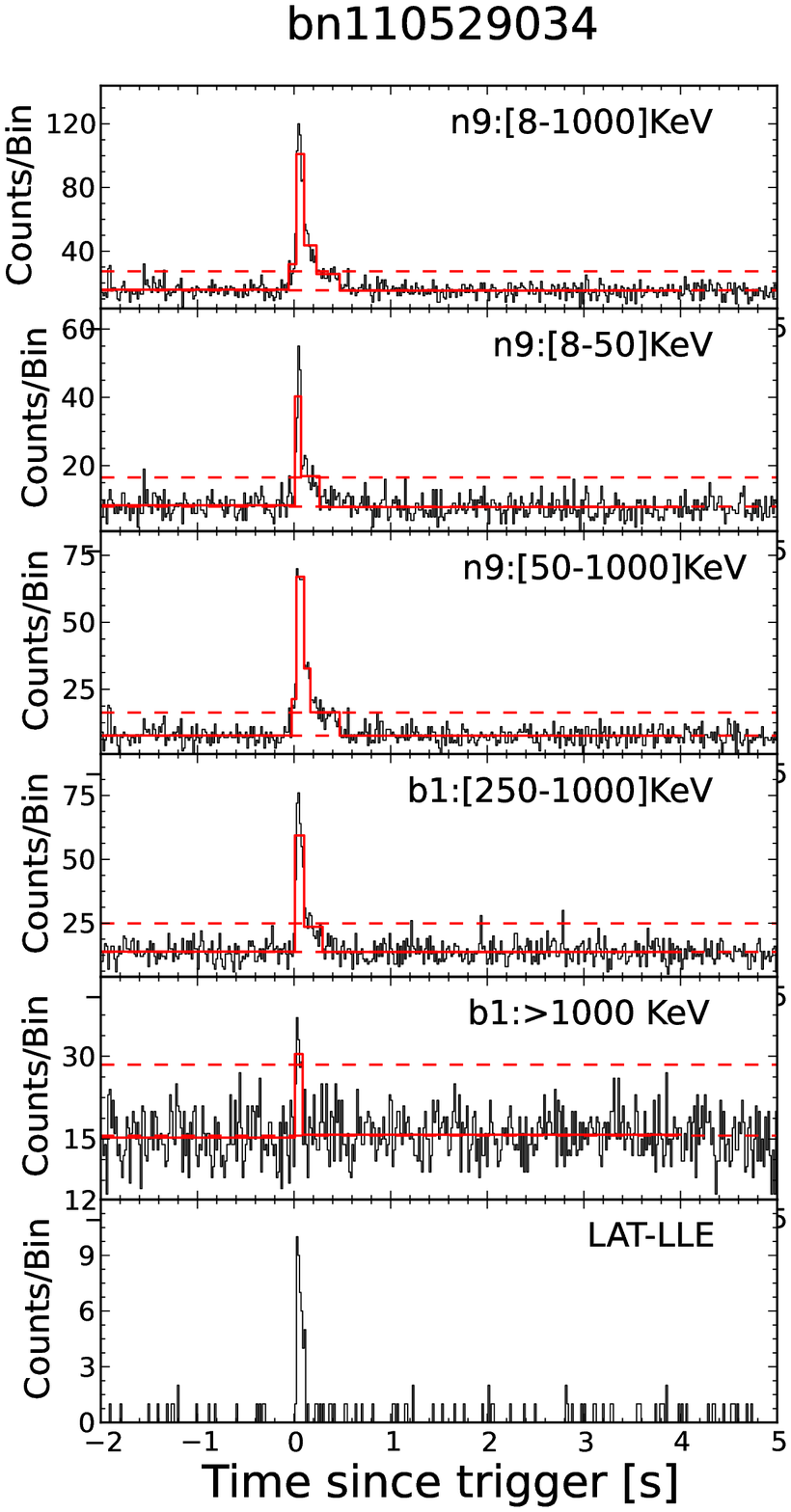}
\hfill
\includegraphics[origin=c,angle=-90,scale=0.6,width=0.4\textwidth,height=0.4\textheight]{./f3gSP.eps}

\center{Fig. \ref{fig:GBMLLE}---Continued.}
\end{figure*}

\begin{figure*}
\includegraphics[origin=c,angle=0,scale=0.6,width=0.4\textwidth,height=0.4\textheight]{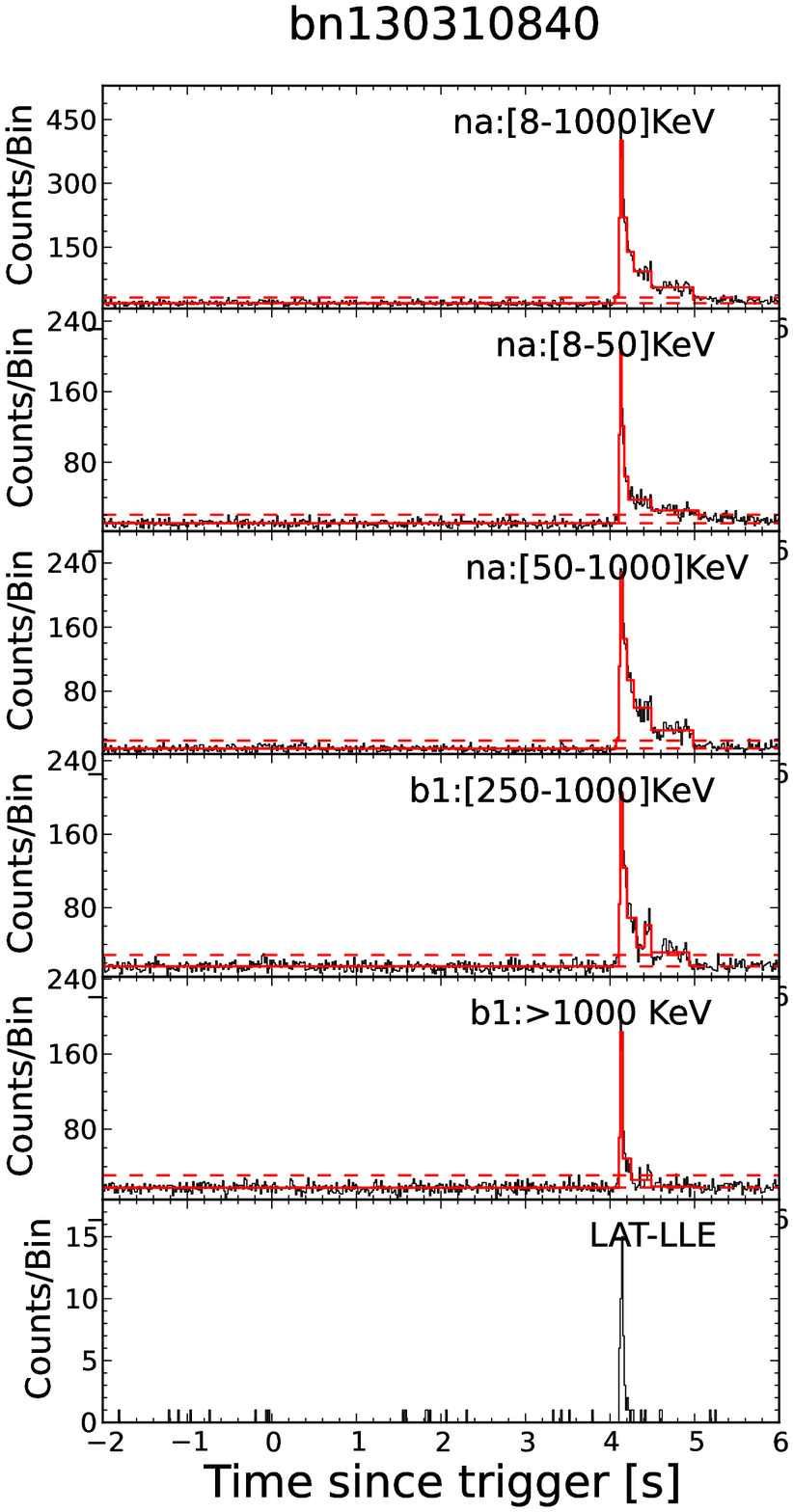}
\includegraphics[origin=c,angle=-90,scale=0.6,width=0.4\textwidth,height=0.4\textheight]{./f3hSP.eps}
\includegraphics[origin=c,angle=0,scale=0.6,width=0.4\textwidth,height=0.4\textheight]{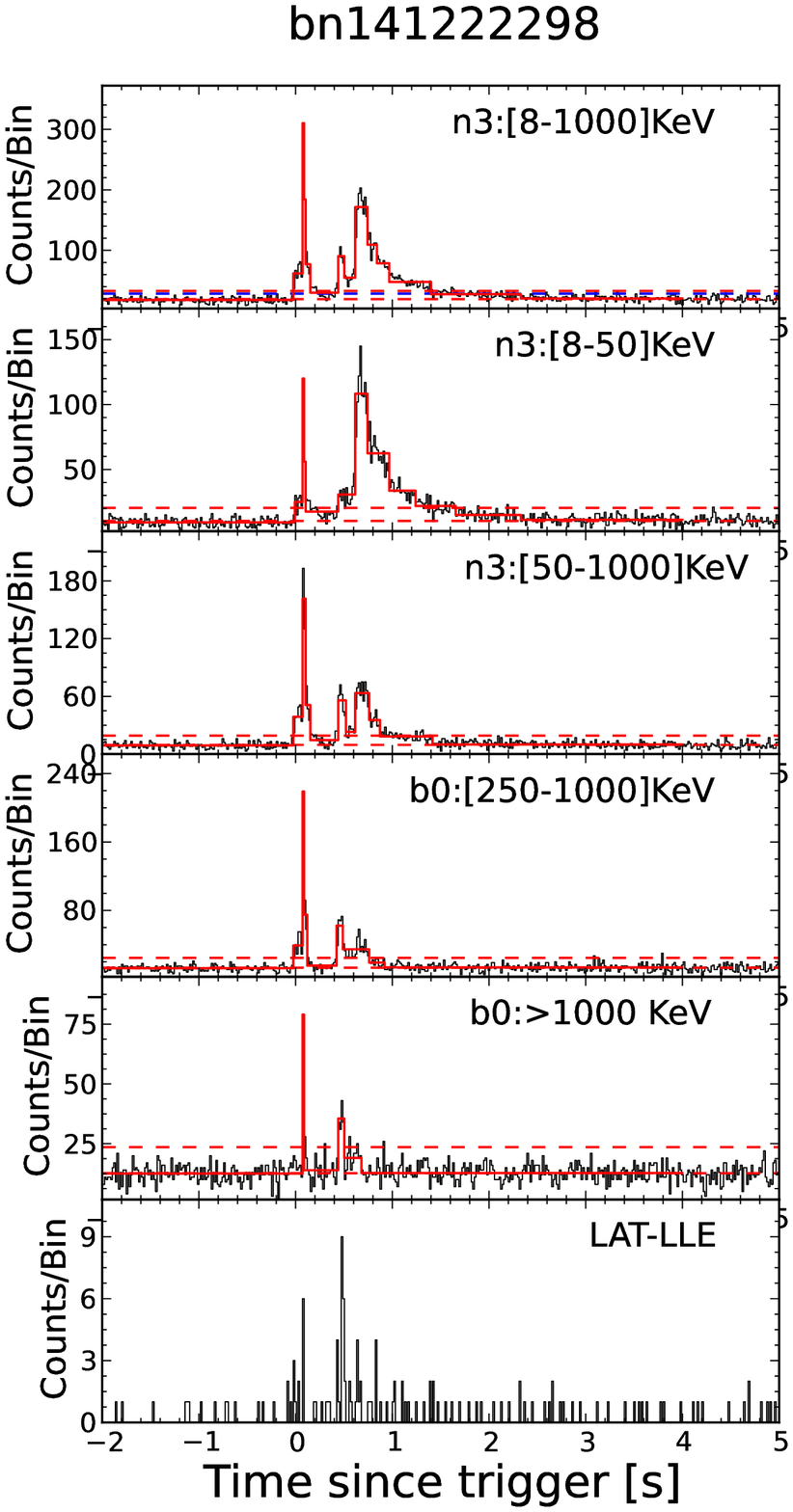}
\hfill
\includegraphics[origin=c,angle=-90,scale=0.6,width=0.4\textwidth,height=0.4\textheight]{./f3iSP.eps}

\center{Fig. \ref{fig:GBMLLE}---Continued.}
\end{figure*}

\begin{figure*}
\includegraphics[origin=c,angle=0,scale=0.6,width=0.4\textwidth,height=0.4\textheight]{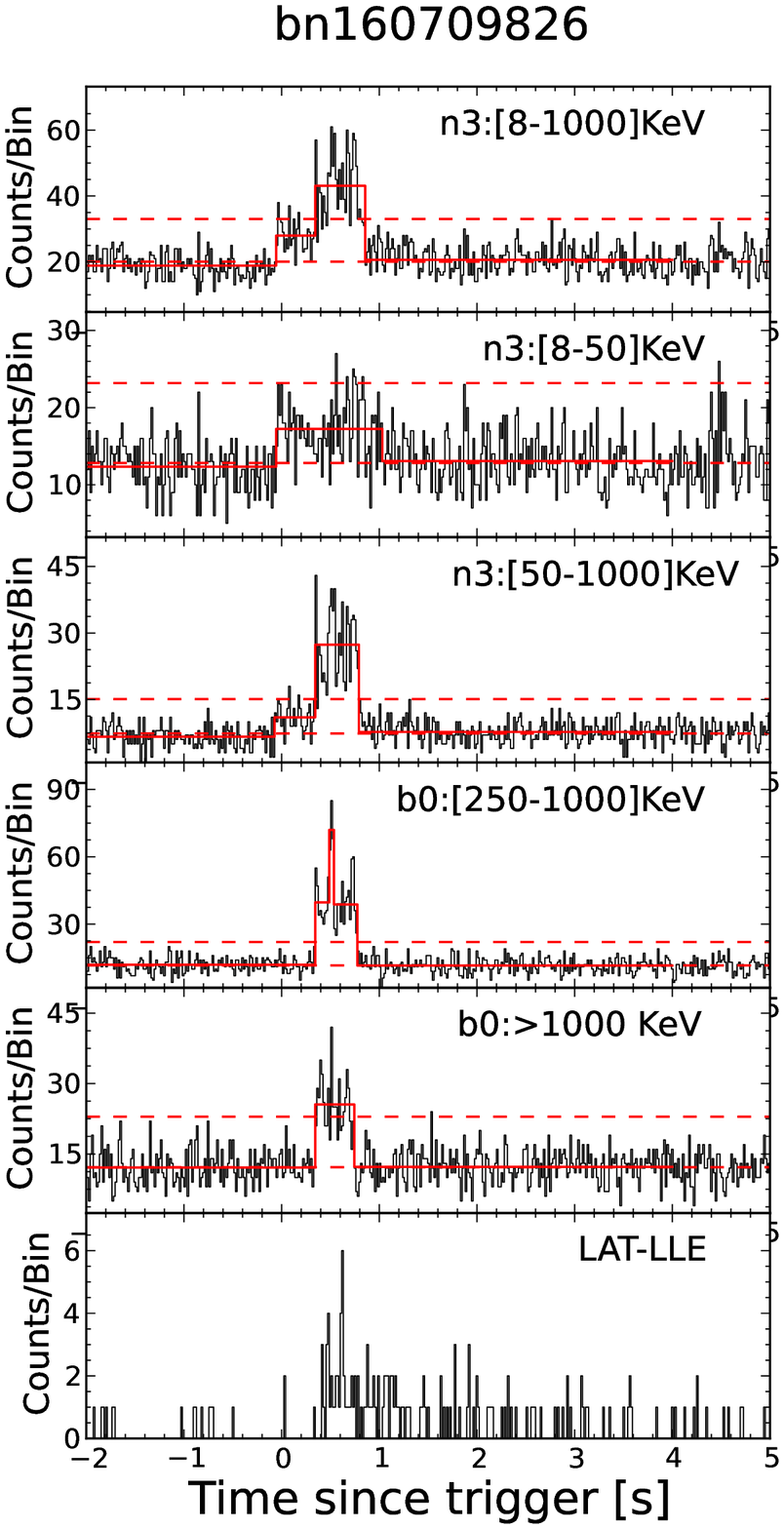}
\includegraphics[origin=c,angle=-90,scale=0.6,width=0.4\textwidth,height=0.4\textheight]{./f3jSP.eps}
\hfill
\center{Fig. \ref{fig:GBMLLE}---Continued.}
\end{figure*}

\clearpage

\begin{figure}
\includegraphics[angle=0,scale=0.6]{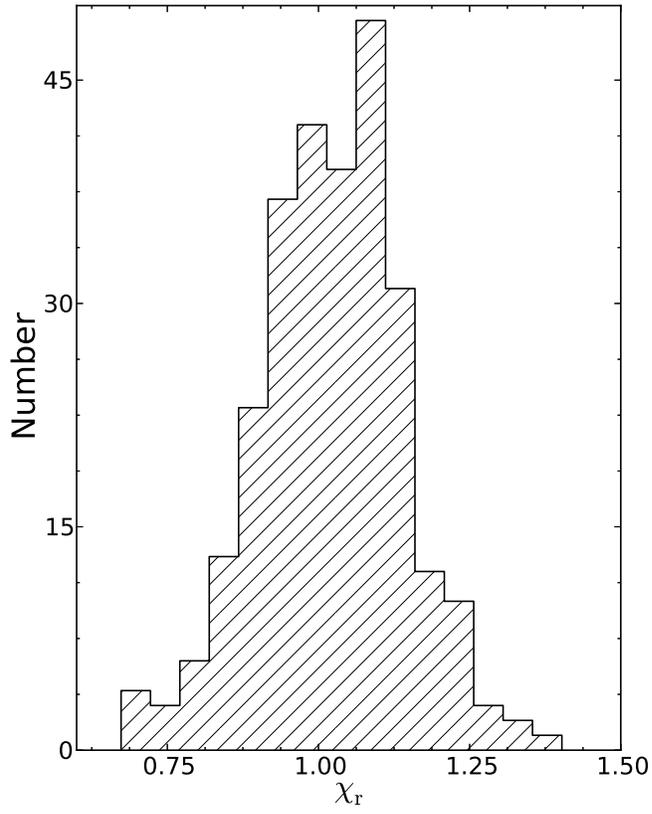}
\caption{Distribution of the reduced $\chi^2$ of our spectral fits.}
\label{fig:goodness}
\hfill
\end{figure}

\begin{figure}
\includegraphics[angle=0,scale=0.6]{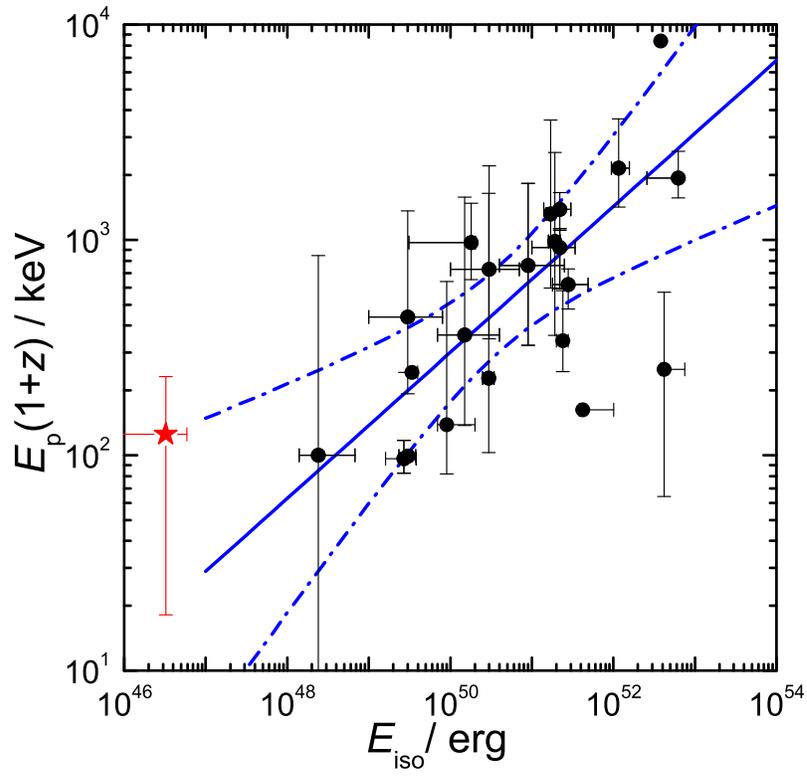}
\caption{$E_{\rm iso}$ as a function of $E_{\rm p}$ in the burst frame for a sample of sGRB taken from Zhang et al. (2009). The red star is GRB 170817A. The solid line is the Spearman linear fit together with its $2\sigma$ confidence level.}
\label{fig:goodness}
\end{figure}


\begin{thebibliography}{}
\bibitem[Abadie et al.(2012)]{2012PhRvD..85h2002A} Abadie, J., Abbott, B.~P., Abbott, R., et al.\ 2012, \prd, 85, 082002


\bibitem[Abadie et al.(2011)]{2011PhRvD..83l2005A} Abadie, J., Abbott, B.~P., Abbott, R., et al.\ 2011, \prd, 83, 122005


\bibitem[Abadie et al.(2010)]{2010PhRvD..81j2001A} Abadie, J., Abbott, B.~P., Abbott, R., et al.\ 2010, \prd, 81, 102001


\bibitem[Abbott et al.(2016)]{2016PhRvL.116f1102A} Abbott, B.~P., Abbott, R., Abbott, T.~D., et al.\ 2016a, Physical Review Letters, 116, 061102


\bibitem[Abbott et al.(2016)]{2016PhRvL.116x1102A} Abbott, B.~P., Abbott, R., Abbott, T.~D., et al.\ 2016b, Physical Review Letters, 116, 241102


\bibitem[Abbott et al.(2016)]{2016PhRvL.116x1103A} Abbott, B.~P., Abbott, R., Abbott, T.~D., et al.\ 2016c, Physical Review Letters, 116, 241103


\bibitem[Abbott et al.(2017)]{2017PhRvD..95f2003A} Abbott, B.~P., Abbott, R., Abbott, T.~D., et al.\ 2017a, \prd, 95, 062003


\bibitem[Abbott et al.(2017)]{2017PhRvL.118v1101A} Abbott, B.~P., Abbott, R., Abbott, T.~D., et al.\ 2017b, Physical Review Letters, 118, 221101


\bibitem[Abbott et al.(2017)]{2017PhRvL.119v1101A} Abbott, B.~P., Abbott, R., Abbott, T.~D., et al.\ 2017c, Physical Review Letters, 119, 161101


\bibitem[Ackermann et al.(2010)]{2010ApJ...716.1178A} Ackermann, M., Asano, K., Atwood, W.~B., et al.\ 2010, \apj, 716, 1178


\bibitem[Ashman et al.(1994)]{1994AJ....108.2348A} Ashman, K.~M., Bird, C.~M., \& Zepf, S.~E.\ 1994, \aj, 108, 2348


\bibitem[Atwood et al.(2009)]{2009ApJ...697.1071A} Atwood, W.~B., Abdo, A.~A., Ackermann, M., et al.\ 2009, \apj, 697, 1071


\bibitem[Band et al.(1993)]{1993ApJ...413..281B} Band, D., Matteson, J., Ford, L., et al.\ 1993, \apj, 413, 281


\bibitem[Barthelmy et al.(2005)]{2005Natur.438..994B} Barthelmy, S.~D., Chincarini, G., Burrows, D.~N., et al.\ 2005, \nat, 438, 994


\bibitem[Berger et al.(2013)]{2013ApJ...774L..23B} Berger, E., Fong, W., \& Chornock, R.\ 2013, \apjl, 774, L23


\bibitem[Berger et al.(2005)]{2005Natur.438..988B} Berger, E., Price, P.~A., Cenko, S.~B., et al.\ 2005, \nat, 438, 988


\bibitem[Berger(2014)]{2014ARA&A..52...43B} Berger, E.\ 2014, \araa, 52, 43


\bibitem[Burrows et al.(2005)]{2005Sci...309.1833B} Burrows, D.~N., Romano, P., Falcone, A., et al.\ 2005, Science, 309, 1833


\bibitem[Connaughton et al.(2016)]{2016ApJ...826L...6C} Connaughton, V., Burns, E., Goldstein, A., et al.\ 2016, \apjl, 826, L6


\bibitem[Dai et al.(2017)]{2017ApJ...838L...7D} Dai, Z.~G., Wang, J.~S., \& Yu, Y.~W.\ 2017, \apjl, 838, L7


\bibitem[Dai et al.(2006)]{2006Sci...311.1127D} Dai, Z.~G., Wang, X.~Y., Wu, X.~F., \& Zhang, B.\ 2006, Science, 311, 1127


\bibitem[Dichiara et al.(2013)]{2013ApJ...777..132D} Dichiara, S., Guidorzi, C., Frontera, F., \& Amati, L.\ 2013, \apj, 777, 132


\bibitem[Eichler et al.(1989)]{1989Natur.340..126E} Eichler, D., Livio, M., Piran, T., \& Schramm, D.~N.\ 1989, \nat, 340, 126


\bibitem[Fan \& Wei(2005)]{2005MNRAS.364L..42F} Fan, Y.~Z., \& Wei, D.~M.\ 2005, \mnras, 364, L42


\bibitem[Fong et al.(2017)]{2017arXiv171005438F} Fong, W., Berger, E., Blanchard, P.~K., et al.\ 2017, arXiv:1710.05438


\bibitem[Fong et al.(2010)]{2010ApJ...708....9F} Fong, W., Berger, E., \& Fox, D.~B.\ 2010, \apj, 708, 9


\bibitem[Gao et al.(2017)]{2017ApJ...837...50G} Gao, H., Zhang, B., L{\"u}, H.-J., \& Li, Y.\ 2017, \apj, 837, 50


\bibitem[Gehrels et al.(2005)]{2005Natur.437..851G} Gehrels, N., Sarazin, C.~L., O'Brien, P.~T., et al.\ 2005, \nat, 437, 851


\bibitem[Goldstein et al.(2017)]{2017arXiv171005446G} Goldstein, A., Veres, P., Burns, E., et al.\ 2017, arXiv:1710.05446


\bibitem[Hallinan et al.(2017)]{2017arXiv171005435H} Hallinan, G., Corsi, A., Mooley, K.~P., et al.\ 2017, arXiv:1710.05435


\bibitem[Hu et al.(2014)]{2014ApJ...789..145H} Hu, Y.-D., Liang, E.-W., Xi, S.-Q., et al.\ 2014, \apj, 789, 145


\bibitem[Keane et al.(2016)]{2016Natur.530..453K} Keane, E.~F., Johnston, S., Bhandari, S., et al.\ 2016, \nat, 530, 453


\bibitem[L{\"u} et al.(2010)]{2010ApJ...725.1965L} L{\"u}, H.-J., Liang, E.-W., Zhang, B.-B., \& Zhang, B.\ 2010, \apj, 725, 1965


\bibitem[L{\"u} et al.(2014)]{2014MNRAS.442.1922L} L{\"u}, H.-J., Zhang, B., Liang, E.-W., Zhang, B.-B., \& Sakamoto, T.\ 2014, \mnras, 442, 1922


\bibitem[Levan et al.(2017)]{2017arXiv171005444L} Levan, A.~J., Lyman, J.~D., Tanvir, N.~R., et al.\ 2017, arXiv:1710.05444


\bibitem[Li \& Paczy{\'n}ski(1998)]{1998ApJ...507L..59L} Li, L.-X., \& Paczy{\'n}ski, B.\ 1998, \apjl, 507, L59


\bibitem[Li et al.(2017)]{2017arXiv171006065L} Li, T., Xiong, S., Zhang, S., et al.\ 2017, arXiv:1710.06065


\bibitem[Liang \& Dai(2004)]{2004ApJ...608L...9L} Liang, E.~W., \& Dai, Z.~G.\ 2004, \apjl, 608, L9


\bibitem[Lorimer et al.(2007)]{2007Sci...318..777L} Lorimer, D.~R., Bailes, M., McLaughlin, M.~A., Narkevic, D.~J., \& Crawford, F.\ 2007, Science, 318, 777


\bibitem[Meegan et al.(1992)]{1992Natur.355..143M} Meegan, C.~A., Fishman, G.~J., Wilson, R.~B., et al.\ 1992, \nat, 355, 143


\bibitem[Meegan et al.(2009)]{2009ApJ...702..791M} Meegan, C., Lichti, G., Bhat, P.~N., et al.\ 2009, \apj, 702, 791-804


\bibitem[Metzger \& Berger(2012)]{2012ApJ...746...48M} Metzger, B.~D., \& Berger, E.\ 2012, \apj, 746, 48


\bibitem[Metzger et al.(2010)]{2010MNRAS.406.2650M} Metzger, B.~D., Mart{\'{\i}}nez-Pinedo, G., Darbha, S., et al.\ 2010, \mnras, 406, 2650


\bibitem[Nakar(2007)]{2007PhR...442..166N} Nakar, E.\ 2007, \physrep, 442, 166


\bibitem[Nicholl et al.(2017)]{2017arXiv171005456N} Nicholl, M., Berger, E., Kasen, D., et al.\ 2017, arXiv:1710.05456


\bibitem[Paczy{\'n}ski(1986)]{1986Natur.324..392P} Paczy{\'n}ski, P.\ 1986, \nat, 324, 392


\bibitem[Paczynski(1991)]{1991AcA....41..257P} Paczynski, B.\ 1991, \actaa, 41, 257


\bibitem[Perna et al.(2016)]{2016ApJ...821L..18P} Perna, R., Lazzati, D., \& Giacomazzo, B.\ 2016, \apjl, 821, L18


\bibitem[Preece et al.(2000)]{2000ApJS..126...19P} Preece, R.~D., Briggs, M.~S., Mallozzi, R.~S., et al.\ 2000, \apjs, 126, 19


\bibitem[Qin et al.(2010)]{2010MNRAS.406..558Q} Qin, S.-F., Liang, E.-W., Lu, R.-J., Wei, J.-Y., \& Zhang, S.-N.\ 2010, \mnras, 406, 558


\bibitem[Qin et al.(2013)]{2013ApJ...763...15Q} Qin, Y., Liang, E.-W., Liang, Y.-F., et al.\ 2013, \apj, 763, 15


\bibitem[Savchenko et al.(2017)]{2017arXiv171005449S} Savchenko, V., Ferrigno, C., Kuulkers, E., et al.\ 2017, arXiv:1710.05449


\bibitem[Sun et al.(2017)]{2017ApJ...835....7S} Sun, H., Zhang, B., \& Gao, H.\ 2017, \apj, 835, 7


\bibitem[Tanvir et al.(2013)]{2013Natur.500..547T} Tanvir, N.~R., Levan, A.~J., Fruchter, A.~S., et al.\ 2013, \nat, 500, 547


\bibitem[Thornton et al.(2013)]{2013Sci...341...53T} Thornton, D., Stappers, B., Bailes, M., et al.\ 2013, Science, 341, 53


\bibitem[Totani(2013)]{2013PASJ...65L..12T} Totani, T.\ 2013, \pasj, 65, L12


\bibitem[Troja et al.(2017)]{2017arXiv171005433T} Troja, E., Piro, L., van Eerten, H., et al.\ 2017, arXiv:1710.05433


\bibitem[Yang et al.(2015)]{2015NatCo...6E7323Y} Yang, B., Jin, Z.-P., Li, X., et al.\ 2015, Nature Communications, 6, 7323


\bibitem[Yu et al.(2013)]{2013ApJ...776L..40Y} Yu, Y.-W., Zhang, B., \& Gao, H.\ 2013, \apjl, 776, L40


\bibitem[Zhang et al.(2017)]{2017arXiv171005851Z} Zhang, B.-B., Zhang, B., Sun, H., et al.\ 2017, arXiv:1710.05851


\bibitem[Zhang et al.(2011)]{2011ApJ...730..141Z} Zhang, B.-B., Zhang, B., Liang, E.-W., et al.\ 2011, \apj, 730, 141


\bibitem[Zhang(2016)]{2016ApJ...827L..31Z} Zhang, B.\ 2016, \apjl, 827, L31


\bibitem[Zhang(2014)]{2014ApJ...780L..21Z} Zhang, B.\ 2014, \apjl, 780, L21


\bibitem[Zhang et al.(2006)]{2006ApJ...642..354Z} Zhang, B., Fan, Y.~Z., Dyks, J., et al.\ 2006, \apj, 642, 354


\bibitem[Zhang et al.(2007)]{2007ApJ...655L..25Z} Zhang, B., Zhang, B.-B., Liang, E.-W., et al.\ 2007, \apjl, 655, L25


\bibitem[Zhang et al.(2009)]{2009ApJ...703.1696Z} Zhang, B., Zhang, B.-B., Virgili, F.~J., et al.\ 2009, \apj, 703, 1696


\bibitem[Zhang et al.(2016)]{2016arXiv160402537Z} Zhang, S.-N., Liu, Y., Yi, S., Dai, Z., \& Huang, C.\ 2016, arXiv:1604.02537


\end{thebibliography}
\end{document}